\documentclass[twocolumn]{autart}

\usepackage{xcolor}
\usepackage{cite}
\usepackage{amsmath,amssymb,amsfonts}
\usepackage{algorithm}
\usepackage{algpseudocode}
\usepackage{graphicx}
\usepackage{textcomp}
\graphicspath{{figure/}}
\newif\ifshowchanges
\showchangesfalse
\ifshowchanges\else\renewcommand{\textcolor}[2]{#2}\fi

\usepackage{tikz,tkz-euclide}
\usetikzlibrary{angles,quotes}

\newtheorem{theorem}{Theorem}
\newtheorem{lemma}{Lemma}
\newtheorem{remark}{Remark}
\newtheorem{assume}{Assumption}
\newtheorem{define}{Definition}
\newtheorem{corollary}{Corollary}
\newtheorem{problem}{Problem}

\usepackage[switch,columnwise]{lineno} %line number
\begin{document}

\begin{frontmatter}

\title{Resilient Angle-Constrained Formation Control and Localization Under DoS Attacks \\(Extended Version)}

\thanks[footnoteinfo]{This paper was not presented at any IFAC meeting. Corresponding author: Lihua Xie.}

\author[NTU]{Nhat-Minh Le-Phan}\ead{phannhat001@e.ntu.edu.sg},
\author[NTU]{Ziliang Lyu}\ead{ziliang.lyu@ntu.edu.sg},
\author[NTU,VinUni]{Lihua Xie}\ead{elhxie@ntu.edu.sg; lihua.x@vinuni.edu.vn}

\address[NTU]{School of Electrical and Electronic Engineering, Nanyang Technological University, Singapore 639798}
\address[VinUni]{NTU--VinUni Joint Research Centre for Robotics and Embodied Intelligence (CREI), VinUniversity, Hanoi, Vietnam 100000}

\begin{keyword}
Formation control; angle rigidity; network localization; denial-of-service attacks; resilient control; distributed observer.
\end{keyword}

\begin{abstract}
This paper studies resilient angle-constrained leader--follower formation control and localization of multi-agent systems under asynchronous denial-of-service (DoS) attacks, which block individual sensing and communication links. To preserve the well-posedness of angle-based localization, an all-or-nothing triangle integrity protocol (ANTIP) is proposed to convert edge-wise attacks into triangle-wise ones, and DoS attacks are classified into localizability-maintained and localizability-broken attacks. First, with relative position measurements over attack-free links, resilient formation tracking laws are designed for leaders with constant velocities and with time-varying velocities comprising translational, rotational, and scaling motions. These laws guarantee asymptotic tracking and finite-time recovery of tracking, respectively, when the duration of localizability-broken attacks is suitably bounded. Second, with only angle measurements and communications, a resilient simultaneous localization and formation control (RSLAF) scheme is developed. It combines a finite-time distributed observer, which estimates the exosystem matrices and states under asynchronous DoS attacks, with a perturbation-based localization law that prevents localization from getting stuck at degenerate configurations. The theoretical results are rigorously proved and illustrated by numerical simulations.
\end{abstract}

\end{frontmatter}
\thispagestyle{plain}   % <- add this line only

\section{Introduction}
\label{sec:introduction}
Formation control of multi-agent systems has attracted significant interest owing to its applications in robotics and autonomous vehicles and its relevance to collective animal behavior. Unlike consensus-based schemes, in which followers maintain predefined positions or orientations relative to a leader, rigidity-based methods specify the desired formation by geometric constraints among neighboring agents, thereby enhancing autonomy and reducing implementation costs \cite{Ahn2020book}. These constraints are typically formulated in terms of relative positions \cite{LAFFERRIERE2005sysl,Fax2004TAC}, distances \cite{krick2008CDC}, ratios of distances \cite{Cao2020TAC}, bearings \cite{Zhao2015tac,trinh2021auto}, or angles \cite{JING2019automatica,Chen2021TAC}, and rigidity theory provides the foundation for analyzing the uniqueness and stability of the desired formation \cite{Ahn2020book}. The type of constraint also determines the admissible maneuvers of the formation: relative positions are invariant only to translations and require a common coordinate frame, distances are invariant to translations and rotations, and bearings are invariant to translations and scalings. In contrast, angle constraints are invariant to translations, rotations, and scalings. This invariance offers additional flexibility in formation design and maneuvering, making angle-constrained formation control an active research area.

Angle-constrained formation control was first studied in \cite{eren2003cdc}, and its theoretical foundation was established by angle rigidity theory \cite{JING2019automatica}. Existing angle-only control laws for single-integrator \cite{chen2021TAC_local} and double-integrator \cite{Chen2022TCNS} agents guarantee only local convergence. To achieve global convergence, \cite{CHEN2022auto} developed triangular angle rigidity, which was first applied to angle-based network localization and subsequently to globally convergent formation control of stationary single-integrator agents \cite{chenTAC2023}. This framework has since been extended to maneuvering double-integrator formations \cite{chen2022ieeecaa}, formations subject to disturbances \cite{PENG2025auto}, and the merging and splitting of formations \cite{LIU2026auto_mergesplit}. Since many applications, such as search and rescue, require agents' positions in a global frame, \cite{CHEN2022auto_localization} addressed simultaneous localization and formation control (SLAF) with angle measurements at the cost of additional inter-agent communication, and \cite{FANG2025auto} extended SLAF to bearing and ratio-of-distance measurements.

Communication is essential in multi-agent systems, where agents interact with their neighbors to accomplish collective objectives. However, communication networks are vulnerable to malicious attacks \cite{TEIXEIRA2015auto}. Denial-of-Service (DoS) attacks are among the most common cyber-attacks against networked systems. They disrupt information exchange by blocking communication channels and causing packet losses. As a result, the system may suffer from reduced performance, failure to achieve the desired coordination objectives, and even instability of the closed-loop system. These challenges have motivated extensive research on resilient control under DoS attacks. The pioneering studies \cite{Persis2015TAC,FENG2017auto} addressed resilient control of linear systems under DoS attacks and established input-to-state stability guarantees. A key feature of these studies is that DoS attacks are characterized only by their frequency and duration, without imposing restrictive structural assumptions on the attack signals. Building on these foundational results, numerous advances have been made in the resilient control of cyber-physical systems under DoS attacks. In the context of multi-agent systems, resilient consensus has been extensively studied for both leaderless and leader--follower networks \cite{Feng2016ijrnc,zuo2022tsmc,zhang2024tcyb,DENG2022auto}. Notably, \cite{Feng2016ijrnc} introduced a classification of DoS attacks into connectivity-maintained and connectivity-broken cases.

Most existing approaches to resilient formation control under DoS attacks rely on consensus-based strategies \cite{zhou2024tcyb,yan2024tcyb,yang2025TSIPN,Deng2020TCNS,Peng2025ijrnc}. In these approaches, resilient distributed observers are used to estimate the leader's state, which is typically generated by an exosystem whose matrices are known to all or only a subset of followers. For example, \cite{Deng2020TCNS,Peng2025ijrnc} design resilient distributed observers that enable all followers to reconstruct the exosystem matrices and asymptotically estimate the leader's trajectory. To improve the convergence speed, finite-time, fixed-time, and prescribed-time resilient leader-following consensus algorithms have been studied in \cite{Ni2022TCSI,Ye2023CSYS,ZHANG2025auto,YE2026auto}. However, many of these works, as well as a large body of existing results \cite{liu2023ijrnc,yan2024tcyb,Ding2024tfuz}, consider synchronous DoS attacks, under which all communication links are disconnected simultaneously. This assumption may be restrictive in practice, since an adversary can target individual links or agents.

Rigidity-based resilient formation control and localization under DoS attacks lag behind consensus-based approaches and remain challenging. To address this gap, this paper develops resilient angle-constrained formation control and localization algorithms that enable agents to track desired time-varying formations in the presence of asynchronous DoS attacks. The main contributions of this work are summarized as follows.
\begin{enumerate}
\item \textcolor{blue}{To the best of our knowledge,} this paper presents the first study of resilient angle-constrained formation control for multi-agent systems subject to DoS attacks. We propose a unified framework built upon two key ingredients. First, to guarantee a well-defined notion of localizability under adversarial network conditions, we introduce the All-or-Nothing Triangle Integrity Protocol (ANTIP). Second, we categorize DoS attacks into two classes, namely localizability-maintained and localizability-broken attacks. This classification allows the impact of DoS attacks to be characterized by the total time during which localizability is lost.
\item Furthermore, this work investigates the problem of resilient simultaneous localization and formation control (RSLAF) under DoS attacks using inter-agent angle measurements and communications. A finite-time resilient distributed observer is first constructed to estimate the reference trajectories. Leveraging these estimates, we develop resilient angle-based localization and control algorithms that allow agents to simultaneously track a desired time-varying formation and achieve self-localization in a global coordinate frame in the presence of DoS attacks.
\item In contrast to \cite{Ni2022TCSI,Ye2023CSYS,ZHANG2025auto,YE2026auto,liu2023ijrnc,yan2024tcyb,Ding2024tfuz}, which assume a uniform attack model over all communication channels, this work considers asynchronous DoS attacks, i.e., attacks that occur independently across inter-agent communication channels. In particular, the proposed resilient distributed observer in Section~\ref{sec:RSLAF} guarantees finite-time estimation of both the exosystem matrices and states under asynchronous DoS attacks. To the best of our knowledge, this result has not been reported in the existing literature.
\end{enumerate}

The remainder of this paper is organized as follows. Section~\ref{sec:background} presents the necessary background on triangular angle rigidity and DoS attacks, followed by the problem formulation in Section~\ref{subsec:prob_form}. Section~\ref{sec:formation} studies the resilient angle-constrained formation tracking problem. Section~\ref{sec:RSLAF} investigates the resilient simultaneous localization and formation control problem. Simulation results are presented in Section~\ref{sec:sim}. Finally, conclusions are drawn in Section~\ref{sec:conclude}.

\textbf{Notation:}
Throughout this paper, $\mathbb{R}$ and $\mathbb{N}$ denote the sets of real and natural numbers, respectively\textcolor{blue}{, and $|\cdot|$ denotes the absolute value of a scalar or the cardinality of a set}. For $\mathbf{x}\in\mathbb{R}^n$, $\|\mathbf{x}\|=\sqrt{\mathbf{x}^\top\mathbf{x}}$, $\|\mathbf{x}\|_1$, and $\|\mathbf{x}\|_\infty$ denote its Euclidean, $1$-, and infinity norms, and $\mathrm{diag}(\mathbf{x})$ denotes the diagonal matrix with diagonal entries from $\mathbf{x}$. For a matrix, $\|\cdot\|$ and $\|\cdot\|_F$ denote the induced $2$-norm and the Frobenius norm, and, if it is symmetric, $\lambda_{\max}(\cdot)$ and $\lambda_{\min}(\cdot)$ denote its maximum and minimum eigenvalues. Moreover, $\mathrm{blkdiag}\{\cdot\}$, $\mathrm{null}(\cdot)$, $\mathrm{span}(\cdot)$, $\otimes$, and $\mathrm{sgn}(\cdot)$ denote a block-diagonal matrix, the null space, the span of a set of vectors, the Kronecker product, and the signum function, respectively. The operator $\operatorname{vec}(\cdot)$ stacks the columns of a matrix into a column vector, and $M_p^q(\cdot)$ denotes its inverse, i.e., $\mathbf{M}=M_p^q(\mathbf{x})\in\mathbb{R}^{p\times q}$ satisfies $\operatorname{vec}(\mathbf{M})=\mathbf{x}$ for $\mathbf{x}\in\mathbb{R}^{pq}$. In addition, $\mathcal{R}(\theta)=\begin{bmatrix}\cos\theta&-\sin\theta\\ \sin\theta&\cos\theta\end{bmatrix}$ denotes the rotation matrix, and $D_+$ denotes the right-hand derivative. Finally, $\mathbf{1}$, $\mathbf{0}$, and $\mathbf{I}$ denote the all-ones vector, the zero vector, and the identity matrix of appropriate dimensions.
\section{Preliminaries and Background}\label{sec:background}
This section reviews the preliminaries on graph theory, triangular angle rigidity, and DoS attacks relevant to the problems considered in this paper.
\subsection{Graph Theory and Triangular Angle Rigidity}
In this paper, agent interactions are modeled by a communication graph $\mathcal{G} = (\mathcal{V}, \mathcal{E})$, where $\mathcal{V} = \{1,2,\ldots,n\}$ is the set of vertices (agents) and $\mathcal{E} \subseteq \mathcal{V} \times \mathcal{V}$ is the set of edges with $|\mathcal{E}| = m$. If the interconnections are symmetric, i.e., $(i,j) \in \mathcal{E} \iff (j,i) \in \mathcal{E}$, then $\mathcal{G}$ is said to be \emph{undirected}. The set of neighbors of agent $i$, namely all $j$ such that $(i,j) \in \mathcal{E}$, is denoted by $\mathcal{N}_i$. The \emph{adjacency matrix} $\mathbf{Ad} = [a_{ij}] \in \mathbb{R}^{n\times n}$ associated with $\mathcal{G}$ is defined such that $a_{ij} = 1$ if $(i,j) \in \mathcal{E}$ and $a_{ij} = 0$ otherwise. Thus, for an undirected graph, it holds that $a_{ij} = a_{ji}$. Given an arbitrary indexing and orientation of the edges in $\mathcal{E}$, the \emph{incidence matrix} $\mathbf{H} = [h_{kj}] \in \mathbb{R}^{m \times n}$ is defined such that $h_{ki} = -1$ if $i$ is the start vertex of edge $e_k$, $h_{ki} = +1$ if $i$ is the end vertex of $e_k$, and $h_{ki} = 0$ otherwise\footnote{In this paper, the notations $(i,j)$ and $k$ are used interchangeably to represent the corresponding edge.}. The \emph{Laplacian matrix} associated with an undirected graph $\mathcal{G}$ is defined as $\boldsymbol{\mathcal{L}} = \mathbf{H}^{\top}\mathbf{H}\in \mathbb{R}^{n\times n}$, which is positive semidefinite. An undirected graph $\mathcal{G}$ is said to be \emph{connected} if any two vertices in $\mathcal{V}$ are connected via a sequence of edges in $\mathcal{G}$.

% \textcolor{blue}{We adopt triangular angle rigidity \cite{CHEN2022auto}, whose angle constraints are linear in the agents' positions and thus enable globally convergent formation control and localization.}
\textcolor{blue}{We adopt triangular angle rigidity \cite{CHEN2022auto}, under which localization and formation control reduce to linear problems.} Consider a group of $n$ agents, with $n \geq 3$, interacting over a graph $\mathcal{G}$ in the plane. Each agent $i \in \mathcal{V}$ has position $\mathbf{p}_i = [x_i,y_i]^\top \in \mathbb{R}^2$. A formation is characterized by the pair $(\mathcal{G},\mathbf{p})$, where $\mathbf{p} = [\mathbf{p}_1^\top,\mathbf{p}_2^\top,\ldots,\mathbf{p}_n^\top]^\top \in \mathbb{R}^{2n}$ denotes the configuration vector of all agents. 
%In this paper, the notations $(i,j)$ and $k$ are used interchangeably to denote the corresponding edge.
For three distinct agents $k,i,$ and $j$, the interior angle $\alpha_{kij} \in [0,2\pi)$ formed by agents $k,i,$ and $j$ is defined as \cite{CHEN2022auto}
\begin{equation}\label{eq:angle_def}
\alpha_{kij} = \begin{cases} 
\arccos \left( \mathbf{g}_{ij}^\top \mathbf{g}_{ik} \right), \text{if } \mathbf{g}_{ij}^\top \mathcal{R}(\pi/2)\mathbf{g}_{ik} \ge 0, \\ 
2\pi - \arccos \left( \mathbf{g}_{ij}^\top \mathbf{g}_{ik} \right), \text{otherwise.} 
\end{cases}
\end{equation}
Here, $\mathbf{g}_{ij} = \frac{\mathbf{p}_j - \mathbf{p}_i}{\|\mathbf{p}_i - \mathbf{p}_j\|}$ denotes the \emph{bearing vector} from agent $i$ to agent $j$. By definition, $\alpha_{kij}$ denotes the angle formed by rotating the vector $\mathbf{g}_{ik}$ to $\mathbf{g}_{ij}$ in the counterclockwise direction. 
%It can be observed that $\alpha_{kij}$ is invariant with respect to the local coordinate system $^i\Sigma$ of each agent $i$. Therefore, it can be locally measured by each agent, even when the agents' coordinate systems are not synchronized. 

For a vertex set $\mathcal{V}$, a triplet $(k,i,j)$ of pairwise distinct vertices represents the angle constraint $\alpha_{kij}$, and an \emph{angle set} $\mathcal{A}\subseteq\mathcal{V}\times\mathcal{V}\times\mathcal{V}$ is a set of such triplets. The combination of $\mathcal{V}$, $\mathcal{A}$, and a configuration $\mathbf{p}$, denoted by $\mathbb{A}(\mathcal{V},\mathcal{A},\mathbf{p})$, is called an \emph{angularity}. If $(j,k,i),(k,i,j)\in\mathcal{A}$ for every $(i,j,k)\in\mathcal{A}$, then $\mathcal{A}$ is called a \emph{triangular angle set} and $\mathbb{A}(\mathcal{V},\mathcal{A},\mathbf{p})$ a \emph{triangular angularity}. Moreover, let $\bar{\mathcal{A}}\subseteq\mathcal{A}$ contain exactly one of $(i,j,k)$, $(j,k,i)$, and $(k,i,j)$ for each triangle, so that each triangle is represented once.
% \textcolor{red}{Two angularities $\mathbb{A}(\mathcal{V},\mathcal{A},\mathbf{p})$ and 
% $\mathbb{A}'(\mathcal{V},\mathcal{A},\mathbf{p}')$, defined over the same vertex set $\mathcal{V}$ and angle set $\mathcal{A}$, are said to be \emph{equivalent} if
% $\alpha_{kij} = \alpha'_{kij}$ for all $(i,k,j)\in \mathcal{A}$. 
% They are said to be \emph{congruent} if
% $\alpha_{kij} = \alpha'_{kij}$ for all $(i,k,j)\in \mathcal{V}\times \mathcal{V}\times \mathcal{V}$.}
% \begin{define}[Angle Rigidity]
% \textcolor{red}{definition}    
% \end{define}

We consider a triangular angularity consisting of $q$ triangles among $n$ agents. For each nondegenerate\footnote{A triangle is nondegenerate if its three vertices are not collinear.} triangle $\blacktriangle ijk$, the positions $\mathbf{p}_i$, $\mathbf{p}_j$, and $\mathbf{p}_k$ satisfy the following angle-induced linear equation \cite{CHEN2022auto}
\begin{equation}\label{eq:f_i}
\begin{split}
&f_i^{\blacktriangle ijk}(\alpha, \mathbf{p}) = \mathbf{A}_i^{\blacktriangle ijk}(\alpha)\mathbf{p}_i + \mathbf{A}_j^{\blacktriangle ijk}(\alpha)\mathbf{p}_j + \mathbf{A}_k^{\blacktriangle ijk}(\alpha)\mathbf{p}_k \\
& = \sin \alpha_{jki}(\mathbf{p}_i - \mathbf{p}_k) - \sin \alpha_{ijk} \mathcal{R}^\top (\alpha_{kij})(\mathbf{p}_i - \mathbf{p}_j) = \mathbf{0}_2,
\end{split}
\end{equation}
where $\alpha$ denotes the interior angles of the triangle $\blacktriangle ijk$, and the coefficient matrices $\mathbf{A}_i^{\blacktriangle ijk}$, $\mathbf{A}_j^{\blacktriangle ijk}$, and $\mathbf{A}_k^{\blacktriangle ijk} \in \mathbb{R}^{2\times 2}$ are functions of $\alpha$ defined as
\begin{align*}
\mathbf{A}_i^{\blacktriangle ijk}(\alpha) & = \left( \sin \alpha_{jki} I_2 - \sin \alpha_{ijk} \mathcal{R}^\top (\alpha_{kij}) \right), \\
\mathbf{A}_j^{\blacktriangle ijk}(\alpha) & = \sin \alpha_{ijk} \mathcal{R}^\top (\alpha_{kij}), \\
\mathbf{A}_k^{\blacktriangle ijk}(\alpha) & = -\sin \alpha_{jki} I_2.
\end{align*}
Similarly, all angle-induced linear equations associated with a triangular angularity $\mathbb{A}(\mathcal{V},\mathcal{A},\mathbf{p})$, where the configuration $\mathbf{p}$ is generic\footnote{A position vector $\mathbf{p}$ is said to be generic if its components are algebraically independent~\cite{Chen2021TAC}.}, can be expressed in a compact form 
$$\mathbf{R}_{\mathcal{A}}(\alpha)\mathbf{p} = \mathbf{0}.$$
The matrix $\mathbf{R}_{\mathcal{A}} \in \mathbb{R}^{2q \times 2n}$ is called the \emph{triangular angle rigidity matrix}, which can be expressed as follows \cite{CHEN2022auto}
\begin{equation}\label{eq:R_A}
\scalebox{0.85}{$\displaystyle
\renewcommand{\arraystretch}{1}
\bordermatrix{
    & \dots & i & \dots & j & \dots & k & \dots \cr
    1\text{st}\blacktriangle  & \dots & \dots & \dots & \dots & \dots & \dots & \dots \cr
    \dots & \dots & \dots & \dots & \dots & \dots & \dots & \dots \cr
    \blacktriangle ijk & \mathbf{0} & \mathbf{A}_i^{\blacktriangle ijk} & \mathbf{0} & \mathbf{A}_j^{\blacktriangle ijk} & \mathbf{0} & \mathbf{A}_k^{\blacktriangle ijk} & \mathbf{0} \cr
    \dots & \dots & \dots & \dots & \dots & \dots & \dots & \dots \cr
    q\text{th}\blacktriangle  & \dots & \dots & \dots & \dots & \dots & \dots & \dots \cr
}.$}
\end{equation}
Here, the row blocks correspond to the triangles in $\mathcal{A}$, whereas the column blocks correspond to the vertices in $\mathcal{V}$. The maximum rank of $\mathbf{R}_{\mathcal{A}}(\alpha)$ is $2n-4$. The notion of \emph{triangular angle rigidity} is formally defined as follows.
\begin{define}[Triangular angle rigidity \cite{CHEN2022auto}]
A triangular angularity $\mathbb{A}(\mathcal{V},\mathcal{A},\mathbf{p})$ with generic $\mathbf{p}$ is said to be triangularly angle rigid if $\text{rank}(\mathbf{R}_{\mathcal{A}}(\alpha)) = 2n-4$.
\end{define}
\begin{remark}
One can define the \emph{triangular angle function} as
\[
f_{\mathcal{A}}(\mathbf{p}) := [\dots,(f_i^{\blacktriangle ijk}(\alpha, \mathbf{p}))^\top,\dots]^\top \in \mathbb{R}^{2q}.
\]
Accordingly, the triangular angle rigidity matrix can be expressed as
$\mathbf{R}_{\mathcal{A}}(\alpha) = \frac{\partial f_{\mathcal{A}}(\mathbf{p})}{\partial \mathbf{p}}$.
\textcolor{blue}{Since scaling, rotational, and translational motions preserve $f_{\mathcal{A}}$, $\text{null}(\mathbf{R}_{\mathcal{A}}(\alpha)) \supseteq \text{span}\{\mathbf{1}_n\otimes \mathbf{I}_2,\bar{\mathbf{J}}_n(\mathbf{p}-\mathbf{1}_n\otimes\bar{\mathbf{p}}),\mathbf{p}-\mathbf{1}_n\otimes\bar{\mathbf{p}}\}$, where $\bar{\mathbf{J}}_n=\mathbf{I}_n\otimes \mathcal{R}(\pi/2)$ and $\bar{\mathbf{p}}=\frac{1}{n}\sum_{i=1}^n\mathbf{p}_i$ is the centroid of $\mathbf{p}$. Triangular angle rigidity ensures equality, i.e., these are the only infinitesimal motions that preserve $f_{\mathcal{A}}$.}
\end{remark}
% \begin{remark}
% It was shown in \cite{CHEN2022auto} that triangular angle rigidity implies global angle rigidity for generic configurations \cite{Chen2021TAC}. The key difference between the two notions is that triangular angle rigidity requires $\mathcal{A}$ to be a triangular angle set, whereas angle rigidity does not.
% \end{remark}
\subsection{Target Formation and Localizability}\label{subsec:formation}
\textcolor{blue}{Among the $n$ agents, the first $n_l$, with $2\leq n_l\leq n-1$, are leaders and the remaining $n_f=n-n_l$ are followers.} Their interactions are described by a leader--follower graph $\mathcal{G}$, whose edges from followers to leaders are directed and whose edges among followers are undirected. Let $\mathcal{G}_f=(\mathcal{V}_f,\mathcal{E}_f)$ denote the subgraph induced by the followers, with $m_f=|\mathcal{E}_f|$ edges, incidence matrix $\mathbf{H}_f\in\mathbb{R}^{m_f\times n_f}$, and Laplacian matrix $\boldsymbol{\mathcal{L}}_f\in\mathbb{R}^{n_f\times n_f}$. For each follower $i$, let $b_i=1$ if it receives information from at least one leader and $b_i=0$ otherwise, and define $\boldsymbol{\mathcal{H}}=\boldsymbol{\mathcal{L}}_f+\boldsymbol{\mathcal{D}}_f$ with $\boldsymbol{\mathcal{D}}_f=\mathrm{diag}(b_{n_l+1},\dots,b_n)$. The following lemma is a direct extension of the well-known result in \cite{DAS2010auto}.
\begin{lemma}\label{lmm:connectivity}
The matrix $\boldsymbol{\mathcal{H}}$ is positive definite if and only if every connected component of $\mathcal{G}_f$ contains at least one follower $i$ with $b_i=1$.
\end{lemma}
\begin{pf}
\textcolor{blue}{Relabel the followers so that those in the same connected component of $\mathcal{G}_f$ have consecutive indices. Since $\mathcal{G}_f$ has no edge between different connected components, the entries of $\boldsymbol{\mathcal{L}}_f$ between followers in different components are zero, and $\boldsymbol{\mathcal{D}}_f$ is diagonal. Hence, $\boldsymbol{\mathcal{H}}$ is block-diagonal, with one block $\boldsymbol{\mathcal{L}}_c+\boldsymbol{\mathcal{D}}_c$ per connected component $c$ of $\mathcal{G}_f$, where $\boldsymbol{\mathcal{L}}_c$ is the Laplacian of $c$ and $\boldsymbol{\mathcal{D}}_c$ collects the corresponding $b_i$. Each block is positive definite if and only if $\boldsymbol{\mathcal{D}}_c\neq\mathbf{0}$ \cite{DAS2010auto}, which completes the proof.}
\hfill$\blacksquare$
\end{pf}

The positions of the leaders and followers are denoted by $\mathbf{p}^L = [\mathbf{p}_1^\top, \dots, \mathbf{p}_{n_l}^\top]^\top$ and $\mathbf{p}^F = [\mathbf{p}_{n_l+1}^\top, \dots, \mathbf{p}_n^\top]^\top$, respectively. The target formation is defined as follows.
\begin{define}[Target formation]
The target formation $(\mathcal{G},\mathbf{p}^*)$, where $\mathbf{p}^*=[\mathbf{p}_1^{*\top},\dots,\mathbf{p}_n^{*\top}]^\top$ denotes the desired configuration, is defined by
\begin{enumerate}
    \item Leaders' positions: $\mathbf{p}_i(t)=\mathbf{p}_i^*(t), \forall i=1,\dots,n_l.$
    \item Desired angle constraints: $\alpha_{kij}^*, \forall (k,i,j)\in \mathcal{A}$.
\end{enumerate}
\end{define}

\textcolor{blue}{Let $\alpha^*\in\mathbb{R}^{3q}$ denote the vector of interior angles of $\mathbf{p}^*$, given by \eqref{eq:angle_def} for all $(k,i,j)\in\mathcal{A}$. In addition, the following assumption is made.}
\begin{assume}\label{assume:desired_formation}
The target angularity $\mathbb{A}(\mathcal{V}, \mathcal{A}, \mathbf{p}^*)$ is triangularly angle rigid.
\end{assume}
% \begin{assume}\label{assume:desired_formation}
% The target angularity $\mathbb{A}(\mathcal{V}, \mathcal{A}, \mathbf{p}^*)$ is triangularly angle rigid.
% \end{assume}

Note that Assumption~\ref{assume:desired_formation} implies that $\boldsymbol{\mathcal{H}}$ is positive definite. Indeed, triangular angle rigidity implies that $\mathcal{G}$ is connected, so every connected component of $\mathcal{G}_f$ contains at least one follower adjacent to a leader, and the claim follows from Lemma~\ref{lmm:connectivity}. To study the \emph{localizability} of the target angularity, we introduce the following matrix
\begin{equation}\label{eq:lapcacian}
\begin{aligned}
\mathbf{L}_{\mathcal{A}}(\alpha^*) =  \mathbf{R}_{\mathcal{A}}^\top(\alpha^*) \mathbf{R}_{\mathcal{A}} (\alpha^*) 
=
\begin{bmatrix}
\mathbf{L}_{ll}(\alpha^*) & \mathbf{L}_{fl}^\top(\alpha^*) \\
\mathbf{L}_{fl}(\alpha^*) & \mathbf{L}_{ff}(\alpha^*)    
\end{bmatrix},
\end{aligned}
\end{equation}
where $\mathbf{L}_{ll}(\alpha^*) \in \mathbb{R}^{2n_l\times2n_l}, \mathbf{L}_{lf} (\alpha^*)=\mathbf{L}_{fl}^\top (\alpha^*) \in \mathbb{R}^{2n_l\times2n_f},$ and $\mathbf{L}_{ff}(\alpha^*) \in \mathbb{R}^{2n_f\times2n_f}$. 
\begin{lemma}\label{lmm:desired_formation}\cite{chenTAC2023}
Under Assumption~\ref{assume:desired_formation}, the matrix $\mathbf{L}_{ff}(\alpha^*)$ is positive definite. Moreover, the desired positions of the followers can be determined from the desired angles $\alpha^*$ and the leaders’ positions as 
$$\mathbf{p}^{F*}(t) = -\mathbf{L}_{ff}^{-1}(\alpha^*) \mathbf{L}_{fl}(\alpha^*) \mathbf{p}^L(t).$$
\end{lemma}

\textcolor{blue}{The target angularity $\mathbb{A}(\mathcal{V},\mathcal{A},\mathbf{p}^*)$ is said to be \emph{localizable} if, for the given $\mathbf{p}^L$, $\mathbf{p}^{F*}$ is the unique solution of $\mathbf{R}_{\mathcal{A}}(\alpha^*)\mathbf{p}=\mathbf{0}$. For generic $\mathbf{p}^*$, this holds if and only if $\mathbf{L}_{ff}(\alpha^*)$ is positive definite \cite{CHEN2022auto}. Hence, by Lemma~\ref{lmm:desired_formation}, triangular angle rigidity is a sufficient topological condition for localizability.} Notably, this notion differs from the localizability of the actual angularity $\mathbb{A}(\mathcal{V}, \mathcal{A}, \mathbf{p})$.

\subsection{Denial-of-Service Attack and All-or-Nothing Triangle Integrity Protocol}
Denial-of-Service (DoS) attacks block data transmission over communication links or jam agents' active sensors, thereby compromising the availability of information. We assume that DoS attacks on different edges $(i,j)\in\mathcal{E}$ occur independently, and that edges $(i,j)$ and $(j,i)$ are always attacked together. The resulting network is modeled as a switching graph $\mathcal{G}^{\sigma(t)}=(\mathcal{V},\mathcal{E}^{\sigma(t)})$, where $\sigma(t)$ is a switching signal and $\mathcal{E}^{\sigma(t)}\subseteq\mathcal{E}$ contains the attack-free edges. Accordingly, $a_{ij}^{\sigma(t)}=a_{ij}$ and $b_i^{\sigma(t)}=b_i$ if edge $(i,j)$ and the link between agent $i$ and the leaders, respectively, are attack-free, and $a_{ij}^{\sigma(t)}=0$ and $b_i^{\sigma(t)}=0$ otherwise.

It can be seen that, under DoS attacks, the resulting matrix $\boldsymbol{\mathcal{H}}^{\sigma(t)} = \boldsymbol{\mathcal{L}}_f^{\sigma(t)} + \boldsymbol{\mathcal{D}}_f^{\sigma(t)}$ may lose positive definiteness due to possible loss of connectivity in 
$\mathcal{G}^{\sigma(t)}$. Accordingly, \cite{Feng2016ijrnc} categorizes DoS attacks based on their impact on network connectivity as follows.
% \begin{define}[Connectivity-maintained attacks]
% An attack is referred to as a \emph{connectivity-maintained attack} if the resulting matrix $\boldsymbol{\mathcal{H}}^{\sigma(t)}$ is positive definite. 
% \end{define}

% \begin{define}[Connectivity-broken attacks]
% An attack is referred to as a \emph{connectivity-broken attack} if the resulting matrix $\boldsymbol{\mathcal{H}}^{\sigma(t)}$ is positive semidefinite but not positive definite.
% \end{define}
\textcolor{blue}{
\begin{define}\textbf{(Connectivity-maintained/broken attacks)}
An attack is referred to as a \emph{connectivity-maintained attack} if the resulting matrix $\boldsymbol{\mathcal{H}}^{\sigma(t)}$ is positive definite. Otherwise, it is referred to as a \emph{connectivity-broken attack}.
\end{define}}
% \subsubsection{All-or-Nothing Triangle Integrity Protocol}

Nevertheless, the above classification of DoS attacks is inadequate for angle-constrained formation control and localization, since the target formation is defined by triangular constraints. In particular, the loss of an edge renders both the triangular angle rigidity matrix and the localizability of the target angularity ill-defined. To address these issues, we introduce a communication protocol termed the \emph{All-or-Nothing Triangle Integrity Protocol} (ANTIP), defined as follows.

\begin{define}\textbf{(All-or-Nothing Triangle Integrity Protocol)}
The formation is said to follow the All-or-Nothing Triangle Integrity Protocol (ANTIP) if, whenever a DoS attack blocks an edge $(i,j)$, agents $i$ and $j$ broadcast an alarm signal to every follower agent $k$ such that $(i,j,k)\in\mathcal{A}$. Upon receiving the alarm, the constraints associated with the triangle $\blacktriangle  ijk$ are deactivated by setting $\mathbf{A}_i^{\blacktriangle ijk}(\alpha) = \mathbf{A}_j^{\blacktriangle ijk}(\alpha) = \mathbf{A}_k^{\blacktriangle ijk}(\alpha) = \mathbf{0}_{2\times 2}$. 
\end{define}

% \begin{assume}
% The formation under consideration operates under the ANTIP protocol.
% \end{assume}
\begin{figure}
\centering
\begin{tikzpicture}[
roundnode/.style={circle, draw=black, thick, minimum size=2mm,inner sep= 0.25mm},
squarednode/.style={rectangle, draw=red!60, fill=red!5, very thick, minimum size=5mm},
]
    \begin{scope}[xshift=-1.8cm]
        \node[roundnode] (a1) at (0,0) {$1$};
        \node[roundnode] (a2) at (1,0) {$2$};
        \node[roundnode] (a3) at (2,0) {$3$};
        \node[roundnode] (a4) at (0,1) {$4$};
        \node[roundnode] (a5) at (1,1) {$5$};
        \node[roundnode] (a6) at (2,1) {$6$};
        \draw[thick] (a1)--(a2);
        \draw[thick] (a6)--(a5)--(a4)--(a1);
        \draw[thick] (a4)--(a2) (a5)--(a2) (a5)--(a3);
        \draw[magenta, thick, dash pattern=on 2pt off 1pt] (a2)--(a3)--(a6);
        \pic[draw=black, thick, angle radius=3mm] {angle=a3--a2--a5};
        \pic[draw=black, thick, angle radius=3.5mm] {angle=a2--a5--a3};
        \pic[draw=black, thick, angle radius=3.5mm] {angle=a5--a3--a2};
        \pic[draw=black, thick, angle radius=3mm] {angle=a5--a6--a3};
        \pic[draw=black, thick, angle radius=3mm] {angle=a3--a5--a6};
        \pic[draw=black, thick, angle radius=3mm] {angle=a6--a3--a5};
        \node at (1,-0.45) {(a)};
    \end{scope}
    \begin{scope}[xshift=1.8cm]
        \node[roundnode] (b1) at (0,0) {$1$};
        \node[roundnode] (b2) at (1,0) {$2$};
        \node[roundnode] (b3) at (2,0) {$3$};
        \node[roundnode] (b4) at (0,1) {$4$};
        \node[roundnode] (b5) at (1,1) {$5$};
        \node[roundnode] (b6) at (2,1) {$6$};
        \draw[thick] (b1)--(b2);
        \draw[thick] (b5)--(b4)--(b1);
        \draw[thick] (b4)--(b2) (b5)--(b2);
        \node at (1,-0.45) {(b)};
    \end{scope}
\end{tikzpicture}
\caption{ANTIP illustration: (a) Initial graph with attacked edges $(2,3)$ and $(3,6)$; the arcs indicate the interior angles of triangles $\blacktriangle 325$ and $\blacktriangle 356$. Agents $2$, $3$, and $6$ detect the attacks and broadcast alarm signals to agent $5$. (b) Graph after ANTIP activation, with triangles $\blacktriangle 325$ and $\blacktriangle 356$ deactivated.}
\label{fig:ANTIP}
\end{figure}
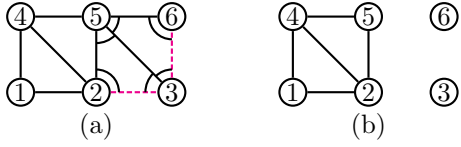

% \begin{algorithm}[t]
% \caption{All-or-Nothing Triangle Integrity Protocol (ANTIP)}
% \label{alg:ANTIP}
% \begin{algorithmic}[1]
% \State Each follower agent $i$ maintains the local triangle set
% $\mathcal{A}_i=\{\blacktriangle ijk\mid(i,j,k)\in\mathcal{A}\}$.
% \State Continuously monitor the status of incident communication links and incoming alarm messages.

% \Statex
% \State \textbf{Upon detecting a DoS attack on edge $(i,j)$:}
% \ForAll{$\blacktriangle ijk\in\mathcal{A}_i$}
%     \State Broadcast \texttt{ALARM}$(\blacktriangle ijk)$ to agent $k$.
%     \State Set $\mathbf{A}_i^{\blacktriangle ijk}(\alpha)\gets\mathbf{0}_{2\times2}$.
% \EndFor

% \Statex
% \State \textbf{Upon receiving \texttt{ALARM}$(\blacktriangle ijk)$:}
% \State Set $\mathbf{A}_i^{\blacktriangle ijk}(\alpha)\gets\mathbf{0}_{2\times2}$.
% \end{algorithmic}
% \end{algorithm}
\begin{algorithm}[t]
\caption{All-or-Nothing Triangle Integrity Protocol (ANTIP)}
\label{alg:ANTIP}
\begin{algorithmic}[1]
\State Each agent $i$ maintains the local triangle set $\mathcal{A}_i=\{\blacktriangle ijk\mid(i,j,k)\in\mathcal{A}\}$.
\Loop
\ForAll{$\blacktriangle ijk\in\mathcal{A}_i$}
\If{the attack status of edge $(i,j)$ (resp. $(i,k)$) changes}
\State Send \texttt{ALARM} if attacked, or \texttt{RESTORE} if attack-free, to agent $k$ (resp. $j$).
\EndIf
\If{an edge of $\blacktriangle ijk$ is attacked, as detected or alarmed}
\State Set $\mathbf{A}_i^{\blacktriangle ijk}(\alpha)\gets\mathbf{0}_{2\times2}$.
\Else
\State Restore $\mathbf{A}_i^{\blacktriangle ijk}(\alpha)$.
\EndIf
\EndFor
\EndLoop
\end{algorithmic}
\end{algorithm}

Algorithm~\ref{alg:ANTIP} describes the ANTIP, and Fig.~\ref{fig:ANTIP} illustrates its operation. Equivalently, ANTIP removes the row block of triangle $\blacktriangle ijk$ from the triangular angle rigidity matrix \eqref{eq:R_A} whenever at least one of its edges is attacked. If two or more edges of $\blacktriangle ijk$ are attacked, every vertex detects an attack on an incident edge, so the triangle is deactivated without any alarm transmission. In this way, edge-wise attacks are converted into triangle-wise attacks, \textcolor{blue}{which preserves the well-definedness of the localizability of} $\mathbb{A}(\mathcal{V},\mathcal{A}^{\sigma(t)},\mathbf{p}^*)$, where $\mathcal{A}^{\sigma(t)}\subseteq\mathcal{A}$ denotes the set of triplets not eliminated by ANTIP. \textcolor{blue}{Throughout this paper, all agents execute ANTIP.} Although ANTIP is actively enforced by each agent, in the subsequent discussion we may, with a slight abuse of terminology, refer to a DoS attack as being subject to ANTIP. This convention is adopted solely for notational simplicity and does not affect the analysis.

For the nominal formation $(\mathcal{G},\mathbf{p})$ with a target formation satisfying Assumption~\ref{assume:desired_formation}, DoS attacks subject to ANTIP are classified into the following two classes.

% \begin{define}[Localizability-maintained attacks]
% An attack is referred to as a \emph{localizability-maintained attack} if the resulting angularity $\mathbb{A}(\mathcal{V}, \mathcal{A}^{\sigma(t)}, \mathbf{p}^*)$ remains localizable. 
% \end{define}

% \begin{define}[Localizability-broken attacks]
% An attack is referred to as a \emph{localizability-broken attack} if the resulting angularity $\mathbb{A}(\mathcal{V}, \mathcal{A}^{\sigma(t)}, \mathbf{p}^*)$ is no longer localizable.
% \end{define}
\textcolor{blue}{
\begin{define}\textbf{(\textcolor{blue}{Localizability-maintained/broken attacks})}
An attack is referred to as a \emph{localizability-maintained attack} if the resulting angularity $\mathbb{A}(\mathcal{V}, \mathcal{A}^{\sigma(t)}, \mathbf{p}^*)$ remains localizable, i.e., $\mathbf{L}^{\sigma(t)}_{ff}(\alpha^*)$ is positive definite. Otherwise, it is referred to as a \emph{localizability-broken attack}.
\end{define}}

It is straightforward to verify that any localizability-maintained attack is also connectivity-maintained. In addition, let us denote two sets 
$S = \{(\mathcal{G}^{\sigma(t)},\mathbf{p}) \mid \mathbf{L}^{\sigma(t)}_{ff}(\alpha^*)\text{ is positive definite}\}$
and
$A = \{(\mathcal{G}^{\sigma(t)},\mathbf{p}) \mid \mathbf{L}^{\sigma(t)}_{ff}(\alpha^*)\text{ is singular}\}$, 
representing the sets of resulting formations that are under localizability-maintained and localizability-broken attacks, respectively. Note that, for simplicity, the attack-free formation $(\mathcal{G},\mathbf{p})$ is also treated as a member of the set $S$. For subsequent analysis, let $\underline{\lambda} > 0$ denote the minimum eigenvalue of $\mathbf{L}^{\sigma(t)}_{ff}(\alpha^*)$\footnote{For notational simplicity, we hereafter write $\mathbf{L}^{\sigma(t)}_{\mathcal{A}}$ and $\mathbf{R}^{\sigma(t)}_{\mathcal{A}}$ for $\mathbf{L}^{\sigma(t)}_{\mathcal{A}}(\alpha^*)$ and $\mathbf{R}^{\sigma(t)}_{\mathcal{A}}(\alpha^*)$, respectively, and similarly for their blocks.} taken over all localizability-maintained formations \textcolor{blue}{$(\mathcal{G}^{\sigma(t)},\mathbf{p}^*) \in S$}.

To characterize the influence of DoS attacks on the system, we first define the timing of the events. Let $\{t_r^{A}\}_{r \in \mathbb{N}}$ and $\{\Delta^{A}_r\}_{r \in \mathbb{N}}$ denote the time instant and duration of the $r$-th localizability-broken attack, i.e., $(\mathcal{G}^{\sigma(t)},\mathbf{p})\in A$, $\forall t \in [t_r^A, t_r^A+\Delta_r^A)$. %Similarly, the starting time instants and durations of the $r$-th rigidity-maintained attacks are denoted by $\{t_r^{S}\}_{r\in\mathbb{N}}$ and $\{\Delta_r^{S}\}_{r\in\mathbb{N}}$, respectively. 
Moreover, let $\Theta^A(\tau,t) = [\tau,t)\cap\bigl(\bigcup_{r\in\mathbb{N}} [t_r^{A},\,t_r^{A}+\Delta_r^{A})\bigr)$ denote the union of durations corresponding to localizability-broken attacks over the interval $[\tau,t)$. Conversely, the set of time instants without attacks or with localizability-maintained attacks is given by ${\Xi}^A(\tau,t) = [\tau,t)\setminus \Theta^A(\tau,t)$.
The following assumption is widely adopted to characterize the duration of DoS attacks \cite{Persis2015TAC,zhang2022TAC}.

\begin{assume}[DoS duration]\label{duration}
There exist constants $\epsilon \geq 0$ and $\tau_d > 1$ such that for any $t \geq \tau \geq 0$
$$|\Theta^A(\tau,t)|\leq\epsilon+\frac{t-\tau}{\tau_d}.$$
\end{assume}

It is worth noting that the proposed general DoS attack model captures a wide range of jamming behaviors, including trivial, periodic, random, and protocol-aware attacks. 
\begin{corollary}\label{cor:longest}
Suppose that Assumption~\ref{duration} holds. Then, the duration of each localizability-broken attack cannot be longer than $\Delta_{\max} = \frac{\epsilon}{1-\frac{1}{\tau_d}}$.    
\end{corollary}
\begin{pf}
Consider the $r$-th localizability-broken attack. If $\Delta_r^A$ is finite, then $\Theta^A(t_r^A,t_r^A+\Delta_r^A)=[t_r^A,t_r^A+\Delta_r^A)$, and Assumption~\ref{duration} yields $\Delta_r^A\leq\epsilon+\frac{\Delta_r^A}{\tau_d}$, i.e., $\Delta_r^A\leq\frac{\epsilon}{1-\frac{1}{\tau_d}}=\Delta_{\max}$. If $\Delta_r^A$ were infinite, the same argument applied to $[t_r^A,t_r^A+T)$ would give $T\leq\Delta_{\max}$ for all $T>0$, which is impossible. This completes the proof.
\hfill$\blacksquare$
\end{pf}
\section{Problem Formulation}\label{subsec:prob_form}
% Consider a system of $n$ autonomous agents in the $d$-dimensional Euclidean space $\mathbb{R}^d$ ($d \ge 2$), whose positions are denoted by $\mathbf{p}_1, \dots, \mathbf{p}_n \in \mathbb{R}^d$. We assume that $2 \le n_l < n$ of these agents are leaders, while the remaining $n_f = n - n_l$ agents are followers. 
We consider each follower $i \in \mathcal{V}_f$ to be governed by single-integrator dynamics
\begin{equation}\label{eq:single}
    \dot{\mathbf{p}}_i(t) = \mathbf{u}_i(t), \quad i=n_l+1,\dots,n,
\end{equation}
where $\mathbf{u}_i(t) \in \mathbb{R}^2$ is the control input. The two problems considered in this paper are formulated as follows.

% {Stop}
% In other words, the leaders are assumed to move with a uniform velocity $\mathbf{v}_0(t)$, which is unknown to the followers.
% When the leaders' velocity is constant, we have $\dot{\mathbf{v}}_0 = \mathbf{0}_d$. 
\subsubsection{Problem 1: Resilient Formation Tracking}
\textcolor{blue}{For this problem, we first make the following assumptions on the followers' measurements and the leaders' motion.}
\begin{assume}\label{assume:sensor}
% Sensing and communication among agents are continuous. 
If edge $(i,j)$ is not under attack, follower $i$ can measure the relative position $\mathbf{p}_i-\mathbf{p}_j$ to its neighbor $j$.
\end{assume}
% \textcolor{green}{In addition, the leaders' motion satisfies the following assumption.}
\begin{assume}\label{assume:leader}
The leaders move with the reference velocity $\mathbf{v}^L = \dot{\mathbf{p}}^L = [\mathbf{v}_1^\top, \dots, \mathbf{v}_{n_l}^\top]^\top$, which is bounded, uniformly continuous, and belongs to the set of triangular angle function--preserving motions, i.e., translational, rotational, and scaling motions. In other words, $\mathbf{v}^L(t) \in \text{span}\{\mathbf{1}_{n_l}\otimes \mathbf{I}_2,\bar{\mathbf{J}}_{n_l}(\mathbf{p}^L-\mathbf{1}_{n_l}\otimes\bar{\mathbf{p}}^L),\mathbf{p}^L-\mathbf{1}_{n_l}\otimes\bar{\mathbf{p}}^L\}$ with $\bar{\mathbf{J}}_{n_l}=\mathbf{I}_{n_l}\otimes\mathcal{R}(\pi/2)$ and $\bar{\mathbf{p}}^L=\frac{1}{n_l}\sum_{i=1}^{n_l}\mathbf{p}_i$.   
\end{assume}

Then, the resilient formation tracking problem is stated as follows.
\begin{problem}\label{prob1}
Given that Assumptions~\ref{assume:desired_formation}--\ref{assume:leader} hold, the objective is to design distributed resilient control laws $\mathbf{u}_i$ such that $\mathbf{p}(t) \to \mathbf{p}^*(t)$ as $t \to +\infty$.
\end{problem}
% \begin{assume}
% When an edge is free from attacks, each agent may measure relative positions, relative velocities, and/or its own velocity.
% \end{assume}

\textcolor{blue}{Note that} asymptotic tracking cannot be guaranteed at all times under DoS attacks. For instance, if all edges of $\mathcal{G}$ are attacked while the leaders change their velocity, the followers are disconnected from the leaders and cannot track their desired positions. Hence, we adopt the following notion of finite-time recovery \cite{NhatMinh2026TCNS}.
\begin{define}\textbf{(Finite-time recovery of formation tracking)}\label{def:recovery}
A control law $\mathbf{u}_i$ is said to achieve resilient finite-time recovery of formation tracking with recovery time $T(\cdot)$ if, under DoS attacks, $\|\mathbf{p}(t)-\mathbf{p}^*(t)\|$ remains bounded for all $t\geq0$. Moreover, for any $\tau\geq0$ with $\mathbf{p}(\tau)\neq\mathbf{p}^*(\tau)$, there exists $t^*\in[\tau,\tau+T(\mathbf{p}(\tau))]$ such that $\mathbf{p}(t^*)=\mathbf{p}^*(t^*)$.
\end{define}
% \begin{define}[Fixed-time recovery of formation tracking] 
% A control law is said to solve the fixed-time recovery of formation tracking if it solves finite-time recovery of formation tracking with a fixed recovery time $T$.
% \end{define}
\subsubsection{Problem 2: Resilient Simultaneous Localization and Formation Control}
In contrast to Problem 1, relative position measurements are unavailable in this problem. Consequently, the agents must communicate with their neighbors to localize their global positions while simultaneously tracking the target formation. Furthermore, the leaders' trajectories are assumed to satisfy the following condition.
\begin{assume}\label{assume:exo}
The leaders' velocities are given by $\mathbf{v}^L(t)=\mathbf{1}_{n_l}\otimes \mathbf{h}(t)$, where $\mathbf{h}(t)\in \mathbb{R}^2$ denotes the common bounded translational velocity generated by the following exosystem
\begin{equation}\label{eq:exo}
\begin{aligned}
\dot{\boldsymbol{\zeta}}(t) &= \mathbf{S} \boldsymbol{\zeta}(t),\\
\mathbf{h}(t)&=\mathbf{F}\boldsymbol{\zeta}(t),
\end{aligned}
\end{equation}
where $\boldsymbol{\zeta}(t)\in\mathbb{R}^l$ is the state of the exosystem, while $\mathbf{S}$ and $\mathbf{F}$ are constant matrices of appropriate dimensions.
\end{assume}

The leader system \eqref{eq:exo} can generate a broad class of signals, including polynomials of arbitrary degree and trigonometric polynomials with arbitrary amplitudes and phases \cite{huang2025TAC}. 
\begin{assume}\label{assume:communicate}
The dimension $l$ of the exosystem state is known to all followers. Each leader $i$ can transmit $\boldsymbol{\zeta}(t)$, $\mathbf{h}(t)$, $\mathbf{p}_i(t)$, $\mathbf{S}$, and $\mathbf{F}$ to its neighboring followers, and each follower $i$ can measure $\alpha_{kij}$ for every $(k,i,j)\in\mathcal{A}^{\sigma(t)}$, i.e., for every triangle $\blacktriangle ijk$ not deactivated by ANTIP. Moreover, neighboring followers can exchange their measured angles and local estimates of $\mathbf{S}$, $\mathbf{F}$, $\boldsymbol{\zeta}$, and their desired and actual positions over attack-free links.
\end{assume} 

We now formally state the resilient simultaneous localization and formation control (RSLAF) problem.
% \begin{problem}\label{prob2}
% Suppose that Assumptions~\ref{assume:desired_formation}--\ref{duration} and \ref{assume:exo}--\ref{assume:communicate} hold. The localization task is to design estimator $\hat{\mathbf{p}}_i(t)$ such that
% \begin{equation*}
% \lim_{t \rightarrow \infty}(\hat{\mathbf{p}}_i(t)-\mathbf{p}_i(t)) = \mathbf{0}, \forall i \in \mathcal{V}_f.
% \end{equation*}
% The formation control task is to design control input $\mathbf{u}_i(t)$ such that 
% \begin{equation*}
% \lim_{t \rightarrow \infty}(\mathbf{p}_i(t)-\mathbf{p}_i^*(t)) = \mathbf{0}, \forall i \in \mathcal{V}_f.
% \end{equation*}
% The aim of RSLAF problem is to simultaneously achieve localization and formation control tasks under DoS attacks.
% \end{problem}
\begin{problem}\label{prob2}
Suppose that Assumptions~\ref{assume:desired_formation}, \ref{duration}, \ref{assume:exo}, and \ref{assume:communicate} hold. The RSLAF problem is to design, for each follower $i\in\mathcal{V}_f$, a position estimator $\hat{\mathbf{p}}_i(t)$ and a control input $\mathbf{u}_i(t)$ such that, under DoS attacks, the localization and formation control tasks are achieved simultaneously, i.e.,
\begin{equation*}
\begin{aligned}
\lim_{t\to\infty}\big(\hat{\mathbf{p}}_i(t)-\mathbf{p}_i(t)\big)&=\mathbf{0},\\
\lim_{t\to\infty}\big(\mathbf{p}_i(t)-\mathbf{p}_i^*(t)\big)&=\mathbf{0}.
\end{aligned}
\end{equation*}
\end{problem}

% Finally, the following lemma will be useful in the subsequent analysis.
% \begin{lemma}[Schur complement lemma \cite{LMI94}]
% Let $Q$ and $R$ be two symmetric matrices, and let $U$ be a matrix of suitable dimensions. Then $\begin{bmatrix}
% Q &U\\U^\top &R
% \end{bmatrix} < 0$ if and only if one of the following conditions holds
% \begin{enumerate}
% \item $Q < 0,\quad R - U^\top Q^{-1} U < 0$,
% \item $R < 0,\quad Q - U R^{-1} U^\top < 0$.
% \end{enumerate}
% \end{lemma}

% \begin{lemma}[Young’s inequality for matrices \cite{horn94matrix}]\label{young}
% Let $\mathcal{B}_1 \in \mathbb{R}^{m \times n}$ and $\mathcal{B}_2 \in \mathbb{R}^{m \times m}$ be two real matrices. 
% Then, for any $\phi \in \mathbb{R}^{n}$, $\psi \in \mathbb{R}^{m}$, and any $\nu > 0$, the following inequality holds
% \begin{equation*}
% 2\,\phi^\top \mathcal{B}_1^\top \mathcal{B}_2 \psi
% \;\le\;
% \nu\,\phi^\top \mathcal{B}_1^\top \mathcal{B}_1 \phi
% +
% \nu^{-1}\,\psi^\top \mathcal{B}_2^\top \mathcal{B}_2 \psi .
% \end{equation*} 
% In particular, when $\mathcal{B}_1 = \mathcal{B}_2$ and $\nu = 1$, letting $\mathcal{B} = \mathcal{B}_1^\top \mathcal{B}_1$, one obtains
% \begin{equation*}
% \phi^\top \mathcal{B} \psi
% \;\le\;
% \frac{1}{2}\phi^\top \mathcal{B} \phi
% +
% \frac{1}{2}\psi^\top \mathcal{B} \psi .
% \end{equation*}
% \end{lemma}
%%%%%%%%%%%%%%%%%%%%%%%%%%%%%%%%%%%%
\section{Angle-Constrained Resilient Formation Control} \label{sec:formation}
This section addresses Problem~\ref{prob1} by developing two control laws: one for constant-velocity leaders and the other for time-varying-velocity leaders.
%%%%%%%%%%%%%%
\subsection{Moving Leaders With Constant Velocity}
Let the leaders move at a constant velocity, i.e., $\mathbf{v}^L(t)=\mathbf{1}_{n_l}\otimes \mathbf{h}(t)$ and $\dot{\mathbf{h}}(t) = \mathbf{0}_2$. The resilient \textcolor{blue}{PI-type} control law for each follower is proposed as follows
\begin{equation}\label{eq:single_PI}
\begin{aligned}
\mathbf{u}_i(t) &= -k_P\mathcal{F}_i^{\sigma(t)}(\alpha^*,\mathbf{p}(t))-k_I\boldsymbol{\eta}_i(t),\\
\dot{\boldsymbol{\eta}}_i(t) &= \mathcal{F}_i^{\sigma(t)}(\alpha^*,\mathbf{p}(t)),
\end{aligned}
\end{equation}
where $k_P, k_I>0$ are control gains, and the expansion of $\mathcal{F}_i^{\sigma(t)}(\alpha^*,\mathbf{p})$ has the following form
\begin{equation}
\begin{aligned}
\mathcal{F}_i^{\sigma(t)}(\alpha^*,\mathbf{p})
&=\sum_{(i,j,k) \in \bar{\mathcal{A}}^{\sigma(t)}} (\mathbf{A}_i^{\blacktriangle  {ijk}}(\alpha^*))^\top f_i^{\blacktriangle  {ijk}}(\alpha^*, \mathbf{p}) \\
&+ \sum_{{(j,i,k)} \in \bar{\mathcal{A}}^{\sigma(t)}} (\mathbf{A}_i^{\blacktriangle  {jik}}(\alpha^*))^\top f_i^{\blacktriangle  {jik}}(\alpha^*, \mathbf{p})\\
&+ \sum_{{(j,k,i)} \in \bar{\mathcal{A}}^{\sigma(t)}} (\mathbf{A}_i^{\blacktriangle  {jki}}(\alpha^*))^\top f_i^{\blacktriangle  {jki}}(\alpha^*, \mathbf{p}).
\end{aligned}
\end{equation}
Here, $\bar{\mathcal{A}}^{\sigma(t)}\subseteq\bar{\mathcal{A}}$ denotes the set of nonredundant three-vertex triplets remaining after applying ANTIP. From (\ref{eq:f_i}), the proposed control law requires only relative position measurements from neighboring agents whose communication links are not under attack. The matrix form of \eqref{eq:single} and (\ref{eq:single_PI}) can be obtained as follows
\begin{equation}\label{eq:single_PI_matrix}
\begin{aligned}
\begin{bmatrix}
  \dot{\mathbf{p}}^{L}\\
  \dot{\mathbf{p}}^{F}  
\end{bmatrix}
&=\begin{bmatrix}
  \mathbf{0}_{2n_l}\\
  \dot{\mathbf{p}}^F - \mathbf{1}_{n_f}\otimes\mathbf{h}
\end{bmatrix}
+\mathbf{1}_n \otimes \mathbf{h}\\
&=-\overline{\mathbf{Z}}\big(k_P\mathbf{L}_{\mathcal{A}}^{\sigma(t)}\mathbf{p}+k_I\boldsymbol{\eta}+\mathbf{1}_n \otimes \mathbf{h}\big)
+\mathbf{1}_n \otimes \mathbf{h},\\
\dot{\boldsymbol{\eta}} &= %\overline{\mathbf{Z}}
\mathbf{L}_{\mathcal{A}}^{\sigma(t)}\mathbf{p},
\end{aligned}
\end{equation}
where 
% $\boldsymbol{\eta}=[\mathbf{0}^\top_{2n_l},\boldsymbol{\eta}_1^\top,\dots,\boldsymbol{\eta}_{n_f}^\top]^\top$ ,
${\mathbf{Z}}=\begin{bmatrix}
\mathbf{0}_{n_l \times n_l} &\mathbf{0}_{n_l \times n_f}\\
\mathbf{0}_{n_f \times n_l} &\mathbf{I}_{n_f}
\end{bmatrix}$ and $\overline{\mathbf{Z}} = {\mathbf{Z}} \otimes \mathbf{I}_2$. Denote $\boldsymbol{\delta}_1(t) = \mathbf{p}^F(t)-\mathbf{p}^{F*}(t)$ and $\textcolor{blue}{\boldsymbol{\delta}_{\eta}}(t) = [\boldsymbol{\eta}^\top_{n_l+1},\dots,\boldsymbol{\eta}_n^\top]^\top+\mathbf{1}_{n_f}\otimes\frac{\mathbf{h}}{k_I}$. By using the fact that $\mathbf{L}_{\mathcal{A}}^{\sigma(t)}\mathbf{p}^* = \mathbf{0}$, and thus $\mathbf{L}_{fl}^{\sigma(t)}\mathbf{p}^L+ \mathbf{L}_{ff}^{\sigma(t)}\mathbf{p}^{F*} = \mathbf{0}_{2n_f}$, the following holds
\begin{equation}
\begin{aligned}
\mathbf{L}_{fl}^{\sigma(t)}\mathbf{p}^L+ \mathbf{L}_{ff}^{\sigma(t)}\mathbf{p}^F &=  \mathbf{L}_{fl}^{\sigma(t)}(\mathbf{p}^L-\mathbf{p}^L)+ \mathbf{L}_{ff}^{\sigma(t)}(\mathbf{p}^F-\mathbf{p}^{F*})\\
&= \mathbf{L}_{ff}^{\sigma(t)}\boldsymbol{\delta}_1.
\end{aligned}
\end{equation}

Hence, the dynamics of $\boldsymbol{\delta}_1(t)$ and $\textcolor{blue}{\boldsymbol{\delta}_{\eta}}(t)$ can be derived from \eqref{eq:single_PI_matrix} as follows
\begin{equation}\label{eq:single_PI_error_dyn}
\begin{aligned}
\begin{bmatrix}
    \dot{\boldsymbol{\delta}}_1\\
    \textcolor{blue}{\dot{\boldsymbol{\delta}}_{\eta}}
\end{bmatrix}
=
\underbrace{\begin{bmatrix}
 -k_P \mathbf{L}_{ff}^{\sigma(t)} &-k_I\mathbf{I}_{2n_f}\\
 \mathbf{L}_{ff}^{\sigma(t)} &\mathbf{0}_{2n_f \times 2n_f}
\end{bmatrix}}_{\mathcal{Q}}
\begin{bmatrix}
    {\boldsymbol{\delta}}_1\\
    \textcolor{blue}{\boldsymbol{\delta}_{\eta}}
\end{bmatrix}.
\end{aligned}
\end{equation}
\begin{theorem}\label{thm:single_PI}
Consider Problem~\ref{prob1} with the leaders moving at constant velocity. Suppose that Assumptions~\ref{assume:desired_formation}--\ref{assume:leader} hold, and that there are no collisions among the agents. 
If $k_P>1$, $k_I>0$, and $\omega>0$ are selected such that
\begin{subequations}\label{eq:cond_single_consant}
\begin{align}
(k_P - 1)\underline{\lambda} &\geq \frac{\omega}{2} + \frac{(\omega - k_I)^2}{4k_I - 2\omega k_P},\label{eq:main_a}\\
k_I &> \frac{\omega k_P}{2}\label{eq:main_b},\\
% (\tau_d-1)\omega &>\frac{2\max \Big\{\frac{k_I \varrho}{2}, \frac{k_I}{2\varrho}-k_I \Big\}}{1 + k_P - \sqrt{(k_P - 1)^2 + 4}}\label{eq:main_c},
(\tau_d-1)\omega &>\frac{2(\sqrt{2}-1)k_I}{1 + k_P - \sqrt{(k_P - 1)^2 + 4}}\label{eq:main_c},
\end{align}
\end{subequations}
\textcolor{blue}{where $\underline{\lambda}>0$ is global information, defined as the minimum eigenvalue of $\mathbf{L}^{\sigma}_{ff}(\alpha^*)$ over all localizability-maintained modes}, then the control law \eqref{eq:single_PI} \textcolor{blue}{achieves that} $\mathbf{p}(t) \to \mathbf{p}^*(t)$ as $t \to \infty$.
\end{theorem}
\begin{pf}
Consider the following common Lyapunov function candidate
\begin{equation*}
V = \frac{1}{2}[ \boldsymbol{\delta}_1^\top \quad
    \textcolor{blue}{\boldsymbol{\delta}_{\eta}}^\top]^\top
\underbrace{\begin{bmatrix}
    \mathbf{I}_{2n_f} &\mathbf{I}_{2n_f}\\
    \mathbf{I}_{2n_f} &k_P\mathbf{I}_{2n_f}
\end{bmatrix}}_{\mathcal{P}}
\begin{bmatrix}
    \boldsymbol{\delta}_1\\
    \textcolor{blue}{\boldsymbol{\delta}_{\eta}}
\end{bmatrix}.
\end{equation*}
Clearly, $V(\boldsymbol{\delta}_1,\textcolor{blue}{\boldsymbol{\delta}_{\eta}})$ is positive definite for $k_P>1$. Furthermore, the minimum eigenvalue of $\mathcal{P}$ can be expressed as
\begin{equation*}
    \lambda_{\min}(\mathcal{P}) 
= \frac{1 + k_P - \sqrt{(k_P - 1)^2 + 4}}{2} > 0.
\end{equation*}
Next, the time derivative of $V$ is given by
\begin{equation}\label{eq:vdot_single_const}
\begin{aligned}
    \dot{V} &= \frac{1}{2}
    [ \boldsymbol{\delta}_1^\top \quad
    \textcolor{blue}{\boldsymbol{\delta}_{\eta}}^\top]
    \big(\mathcal{PQ}+\mathcal{Q^\top P}\big)
\begin{bmatrix}
    \boldsymbol{\delta}_1\\
    \textcolor{blue}{\boldsymbol{\delta}_{\eta}}
\end{bmatrix}\\
&=  [ \boldsymbol{\delta}_1^\top \quad
    \textcolor{blue}{\boldsymbol{\delta}_{\eta}}^\top]
\begin{bmatrix}
    (1-k_P)\mathbf{L}_{ff}^{\sigma(t)} &-\frac{k_I}{2}\mathbf{I}_{2n_f}\\
    -\frac{k_I}{2}\mathbf{I}_{2n_f} &-k_I\mathbf{I}_{2n_f}
\end{bmatrix}
\begin{bmatrix}
    \boldsymbol{\delta}_1\\
    \textcolor{blue}{\boldsymbol{\delta}_{\eta}}
\end{bmatrix}.
\end{aligned}
\end{equation}
\textcolor{blue}{During localizability-maintained attacks, $\mathbf{L}_{ff}^{\sigma(t)} > 0$. Applying the Schur complement \cite{LMI94} to \eqref{eq:main_a}--\eqref{eq:main_b} then yields}
$$\begin{bmatrix} 
(1-k_P)\mathbf{L}_{ff}^{\sigma{(t)}} + \frac{\omega}{2}\mathbf{I}_{2n_f} & \frac{\omega - k_I}{2}\mathbf{I}_{2n_f} \\ 
\frac{\omega - k_I}{2}\mathbf{I}_{2n_f} & (\frac{\omega k_P}{2} - k_I)\mathbf{I}_{2n_f} 
\end{bmatrix} \leq 0,$$
which implies that
\begin{equation}\label{eq:V_single_const}
\dot{V} \leq -\omega V(t) \quad \forall t \in [t_r^A+\Delta_r^A,t_{r+1}^A).
\end{equation}
During localizability-broken attacks, i.e., $t \in [t_r^A,\,t_r^A+\Delta_r^A)$, \eqref{eq:vdot_single_const} yields
\begin{equation}\label{eq:vdot_single_const_2}
\begin{aligned}
\dot{V}&= (1-k_P)\boldsymbol{\delta}_1^\top\mathbf{L}_{ff}^{\sigma(t)}\boldsymbol{\delta}_1-k_I\textcolor{blue}{\boldsymbol{\delta}_{\eta}}^\top\textcolor{blue}{\boldsymbol{\delta}_{\eta}}-k_I\boldsymbol{\delta}_1^\top\textcolor{blue}{\boldsymbol{\delta}_{\eta}}\\
&\leq \frac{k_I \varrho}{2}\boldsymbol{\delta}_1^\top\boldsymbol{\delta}_1+\big(\frac{k_I}{2\varrho}-k_I\big)\textcolor{blue}{\boldsymbol{\delta}_{\eta}}^\top\textcolor{blue}{\boldsymbol{\delta}_{\eta}},
\end{aligned}
\end{equation}
for any $\varrho>0$. The \textcolor{blue}{above} inequality follows from the Cauchy--Schwarz inequality. Consequently, it follows from \eqref{eq:vdot_single_const_2} that
\begin{equation}\label{eq:V_single_const_attack}
\dot{V} \leq \kappa  V(t) \quad \forall t \in [t_r^A,t_{r}^A+\Delta_r^A),
\end{equation}
in which
\begin{align*}
\kappa &= \frac{2}{\lambda_{\min}(\mathcal{P})} {\max} \Big\{\frac{k_I \varrho}{2}, \frac{k_I}{2\varrho}-k_I \Big\}\\
&= \frac{4k_I}{1 + k_P - \sqrt{(k_P - 1)^2 + 4}} {\max} \Big\{\frac{\varrho}{2}, \frac{1}{2\varrho}-1 \Big\}>0.
\end{align*}
For the purpose of control gain design, we choose $\varrho>0$ to minimize
$f(\varrho)=\max\left\{\frac{\varrho}{2},\,\frac{1}{2\varrho}-1\right\}$. The minimum is attained at $\varrho = \sqrt{2} - 1$, leading to
$\kappa = \frac{2(\sqrt{2} - 1)k_I}{1 + k_P - \sqrt{(k_P - 1)^2 + 4}}$.
From \eqref{eq:V_single_const}--\eqref{eq:V_single_const_attack}, for any $t \in [t_r^A+\Delta_r^A,\,t_{r+1}^A)$, we have
\begin{equation}\label{eq:single_const_recursive}
\begin{aligned}
V(t) &\leq V(t_r^A+\Delta_r^A)e^{-\omega(t-t_r^A-\Delta_r^A)}\\
&\leq V(t_r^A)e^{\kappa\Delta_r^A}e^{-\omega(t-t_r^A-\Delta_r^A)}\\
&\leq V(t_{r-1}^A)e^{\kappa\Delta_r^A}e^{-\omega\big((t-t_r^A-\Delta_r^A)+(t_r^A-t_{r-1}^A-\Delta_{r-1}^A)\big)}\\
&\leq \dots\\
&\leq V(0)e^{\kappa|\Theta^A(0,t)|-\omega|\Xi^A(0,t)|}.
\end{aligned}
\end{equation}
The same analysis holds for $t \in [t_r^A,t_r^A+\Delta_r^A)$. Under Assumption~\ref{duration}, we have
\begin{equation}\label{eq:actual}
\begin{aligned}
\omega|\Xi^A(0,t)|-\kappa|\Theta^A(0,t)| &\geq \omega(t-\frac{t}{\tau_d}-\epsilon)-\kappa(\epsilon+\frac{t}{\tau_d})\\
&=-\epsilon(\omega+\kappa)+t(\omega-\frac{\omega+\kappa}{\tau_d}).
\end{aligned}
\end{equation}
Thus, condition \eqref{eq:main_c} implies that 
$\Bigl(\omega - \frac{\omega+\kappa}{\tau_d} \Bigr) > 0$, and consequently 
$\omega |\Xi^A(0,t)| - \kappa |\Theta^A(0,t)| \xrightarrow{t \to \infty} +\infty$, 
which leads to $V(t), \boldsymbol{\delta}_1(t), \textcolor{blue}{\boldsymbol{\delta}_{\eta}}(t) \to 0$ as $t \to \infty$. This completes the proof.
\hfill$\blacksquare$
\end{pf}
% \begin{example}
% By choosing \(k_P = 2\) and \(k_I = \omega = 2\lambda_{\min}\), conditions
% \eqref{eq:main_a}--\eqref{eq:main_b} are satisfied. Consequently,
% \eqref{eq:main_c} yields
% \[
% \tau_d > 1 + \frac{\sqrt{2}-1}{3-\sqrt{5}} \approx 1.542.
% \]
% In other words, with the above choice of control gains, the proposed control
% law ensures resilient asymptotic tracking of the desired formation provided
% that the average fraction of time during which rigidity-broken attacks occur
% does not exceed \(\frac{1}{\tau_d} \approx 64.84\%\).
% \end{example}
\textcolor{blue}{\begin{remark}
The design conditions \eqref{eq:main_a}--\eqref{eq:main_c} depend on both the formation topology, through $\underline{\lambda}$, and the attack characteristics. Specifically, \eqref{eq:main_c} imposes a lower bound on $\tau_d$, which characterizes the admissible class of localizability-broken attacks. A longer average duration of such attacks requires a larger decay rate $\omega$ during localizability-maintained periods. Moreover, \eqref{eq:main_a} and \eqref{eq:main_b} are always feasible. For example, choosing $1<k_P<2$ and $k_I=\omega$ satisfies \eqref{eq:main_b}, and any $\omega\in(0,\,2(k_P-1)\underline{\lambda}]$ then satisfies \eqref{eq:main_a}.
\end{remark}}

% \begin{remark}[Input disturbance handling]
% Suppose there exists an unknown constant input disturbance
% for each follower $\boldsymbol{\zeta}_i \in \mathbb{R}^d$, i.e.,
% $$\dot{\mathbf{p}}_i = \mathbf{u}_i + \boldsymbol{\zeta}_i \quad i=n_l+1,\dots,n.$$
% Then, under the proposed control law, \eqref{eq:single_PI_matrix} becomes
% \begin{equation}
% \begin{aligned}
% \begin{bmatrix}
%   \dot{\mathbf{p}}^{L}\\
%   \dot{\mathbf{p}}^{F}  
% \end{bmatrix}
% &=-k_P\overline{\mathbf{Z}}\mathbf{L}_b^{\sigma(t)}\mathbf{p}-
% \begin{bmatrix}
% \mathbf{0}_{2n_l}\\
% k_I\boldsymbol{\eta}(t)-\boldsymbol{\zeta}+\mathbf{1}_{n_f}\otimes\mathbf{v}_0
% \end{bmatrix}
% +\mathbf{1}_n \otimes \mathbf{v}_0,\\
% \dot{\boldsymbol{\eta}} &= \overline{\mathbf{Z}}\mathbf{L}_b^{\sigma(t)}\mathbf{p},
% \end{aligned}
% \end{equation}
% where $\boldsymbol{\zeta} = \big[ \boldsymbol{\zeta}_{n_l+1}^\top, \ldots, \boldsymbol{\zeta}_n^\top \big]^\top$ is an unknown constant vector. We then define the variable
%     $\boldsymbol{\delta}_2(t)
%     = \boldsymbol{\eta}(t)
%     - \frac{\boldsymbol{\zeta}}{k_I}
%     + \mathbf{1}_{n_f} \otimes \frac{\mathbf{v}_0}{k_I}$,
% and the subsequent analysis proceeds in a similar manner. In other words, the proposed control law effectively compensates for constant input disturbances. It is worth noting that the presence of $\boldsymbol{\zeta}_i$ does not influence the selection of the control parameters.
% \end{remark}

\begin{lemma}[Collision avoidance]\label{lm:single_collision} Suppose that Assumptions~\ref{assume:desired_formation}--\ref{assume:leader} hold and that the control gains satisfy the conditions in Theorem~\ref{thm:single_PI}. Suppose further that there exists a constant $0<D < \min_{i\neq j}\{\|\mathbf{p}_i^*-\mathbf{p}_j^*\|\}$ such that the initial formation of the agents satisfies
\begin{equation}\label{eq:single_collision}
    \|\boldsymbol{\delta}(0)\| \leq\sqrt{\frac{\lambda_{\min}(\mathcal{P})}{\lambda_{\max}(\mathcal{P})}}\frac{\min_{i\neq j}\{\|\mathbf{p}_i^*-\mathbf{p}_j^*\|\}-D}{\sqrt{e^{\epsilon(\omega+\kappa)}}\sqrt{n_f}},
\end{equation}
where $\boldsymbol{\delta}(t) = [\boldsymbol{\delta}_1^\top(t),\, \textcolor{blue}{\boldsymbol{\delta}_{\eta}}^\top(t)]^\top$, and $\mathcal{P}$, $\omega$, $\kappa$ are defined in the proof of Theorem~\ref{thm:single_PI}. Then, under the control law~\eqref{eq:single_PI}, collisions among the agents are fully avoided, i.e.,
$\|\mathbf{p}_i(t) - \mathbf{p}_j(t)\| \geq D$ for all $i,j \in \mathcal{V}$ and all $t$.
\end{lemma}
\begin{pf}
For any two agents $i,j \in \mathcal{V}$, it holds that
\begin{equation}\label{eq:pi_pj_PI}
\begin{aligned}
\|\mathbf{p}_i(t)&-\mathbf{p}_j(t)\|\\
&\geq \|\mathbf{p}_i^*-\mathbf{p}_j^*\|-\|\mathbf{p}_i-\mathbf{p}_i^*\|-\|\mathbf{p}_j-\mathbf{p}_j^*\|\\
&\geq \|\mathbf{p}_i^*-\mathbf{p}_j^*\|-\sum_{k=n_l+1}^n\|\mathbf{p}_k-\mathbf{p}_k^*\|\\
&\geq  \|\mathbf{p}_i^*-\mathbf{p}_j^*\|-\sqrt{n_f}\|\mathbf{p}-\mathbf{p}^*\|\\
% &\geq \|\mathbf{p}_i^*-\mathbf{p}_j^*\|-\sqrt{n_f}\|\boldsymbol{\delta}(t)\|\\
&\geq \|\mathbf{p}_i^*-\mathbf{p}_j^*\|-\sqrt{n_f}\sqrt{\frac{2V(t)}{\lambda_{\min}(\mathcal{P})}}.
\end{aligned}
\end{equation}
The last inequality follows directly from
$V(t)\geq \frac{1}{2}\lambda_{\min}(\mathcal{P})\|\boldsymbol{\delta}(t)\|^2$.
Furthermore, from \eqref{eq:single_const_recursive}--\eqref{eq:actual}, we have
\begin{align*}
V(t) \leq e^{\epsilon(\omega+\kappa)}V(0)\leq \frac{1}{2}e^{\epsilon(\omega+\kappa)}\lambda_{\max}(\mathcal{P})\|\boldsymbol{\delta}(0)\|^2.
\end{align*}
Meanwhile, the condition \eqref{eq:single_collision} implies
\begin{equation}
\begin{aligned}
\|\mathbf{p}_i^*-\mathbf{p}_j^*\|-\sqrt{n_f}\sqrt{\frac{e^{\epsilon(\omega+\kappa)}\lambda_{\max}(\mathcal{P})}{\lambda_{\min}(\mathcal{P})}}\|\boldsymbol{\delta}(0)\| \geq D,
\end{aligned}
\end{equation}
for all $i,j \in \mathcal{V}$. Thus, it follows that $\|\mathbf{p}_i(t)-\mathbf{p}_j(t)\|\geq D$ for all $i,j \in \mathcal{V}$ and all $t$, as a direct consequence.
\hfill$\blacksquare$
\end{pf}
% The intuition behind this lemma is that if the initial formation is sufficiently close to the desired formation, then collisions among agents can be avoided.
%%%%%%%%
\subsection{Moving Leaders With Time-Varying Velocity}\label{sec:single_time_vary}
We now consider the case of maneuvering leaders. Under Assumption~\ref{assume:leader}, the velocity vector of the leaders can be expressed as
$$\mathbf{v}^L(t)=\mathbf{1}_{n_l}\otimes \mathbf{h}(t)+(s(t)\mathbf{I}_{2n_l}+\theta(t)\bar{\mathbf{J}}_{n_l})(\mathbf{p}^L-\mathbf{1}_{n_l}\otimes\bar{\mathbf{p}}^L),$$ where $\mathbf{h}(t) \in \mathbb{R}^2$ denotes the translational velocity component, and $s(t),\theta(t) \in \mathbb{R}$ denote the scaling rate and angular velocity with respect to the leaders' centroid, respectively. Moreover, Assumption~\ref{assume:leader} ensures that $\mathbf{h}(t)$, $s(t)$, and $\theta(t)$ are bounded and uniformly continuous. The matrix form of the considered multi-agent system can thus be obtained as follows 
\begin{equation}\label{eq:single_matrix}
\begin{aligned}
\dot{\mathbf{p}}&=
\begin{bmatrix}
\mathbf{1}_{n_l}\otimes \mathbf{h}+(s\mathbf{I}_{2n_l}+\theta\bar{\mathbf{J}}_{n_l})(\mathbf{p}^L-\mathbf{1}_{n_l}\otimes\bar{\mathbf{p}}^L)\\
\mathbf{u}^F
\end{bmatrix}\\
&=\overline{\mathbf{Z}}
\begin{bmatrix}
\mathbf{0}_{2n_l}\\
\mathbf{u}^F+\boldsymbol{\chi}^F
\end{bmatrix}
+\boldsymbol{\xi}(t)=\overline{\mathbf{Z}}(\mathbf{u}+\boldsymbol{\chi})+\boldsymbol{\xi}(t),
\end{aligned}
\end{equation}
where $\overline{\mathbf{Z}}=\mathbf{Z}\otimes \mathbf{I}_2$, $\mathbf{Z}=\begin{bmatrix} \mathbf{0}_{n_l \times n_l} & \mathbf{0}_{n_l \times n_f} \\ \mathbf{0}_{n_f \times n_l} & \mathbf{I}_{n_f} \end{bmatrix}$, $\mathbf{u}^F=[\mathbf{u}_{n_l+1}^\top,\dots,\mathbf{u}_{n}^\top]^\top$, $\boldsymbol{\chi}(t)=[\mathbf{0}_{2n_l}^\top,(\boldsymbol{\chi}^F)^\top]^\top$, and
\begin{align*}
\boldsymbol{\chi}^F(t)&=-\mathbf{1}_{n_f}\otimes \mathbf{h}-(s\mathbf{I}_{2n_f}+\theta\bar{\mathbf{J}}_{n_f})(\mathbf{p}^{F*}-\mathbf{1}_{n_f}\otimes\bar{\mathbf{p}}^L),\\
\boldsymbol{\xi}(t)&=\mathbf{1}_{n}\otimes \mathbf{h}+(s\mathbf{I}_{2n}+\theta\bar{\mathbf{J}}_{n})(\mathbf{p}^*-\mathbf{1}_{n}\otimes\bar{\mathbf{p}}^L).
\end{align*}
\begin{lemma}\label{cor:desired_velocity}
The desired velocity of the formation is given by $\dot{\mathbf{p}}^{*}(t) = \boldsymbol{\xi}(t)$.
\end{lemma}
\begin{pf}
It can be seen that $\boldsymbol{\xi}(t)\in \text{null}(\mathbf{R}_{\mathcal{A}}(\alpha^*))$. Hence, it follows from (\ref{eq:lapcacian}) that  
$\mathbf{L}_{fl}(\alpha^*)\mathbf{v}^L + \mathbf{L}_{ff}(\alpha^*)\boldsymbol{\xi}^F = \mathbf{0}$,  
where $\boldsymbol{\xi}^F$ denotes the last $2n_f$ entries of $\boldsymbol{\xi}$. In other words, we have $\boldsymbol{\xi}^F=-\mathbf{L}_{ff}^{-1}(\alpha^*) \mathbf{L}_{fl}(\alpha^*) \dot{\mathbf{p}}^L(t)$. Combining this result with Lemma~\ref{lmm:desired_formation}, the proof is completed.
\hfill$\blacksquare$
\end{pf}

It follows directly from Lemma~\ref{cor:desired_velocity} that the desired velocity of the followers is given by $\dot{\mathbf{p}}^{F*}(t) = -\boldsymbol{\chi}^F(t)$. The following control law is proposed for each follower
\begin{equation}\label{eq:single_SMC}
\begin{aligned}
\mathbf{u}_i(t)&=\\
-\gamma &\left[ \sum_{{(i,j,k)} \in \bar{\mathcal{A}}^{\sigma(t)}} (\mathbf{A}_i^{\blacktriangle {ijk}}(\alpha^*))^\top \mathrm{sgn}\big(f_i^{\blacktriangle {ijk}}(\alpha^*, \mathbf{p})\big) \right. \\
& + \sum_{{(j,i,k)} \in \bar{\mathcal{A}}^{\sigma(t)}} (\mathbf{A}_i^{\blacktriangle {jik}}(\alpha^*))^\top \mathrm{sgn}\big(f_i^{\blacktriangle {jik}}(\alpha^*, \mathbf{p})\big) \\
& + \left. \sum_{{(j,k,i)} \in \bar{\mathcal{A}}^{\sigma(t)}} (\mathbf{A}_i^{\blacktriangle {jki}}(\alpha^*))^\top \mathrm{sgn}\big(f_i^{\blacktriangle {jki}}(\alpha^*, \mathbf{p})\big) \right],
\end{aligned}
\end{equation}
where the control parameter $\gamma>0$ will be designed later. Under the above control law, the overall system dynamics can be expressed in the following matrix form
\begin{equation}\label{eq:single_SMC_matrix}
\begin{aligned}
% \begin{bmatrix}
%   \dot{\mathbf{p}}^{L}\\
%   \dot{\mathbf{p}}^{F}  
% \end{bmatrix}
\dot{\mathbf{p}}
&=\overline{\mathbf{Z}}\bigg(-\gamma\big(\mathbf{R}_{\mathcal{A}}^{\sigma(t)}\big)^\top\mathrm{sgn}(\mathbf{R}_{\mathcal{A}}^{\sigma(t)}\mathbf{p}) +\boldsymbol{\chi}\bigg)+\boldsymbol{\xi}(t).
\end{aligned}
\end{equation}
Denote the tracking error as $\boldsymbol{\delta}(t)=\mathbf{p}(t)-\mathbf{p}^*(t)=[\mathbf{0}_{2n_l}^\top,(\boldsymbol{\delta}^F)^\top]^\top$, where $\boldsymbol{\delta}^F=\mathbf{p}^F-\mathbf{p}^{F*}$. Thus, it follows from Lemma~\ref{cor:desired_velocity} that
$$\dot{\boldsymbol{\delta}}=\overline{\mathbf{Z}}\bigg(-\gamma\big(\mathbf{R}_{\mathcal{A}}^{\sigma(t)}\big)^\top\mathrm{sgn}(\mathbf{R}_{\mathcal{A}}^{\sigma(t)}\mathbf{p}) +\boldsymbol{\chi}\bigg).$$

Due to the discontinuity of the control law \eqref{eq:single_SMC}, the solution is understood in the Filippov sense. We introduce the following theorem.
\begin{theorem}\label{thm:single_recovery}
Consider Problem~\ref{prob1}, where the desired velocity of the followers is bounded by a known constant~$\beta$, i.e., $\|\boldsymbol{\chi}(t)\|_{\infty} \leq \beta$. Suppose that Assumptions~\ref{assume:desired_formation}--\ref{assume:leader} hold, and that there is no collision among the agents. Under the control law \eqref{eq:single_SMC}, if the control gain is chosen such that \textcolor{blue}{$\gamma > \frac{\tau_d}{\tau_d - 1} \beta \sqrt{\frac{2n}{\underline{\lambda}}}$}, the following holds
\begin{itemize}
    \item \textcolor{blue}{Under localizability-maintained attacks only,} $\mathbf{p}(t)\to\mathbf{p}^*(t)$ in finite time. 
    \item \textcolor{blue}{In the presence of localizability-broken attacks, $\|\mathbf{p}(t)-\mathbf{p}^*(t)\|$ remains bounded for all $t\ge 0$, and the control law~\eqref{eq:single_SMC} ensures resilient finite-time recovery of formation tracking.}
\end{itemize}
\end{theorem}
\begin{pf}
Consider the following common Lyapunov function candidate $V(t)=\frac{1}{2}(\boldsymbol{\delta}^{F})^\top\boldsymbol{\delta}^F=\frac{1}{2}\boldsymbol{\delta}^\top\boldsymbol{\delta}$. Then, $\dot V \in \dot{\tilde V}$ almost everywhere, 
where $\dot{\tilde V}=\nabla V^\top\mathcal{K}[\dot{\boldsymbol{\delta}}]$ denotes the set-valued Lie derivative of $V$ and $\mathcal{K}[\cdot]$ denotes the Filippov set-valued map \cite{cortes2008}. Keeping in mind that $\overline{\mathbf{Z}}\boldsymbol{\delta}=\boldsymbol{\delta}$ and $\mathbf{R}_{\mathcal{A}}^{\sigma(t)}\mathbf{p}=\mathbf{R}_{\mathcal{A}}^{\sigma(t)}\boldsymbol{\delta}$, the following holds
\begin{equation}
\begin{aligned}
 \dot{\tilde V} &=  \boldsymbol{\delta}^\top\overline{\mathbf{Z}}(-\gamma(\mathbf{R}_{\mathcal{A}}^{\sigma(t)})^\top\mathcal{K}[\mathrm{sgn}(\mathbf{R}_{\mathcal{A}}^{\sigma(t)}\mathbf{p})]+\boldsymbol{\chi})\\
 &=-\gamma \boldsymbol{\delta}^\top\big((\mathbf{R}_{\mathcal{A}}^{\sigma(t)})^\top\mathcal{K}[\mathrm{sgn}(\mathbf{R}_{\mathcal{A}}^{\sigma(t)}\mathbf{p})]\big)+\boldsymbol{\delta}^\top\boldsymbol{\chi}\\
 &=-\gamma\|\mathbf{R}_{\mathcal{A}}^{\sigma(t)}\boldsymbol{\delta}\|_1+\boldsymbol{\delta}^\top\boldsymbol{\chi}.
\end{aligned}
\end{equation}
Here, we used the property of Filippov's differential inclusion, $\mathbf{x}^\top \mathcal{K}[\mathrm{sgn}(\mathbf{x})] = \{ \|\mathbf{x}\|_1 \}$. Thus, $\dot{\tilde V}$ is a singleton and $\dot{V}=\dot{\tilde V}$. By utilizing the standard norm inequality $\|\mathbf{x}\|_2 \le \|\mathbf{x}\|_1 \le \sqrt{d} \|\mathbf{x}\|_2$ for any $\mathbf{x} \in \mathbb{R}^d$, we can further obtain
\begin{equation}\label{eq:26}
\begin{aligned}
\dot{V}  &\leq -\gamma\|\mathbf{R}_{\mathcal{A}}^{\sigma(t)}\boldsymbol{\delta}\|_1+\|\boldsymbol{\delta}\|_1\|\boldsymbol{\chi}\|_{\infty}\\
&\leq
-\gamma\|\mathbf{R}_{\mathcal{A}}^{\sigma(t)}\boldsymbol{\delta}\|+\beta\|\boldsymbol{\delta}\|_1\\
&\leq 
-\gamma(\boldsymbol{\delta}^\top\mathbf L_{\mathcal A}^{\sigma(t)}\boldsymbol{\delta})^{1/2}+\beta\sqrt{2n}\|\boldsymbol{\delta}\|\\
&=-\gamma\big((\boldsymbol{\delta}^F)^\top\mathbf{L}_{ff}^{\sigma(t)}\boldsymbol{\delta}^F\big)^{1/2}+\beta\sqrt{2n}\|\boldsymbol{\delta}\|.
\end{aligned}
\end{equation}
%  Note that, the following always holds
% \begin{align*}
% (\boldsymbol{\delta}^F)^\top\mathbf{L}_{ff}^{\sigma(t)}\boldsymbol{\delta}^F)\geq{\underline{\lambda}}(\mathbf{L}_{ff}(\alpha^*))\|\boldsymbol{\delta}^F\|^2=\lambda_{\min}(\mathbf{L}_{ff}(\alpha^*))\|\boldsymbol{\delta}\|^2.
% \end{align*}
Thus, for $t \in [t_r^A+\Delta_r^A,t_{r+1}^A)$, we have
\begin{equation}
\begin{aligned}
\dot{V} &\leq -\gamma \underline{\lambda}^{1/2}\|\boldsymbol{\delta}^F\|+\beta \sqrt{2n}\|\boldsymbol{\delta}^F\|\\
&=-\underbrace{\sqrt{2}(\gamma \underline{\lambda}^{1/2}-\beta\sqrt{2n})}_{\rho > 0}\frac{\|\boldsymbol{\delta}^F\|}{\sqrt{2}}=-\rho\sqrt{V}.
\end{aligned}
\end{equation}
\textbullet\ Under localizability-maintained attacks only, $\dot V\leq-\rho V^{1/2}$ implies that $V(t)=0$ for all $t\geq T_1=\frac{2V^{1/2}(0)}{\rho}$, which proves the first statement.

\textbullet\ Consider the intervals during which localizability-broken attacks occur. In these intervals, it follows from \eqref{eq:26} that
\begin{equation}\label{eq:V_single_smc_attack}
\begin{aligned}
\dot{V} \leq \underbrace{2\beta\sqrt{n}}_{\kappa}\sqrt{V}=\kappa \sqrt{V}.
\end{aligned}
\end{equation}
For any $\tau$ satisfying $V(\tau)\neq 0$, suppose that there does not exist
$t^*\in[\tau,\tau+T_2]$ such that $V(t^*)=0$. Then, integrating
$\frac{\dot{V}(t)}{\sqrt{V(t)}}$ over the interval $[\tau,t]$ yields
\begin{equation}
V^{1/2}(t) \leq 
\begin{cases}
V^{1/2}(t_r^A) + \dfrac{\kappa}{2} (t - t_r^A),\\
\qquad \qquad \qquad t \in [t_r^A, t_r^A+\Delta_r^A),\\[1mm]
V^{1/2}(t_r^A+\Delta_r^A) - \dfrac{\rho}{2} (t - t_r^A - \Delta_r^A),\\
\qquad \qquad \qquad t \in [t_r^A+\Delta_r^A, t_{r+1}^A).
\end{cases}
\end{equation}
Consequently, the following relation holds
\begin{equation}\label{recursive:SMC}
\begin{aligned}
V&^{1/2}(t) \leq V^{1/2}(t_r^A+{\Delta_r^A})-\frac{\rho}{2}(t-t_r^A-\Delta_r^A)\\
&\leq  V^{1/2}(t_r^A)+ \frac{\kappa}{2}\Delta_r^A-\frac{\rho}{2}(t-t_r^A-\Delta_r^A)\\
&\leq  V^{1/2}(t_{r-1}^A+\Delta_{r-1}^A)+ \frac{\kappa}{2}\Delta_r^A\\
& \quad-\frac{\rho}{2}\left[(t-t_r^A-\Delta_r^A) + (t_r^A-t_{r-1}^A-\Delta_{r-1}^A)\right]\\
&\leq \dots\\
&\leq V^{1/2}(\tau)+\frac{\kappa}{2}|\Theta^A(\tau,t)|-\frac{\rho}{2}|\Xi^A(\tau,t)|\\
&\leq V^{1/2}(\tau)+\frac{\rho+\kappa}{2}\epsilon+(t-\tau)(\frac{\rho+\kappa}{2\tau_d}-\frac{\rho}{2}).
\end{aligned}
\end{equation}
Here, the last inequality follows from Assumption~\ref{duration}. The condition stated in the theorem implies that $\frac{\rho+\kappa}{\tau_d}<\rho$. For $t^*-\tau=T_2:=\big(\frac{\rho}{2}-\frac{\rho+\kappa}{2\tau_d}\big)^{-1}\big(V^{1/2}(\tau)+\frac{\rho+\kappa}{2}\epsilon\big)$, we have $V^{1/2}(t^*)\le0$. However, since $V^{1/2}(\boldsymbol{\delta}^F)$ is positive definite, it follows by contradiction that there exists a finite time $t^* \leq \tau + T_2$ such that $\boldsymbol{\delta}^F(t^*) = \mathbf{0}$. Finally, since the last term in \eqref{recursive:SMC} is nonpositive, $V^{1/2}(t)\leq V^{1/2}(\tau)+\frac{\rho+\kappa}{2}\epsilon$ as long as $V>0$ on $[\tau,t]$. Applying this bound from $\tau=0$ and from every instant at which $V$ leaves zero yields $V^{1/2}(t)\leq V^{1/2}(0)+\frac{\rho+\kappa}{2}\epsilon$ for all $t\geq0$, i.e., $\|\mathbf{p}(t)-\mathbf{p}^*(t)\|$ remains bounded. This completes the proof.
\hfill$\blacksquare$
\end{pf}

% \begin{remark}
% We next establish that the proposed control law ensures fixed-time recovery of the formation tracking objective. For $V(0)\neq0$, the formation error reaches zero within a finite time $t^*\leq T_2$, i.e., $V(t^*)=0$, and remains at zero until the onset of the next localizability-broken attack at $t=t_r^A$. At this instant, $V(t_r^{A-})=0$ and $\dot V(t)\leq\sqrt{V(t)}$. Although $\sqrt{V}$ is not Lipschitz continuous at the origin and thus the zero solution is nonunique, the comparison lemma ensures that $V(t)$ is upper bounded by the maximal nontrivial solution of $\dot y=\sqrt{y}$ initialized at $y(t_r^A)=0$, namely,
% \[
% V(t)\le \frac{(t-t_r^A)^2}{4},
% \quad \forall t\in[t_r^A,\, t_r^A+\Delta_r^A).
% \]
% Thus, $V(t_r^{A+})=0$. Hence, for every localizability-broken attack occurring at time $t_r^A$, the formation recovers within a fixed time upper bounded by $T_2^{*} := \left(\frac{\rho}{2}-\frac{\rho+\kappa}{2\tau_d}\right)^{-1} \frac{\rho+\kappa}{2}\epsilon$. 
% \end{remark}
\begin{lemma}\label{lm:single_collision_SMC}
Suppose that Assumptions~\ref{assume:desired_formation}--\ref{assume:leader} hold and that the control gain satisfies the condition in Theorem~\ref{thm:single_recovery}. Suppose further that the target formation never shrinks below its initial size, i.e., $\int_0^t s(\tau)d\tau\geq0$ for all $t\geq0$, and that there exists a constant $0<D<\min_{i\neq j}\|\mathbf{p}_i^*(0)-\mathbf{p}_j^*(0)\|$ such that the initial agent formation satisfies
\begin{equation}\label{eq:single_collision_SMC}
    \|\boldsymbol{\delta}^F(0)\| \leq\frac{\min_{i\neq j}\{\|\mathbf{p}_i^*{(0)}-\mathbf{p}_j^*{(0)}\|\}-D}{\sqrt{n_f}}-\frac{\rho+\kappa}{\sqrt{2}}\,\epsilon,
\end{equation}
where $\rho$ and $\kappa$ are defined in the proof of Theorem~\ref{thm:single_recovery}. 
Under the control law~\eqref{eq:single_SMC}, collisions among the agents are fully avoided, i.e.,
$\|\mathbf{p}_i(t) - \mathbf{p}_j(t)\| \geq D$ for all $i,j \in \mathcal{V}$ and all $t$.
\end{lemma}
\begin{pf}
Since translations and rotations preserve inter-agent distances, $\|\mathbf{p}_i^*(t)-\mathbf{p}_j^*(t)\|=e^{\int_0^t s(\tau)d\tau}\|\mathbf{p}_i^*(0)-\mathbf{p}_j^*(0)\|\geq\|\mathbf{p}_i^*(0)-\mathbf{p}_j^*(0)\|$ for all $t\geq0$. Hence, by following the same procedure as \eqref{eq:pi_pj_PI}, we obtain
$$\|\mathbf{p}_i(t)-\mathbf{p}_j(t)\| \geq \|\mathbf{p}_i^*{(0)}-\mathbf{p}_j^*{(0)}\|-\sqrt{n_f}\sqrt{2V(t)}.$$
Consider the time interval from $t = 0$ to the first recovery time. For any $t$ within this interval, it follows from~\eqref{recursive:SMC} that
\begin{align*}
V^{1/2}(t) \leq V^{1/2}(0)+\frac{\rho+\kappa}{2}\epsilon\leq \frac{\|\boldsymbol{\delta}^F(0)\|}{\sqrt{2}}+\frac{\rho+\kappa}{2}\epsilon.
\end{align*}
Meanwhile, condition~\eqref{eq:single_collision_SMC} implies that
\[
\|\mathbf{p}_i^*{(0)} - \mathbf{p}_j^*{(0)}\|
- \sqrt{n_f}\left(\|\boldsymbol{\delta}^F(0)\| + \frac{\rho+\kappa}{\sqrt{2}}\,\epsilon\right)
\geq D.
\]
Thus, $\|\mathbf{p}_i(t)-\mathbf{p}_j(t)\|\geq D$ for all $i,j\in\mathcal{V}$ and for all $t$ prior to the first recovery time. If there are no localizability-broken attacks, $\|\boldsymbol{\delta}^F(t)\|$ remains at the origin, which completes the proof. Otherwise, let $t_r^A$ denote the first localizability-broken attack after this instant. Since $V(t_r^{A+})=0$, for all $t$ {between $t_r^A$ and the next recovery time}, we have
$V^{1/2}(t) \leq \frac{\rho+\kappa}{2}\,\epsilon$ and $\|\mathbf{p}_i^*{(0)} - \mathbf{p}_j^*{(0)}\|
- \sqrt{n_f}\frac{\rho+\kappa}{\sqrt{2}}\,\epsilon
\geq D$. Therefore, inter-agent collisions are avoided over this interval, as well as over all subsequent intervals between consecutive recovery times. This completes the proof.
\hfill$\blacksquare$
\end{pf}
\section{Resilient Simultaneous
Localization and Formation Control}\label{sec:RSLAF}
This section addresses Problem~\ref{prob2}. First, a resilient distributed observer and estimation algorithm are developed to enable each follower to estimate its desired trajectory in finite time. Then, resilient localization and formation tracking control laws are proposed. {The following assumptions are additionally imposed.}
{
\begin{assume}[Minimum dwell time]\label{assume:dwell}
There exists a constant $\tau_s>0$ such that any two consecutive switching instants of $\sigma(t)$ are separated by at least $\tau_s$.
\end{assume}
\begin{remark}
Assumption~\ref{assume:dwell} is mild and practically reasonable. The constant $\tau_s$ can be arbitrarily small and does not need to be known, since none of the control parameters shown later depends on it. The assumption only excludes infinitely many switches within a finite time interval, which is not expected in practical systems.
\end{remark}}
\begin{assume}\label{assume:nonsing}
For every localizability-maintained mode $\sigma$, the matrix $\mathbf{L}_{ff}^{\sigma}(\alpha(\mathbf{p}))$ is nonsingular at every configuration $\mathbf{p}$ in which no two agents coincide and all active triangles are nondegenerate.
\end{assume}
\begin{remark}
Assumption~\ref{assume:nonsing} is implicitly required in \cite{CHEN2022auto_localization}. It holds at every collision-free configuration with nondegenerate active triangles if the triangle set of each localizability-maintained mode is constructed by vertex addition \cite{LIU2026auto_mergesplit}, i.e., by successively adding followers, each forming an active triangle with leaders or previously added followers. For example, the formation in Section~\ref{sec:sim} is constructed by adding followers $4$, $3$, $5$, $6$, and $7$ in order.
\end{remark}
\subsection{Finite-Time Resilient Distributed Observer and Target Formation Estimator}
Since not every follower has access to $\boldsymbol{\zeta}(t)$, $\mathbf{h}(t)$, or the system matrices $\mathbf{S}$ and $\mathbf{F}$, the following adaptive distributed observer is proposed for each follower $i$
\begin{subequations}\label{eq:distributed_observer}
\small
\begin{align}
\dot{\mathbf{S}}_{0i} &=
-{\gamma_S}\Bigl[\sum_{j={n_l+1}}^na_{ij}^{\sigma(t)}\mathrm{sgn}\big(a_{ij}^{\sigma(t)}(\mathbf{S}_{0i}-\mathbf{S}_{0j})\big)\notag\\
&\quad \quad \quad \quad+b_i^{\sigma(t)}\mathrm{sgn}\big(b_i^{\sigma(t)}(\mathbf{S}_{0i}-\mathbf{S}_0)\big)\Bigr],\label{eq:observer_a}\\
\dot{\mathbf{F}}_{0i} &=
-{\gamma_F}\Bigl[\sum_{j={n_l+1}}^na_{ij}^{\sigma(t)}\mathrm{sgn}\big(a_{ij}^{\sigma(t)}(\mathbf{F}_{0i}-\mathbf{F}_{0j})\big)\notag\\
&\quad \quad \quad \quad+b_i^{\sigma(t)}\mathrm{sgn}\big(b_i^{\sigma(t)}(\mathbf{F}_{0i}-\mathbf{F}_0)\big)\Bigr],\label{eq:observer_b}\\
\dot{\boldsymbol{\zeta}}_i &= \mathbf{S}_i \boldsymbol{\zeta}_i-{k_\zeta}\Bigl[\sum_{j={n_l+1}}^na_{ij}^{\sigma(t)}(\boldsymbol{\zeta}_i-\boldsymbol{\zeta}_j)+b_i^{\sigma(t)}(\boldsymbol{\zeta}_i-\boldsymbol{\zeta})\Bigr]\notag\\
-&{\gamma_\zeta}\Bigl[\sum_{j={n_l+1}}^na_{ij}^{\sigma(t)}\mathrm{sgn}\big(a_{ij}^{\sigma(t)}(\boldsymbol{\zeta}_i-\boldsymbol{\zeta}_j)\big)+b_i^{\sigma(t)}\mathrm{sgn}\big(b_i^{\sigma(t)}(\boldsymbol{\zeta}_i-\boldsymbol{\zeta})\big)\Bigr],\label{eq:observer_c}\\ 
&\mathbf{h}_i(t)=\mathbf{F}_i(t)\boldsymbol{\zeta}_i(t).\label{eq:observer_d}
\end{align}
\end{subequations}
Here, $\mathbf{S}_0 = \operatorname{vec}(\mathbf{S}) \in \mathbb{R}^{l^2}$ and $\mathbf{F}_0 = \operatorname{vec}(\mathbf{F}) \in \mathbb{R}^{2l}$. Furthermore, {$\mathbf{S}_{0i}(t)$ and $\mathbf{F}_{0i}(t)$} denote follower $i$'s estimates of $\mathbf{S}_0$ and $\mathbf{F}_0$, respectively. In addition, $\mathbf{S}_i = M_l^l(\mathbf{S}_{0i})$ and $\mathbf{F}_i = M_2^l(\mathbf{F}_{0i})$. The variables $\boldsymbol{\zeta}_i(t)$ and $\mathbf{h}_i(t)$ denote follower $i$'s estimates of the exosystem state $\boldsymbol{\zeta}(t)$ and the desired velocity $\mathbf{h}(t)$, respectively. Finally, {$\gamma_S,\gamma_F,k_\zeta,\gamma_\zeta>0$} are design constants to be determined later. Based on the proposed distributed observer, each follower $i$ estimates its desired trajectory $\hat{\mathbf{p}}_i^*(t)$ with the following update law
\begin{equation}\label{eq:target_estimator}
\begin{aligned}
\dot{\hat{\mathbf{p}}}^*_i &= \mathbf{h}_i(t)\\
-\gamma &\left[ \sum_{{(i,j,k)} \in \bar{\mathcal{A}}^{\sigma(t)}} (\mathbf{A}_i^{\blacktriangle {ijk}}(\alpha^*))^\top \mathrm{sgn}\big(f_i^{\blacktriangle {ijk}}(\alpha^*, \hat{\mathbf{p}}^*)\big) \right.\\
 +& \sum_{{(j,i,k)} \in \bar{\mathcal{A}}^{\sigma(t)}} (\mathbf{A}_i^{\blacktriangle {jik}}(\alpha^*))^\top \mathrm{sgn}\big(f_i^{\blacktriangle {jik}}(\alpha^*, \hat{\mathbf{p}}^*)\big) \\
 +& \left. \sum_{{(j,k,i)} \in \bar{\mathcal{A}}^{\sigma(t)}} (\mathbf{A}_i^{\blacktriangle {jki}}(\alpha^*))^\top \mathrm{sgn}\big(f_i^{\blacktriangle {jki}}(\alpha^*, \hat{\mathbf{p}}^*)\big) \right],
\end{aligned}
\end{equation}
where $\hat{\mathbf{p}}^*=[(\mathbf{p}^L)^\top,(\hat{\mathbf{p}}^{F*})^\top]^\top=[(\mathbf{p}^L)^\top,\hat{\mathbf{p}}^*_{n_l+1},\dots,\hat{\mathbf{p}}^*_n]^\top$, and $\gamma>0$ is a constant. 

To analyze the performance of the proposed distributed observer, define $\tilde{\mathbf{S}}_i(t)=\mathbf{S}_i(t)-\mathbf{S}$, $\tilde{\mathbf{F}}_i(t)=\mathbf{F}_i(t)-\mathbf{F}$, $\tilde{\mathbf{S}}_{0i}(t)=\mathbf{S}_{0i}(t)-\mathbf{S}_0$, $\tilde{\mathbf{F}}_{0i}(t)=\mathbf{F}_{0i}(t)-\mathbf{F}_0$, $\tilde{\boldsymbol{\zeta}}_i(t)=\boldsymbol{\zeta}_i(t)-\boldsymbol{\zeta}(t)$, and $\tilde{\mathbf{h}}_i(t)=\mathbf{h}_i(t)-\mathbf{h}(t)$. We obtain
 
\begin{subequations}\label{eq:distributed_observer_error}
\small
\begin{align}
\dot{\tilde{\mathbf{S}}}_{0i} &=
-{\gamma_S}\Bigl[\sum_{j={n_l+1}}^na_{ij}^{\sigma(t)}\mathrm{sgn}\big(a_{ij}^{\sigma(t)}(\tilde{\mathbf{S}}_{0i}-\tilde{\mathbf{S}}_{0j})\big)\notag\\
&\qquad \quad \quad+b_i^{\sigma(t)}\mathrm{sgn}\big(b_i^{\sigma(t)}\tilde{\mathbf{S}}_{0i}\big)\Bigr],\label{eq:observer_error_a}\\
\dot{\tilde{\mathbf{F}}}_{0i} &=
-{\gamma_F}\Bigl[\sum_{j={n_l+1}}^na_{ij}^{\sigma(t)}\mathrm{sgn}\big(a_{ij}^{\sigma(t)}(\tilde{\mathbf{F}}_{0i}-\tilde{\mathbf{F}}_{0j})\big)\notag\\
&\qquad \quad \quad+b_i^{\sigma(t)}\mathrm{sgn}\big(b_i^{\sigma(t)}\tilde{\mathbf{F}}_{0i}\big)\Bigr],\label{eq:observer_error_b}\\
\dot{\tilde{\boldsymbol{\zeta}}}_i &=  \tilde{\mathbf{S}}_i \tilde{\boldsymbol{\zeta}}_i+\mathbf{S} \tilde{\boldsymbol{\zeta}}_i+\tilde{\mathbf{S}}_i \boldsymbol{\zeta}-{k_\zeta}\Bigl[\sum_{j={n_l+1}}^na_{ij}^{\sigma(t)}(\tilde{\boldsymbol{\zeta}}_i-\tilde{\boldsymbol{\zeta}}_j)+b_i^{\sigma(t)}\tilde{\boldsymbol{\zeta}}_i\Bigr]\notag\\
-&{\gamma_\zeta}\Bigl[\sum_{j={n_l+1}}^na_{ij}^{\sigma(t)}\mathrm{sgn}\big(a_{ij}^{\sigma(t)}(\tilde{\boldsymbol{\zeta}}_i-\tilde{\boldsymbol{\zeta}}_j)\big)+b_i^{\sigma(t)}\mathrm{sgn}\big(b_i^{\sigma(t)}\tilde{\boldsymbol{\zeta}}_i\big)\Bigr],\label{eq:observer_error_c}\\ 
&\tilde{\mathbf{h}}_i(t)=\tilde{\mathbf{F}}_i\tilde{\boldsymbol{\zeta}}_i+\mathbf{F}\tilde{\boldsymbol{\zeta}}_i+\tilde{\mathbf{F}}_i \boldsymbol{\zeta}.\label{eq:observer_error_d}
\end{align}
\end{subequations}
\begin{lemma}\label{lm:DOB}
Consider the distributed observer (\ref{eq:distributed_observer}) under Assumptions~\ref{duration}, \ref{assume:exo}, and \ref{assume:communicate}. \textcolor{blue}{Let $\lambda_{\min}^{S}(\boldsymbol{\mathcal{H}}^{\sigma})>0$ denote the minimum eigenvalue of $\boldsymbol{\mathcal{H}}^{\sigma}$ over all localizability-maintained modes $\sigma$. For any $\gamma_S,\gamma_F>0$, if $k_\zeta$, $\gamma_\zeta$, $\omega$, and $\kappa$ are selected such that}
\begin{subequations}\label{eq:cond_distributed_observer}
\begin{align}
 \mathbf{S}+\mathbf{S}^\top  &<\textcolor{blue}{\big(2k_\zeta\lambda_{\min}^{S}(\boldsymbol{\mathcal{H}}^{\sigma})-\omega\big)}\mathbf{I}_{l} \label{eq:DOB_a},\\
\mathbf{S}+\mathbf{S}^\top &<\kappa\mathbf{I}_{l} \label{eq:DOB_b},\\
\tau_d &> \frac{\kappa+\omega}{\omega}\label{eq:DOB_c},\\
\gamma_\zeta &> \frac{1}{\tau_d - 1} \frac{\kappa}{\sqrt{2\lambda_{\min}^{S}(\boldsymbol{\mathcal{H}}^{\textcolor{blue}{\sigma}})}}\label{eq:DOB_d},
\end{align}
\end{subequations}
\textcolor{blue}{then} each follower estimates the leaders' velocity in finite time, i.e., there exists a finite time $T_{DOB}>0$ such that $\tilde{\boldsymbol{\zeta}}_i(t)=\mathbf{0}$ and $\tilde{\mathbf{h}}_i(t)=\mathbf{0}$ for all \textcolor{blue}{followers} $i$ and all $t\ge T_{DOB}$.
\end{lemma}
\begin{pf}
As the right-hand sides of \eqref{eq:observer_error_a}--\eqref{eq:observer_error_c} are discontinuous, their solutions are understood in the Filippov sense. The proof consists of three parts. First, we establish the finite-time stability of $\tilde{\mathbf{S}}_{0i}$ and $\tilde{\mathbf{F}}_{0i}$ with settling time $T_1$. Next, we prove that the observer errors $\tilde{\boldsymbol{\zeta}}_i$ and $\tilde{\mathbf{h}}_i$ remain bounded over the interval $[0,T_1)$, thereby ruling out finite-time escape. Lastly, we prove that the observer errors $\tilde{\boldsymbol{\zeta}}_i$ and $\tilde{\mathbf{h}}_i$ converge to zero in finite time $T_{DOB}$.

\textbullet\ \textbf{Part 1:} Finite-time convergence of $\tilde{\mathbf{S}}_i$ and $\tilde{\mathbf{F}}_i$.

Define $\tilde{\mathbf{S}}=[\tilde{\mathbf{S}}_{0n_l+1}^\top,\dots,\tilde{\mathbf{S}}_{0n}^\top]^\top$. From (\ref{eq:observer_error_a}), it follows that
\begin{equation}
\begin{aligned}
\dot{\tilde{\mathbf{S}}}=& -{\gamma_S}(\mathbf{H}_f^{\sigma(t)}\otimes \mathbf{I}_{l^2})^\top\mathrm{sgn}\big((\mathbf{H}_f^{\sigma(t)}\otimes \mathbf{I}_{l^2})\tilde{\mathbf{S}}\big)\\
&-{\gamma_S}(\boldsymbol{\mathcal{D}}_f^{\sigma(t)}\otimes \mathbf{I}_{l^2})^\top\mathrm{sgn}\big((\boldsymbol{\mathcal{D}}_f^{\sigma(t)}\otimes \mathbf{I}_{l^2})\tilde{\mathbf{S}}\big).
\end{aligned}
\end{equation}
Consider the Lyapunov function {candidate} $V_1 = \frac{1}{2}\tilde{\mathbf{S}}^\top \tilde{\mathbf{S}}$. We have
\begin{equation}
\begin{aligned}
 \dot{\tilde V}_1 =&  -{\gamma_S}\tilde{\mathbf{S}}^\top(\mathbf{H}_f^{\sigma(t)}\otimes \mathbf{I}_{l^2})^\top\mathcal{K}[\mathrm{sgn}\big((\mathbf{H}_f^{\sigma(t)}\otimes \mathbf{I}_{l^2})\tilde{\mathbf{S}}\big)]\\
 &-{\gamma_S}\tilde{\mathbf{S}}^\top(\boldsymbol{\mathcal{D}}_f^{\sigma(t)}\otimes \mathbf{I}_{l^2})^\top\mathcal{K}[\mathrm{sgn}\big((\boldsymbol{\mathcal{D}}_f^{\sigma(t)}\otimes \mathbf{I}_{l^2})\tilde{\mathbf{S}}\big)]\\
 =&-{\gamma_S}\|(\mathbf{H}_f^{\sigma(t)}\otimes \mathbf{I}_{l^2})\tilde{\mathbf{S}}\|_1-{\gamma_S}\|(\boldsymbol{\mathcal{D}}_f^{\sigma(t)}\otimes \mathbf{I}_{l^2})\tilde{\mathbf{S}}\|_1.
\end{aligned}
\end{equation}
Hence, $\dot{\tilde V}_1$ is a singleton and $\dot{V}_1=\dot{\tilde V}_1$. Using the norm inequality, we can further obtain
\begin{equation}
\begin{aligned}
\dot{V}_1 \leq& -{\gamma_S}\|(\mathbf{H}_f^{\sigma(t)}\otimes \mathbf{I}_{l^2})\tilde{\mathbf{S}}\|-{\gamma_S}\|(\boldsymbol{\mathcal{D}}_f^{\sigma(t)}\otimes \mathbf{I}_{l^2})\tilde{\mathbf{S}}\|\\
\leq&-{\gamma_S}\big(\|(\mathbf{H}_f^{\sigma(t)}\otimes \mathbf{I}_{l^2})\tilde{\mathbf{S}}\|^2+\|(\boldsymbol{\mathcal{D}}_f^{\sigma(t)}\otimes \mathbf{I}_{l^2})\tilde{\mathbf{S}}\|^2\big)^{1/2}\\
=&-{\gamma_S}\sqrt{
\tilde{\mathbf{S}}^\top
\left[
\bigl(
(\mathbf{H}_f^{\sigma(t)})^\top\mathbf{H}_f^{\sigma(t)}
+
(\boldsymbol{\mathcal{D}}_f^{\sigma(t)})^\top\boldsymbol{\mathcal{D}}_f^{\sigma(t)}
\bigr)
\otimes
\mathbf{I}_{l^2}
\right]
\tilde{\mathbf{S}}
}\\
=&-{\gamma_S}\sqrt{
\tilde{\mathbf{S}}^\top
(\boldsymbol{\mathcal{H}}^{\sigma(t)}\otimes
\mathbf{I}_{l^2})
\tilde{\mathbf{S}}
}.
\end{aligned}
\end{equation}
Thus, for $t \in [t_r^A+\Delta_r^A,t_{r+1}^A)$, we have
\begin{equation}
\begin{aligned}
\dot{V}_1 \leq -{\gamma_S} \sqrt{\lambda_{\min}^S(\boldsymbol{\mathcal{H}}^{\sigma(t)})}\|\tilde{\mathbf{S}}\|= -{\gamma_S} \sqrt{2\lambda_{\min}^S(\boldsymbol{\mathcal{H}}^{\sigma(t)})}V^{1/2}_1.
\end{aligned}
\end{equation}
Since localizability implies {that every connected component of $\mathcal{G}_f^{\sigma(t)}$ contains at least one follower connected to a leader}, it follows from Lemma~\ref{lmm:connectivity} that $\lambda_{\min}^{S}(\boldsymbol{\mathcal{H}}^{\sigma(t)})>0$. Meanwhile, under localizability-broken attacks, it holds that $\dot{V}_1 \leq 0$. Thus, by following the same procedure as in the proof of Theorem~\ref{thm:single_recovery} with $\kappa = 0$, it can be shown that $V_1$ is bounded and converges to zero in finite time $T_{1,\mathbf{S}}$. Moreover, as $\dot{V}_1 \leq 0$ during localizability-broken attacks, $V_1$ cannot increase after reaching zero and therefore remains at zero thereafter. Similarly, by using the same approach, one can prove the finite-time convergence of $\tilde{\mathbf{F}}_{0i}(t)$, i.e., there exists a finite settling time $T_{1,\mathbf{F}} \geq 0$ such that $\lim_{t\rightarrow T_{1,\mathbf{F}}}\tilde{\mathbf{F}}_{0i}(t) = \mathbf{0}$ and $\tilde{\mathbf{F}}_{0i}(t) = \mathbf{0}$ for $t \geq T_{1,\mathbf{F}}$. Hence, the finite settling time of \eqref{eq:observer_error_a}--\eqref{eq:observer_error_b} can be chosen as $T_1=\max\{T_{1,\mathbf{S}},\,T_{1,\mathbf{F}}\}$.

\textbullet\ \textbf{Part 2:} Boundedness of $\tilde{\boldsymbol{\zeta}}_i(t)$ and $\tilde{\mathbf{h}}_i(t)$.

Let $\tilde{\boldsymbol{\zeta}} = [\tilde{\boldsymbol{\zeta}}_{n_l+1}^\top,\dots,\tilde{\boldsymbol{\zeta}}_{n}^\top]^\top$. From \eqref{eq:observer_error_c}, we obtain
\begin{equation}
\begin{aligned}
\dot{\tilde{\boldsymbol{\zeta}}} =&\ \ \tilde{\mathbf{S}}^d\tilde{\boldsymbol{\zeta}} +\tilde{\mathbf{S}}^d\bar{\boldsymbol{\zeta}}+ \big(\mathbf{I}_{n_f}\otimes \mathbf{S}-{k_\zeta}(\boldsymbol{\mathcal{H}}^{\sigma(t)}\otimes
\mathbf{I}_{l})\big)\tilde{\boldsymbol{\zeta}}\\
& -{\gamma_\zeta}(\mathbf{H}_f^{\sigma(t)}\otimes \mathbf{I}_{l})^\top\mathrm{sgn}\big((\mathbf{H}_f^{\sigma(t)}\otimes \mathbf{I}_{l})\tilde{\boldsymbol{\zeta}}\big)\\
&-{\gamma_\zeta}(\boldsymbol{\mathcal{D}}_f^{\sigma(t)}\otimes \mathbf{I}_{l})^\top\mathrm{sgn}\big((\boldsymbol{\mathcal{D}}_f^{\sigma(t)}\otimes \mathbf{I}_{l})\tilde{\boldsymbol{\zeta}}\big),
\end{aligned}
\end{equation}
where $\tilde{\mathbf{S}}^d = \mathrm{blkdiag}\{\tilde{\mathbf{S}}_{n_l+1},\dots,\tilde{\mathbf{S}}_n\}$ and $\bar{\boldsymbol{\zeta}}=\mathbf{1}_{n_f}\otimes \boldsymbol{\zeta}$. We consider the Lyapunov function {candidate} $V_2 = \frac{1}{2}\tilde{\boldsymbol{\zeta}}^\top\tilde{\boldsymbol{\zeta}}$. Then
\begin{equation}
\begin{aligned}
\dot{V}_2 &\leq \tilde{\boldsymbol{\zeta}}^\top\tilde{\mathbf{S}}^d\tilde{\boldsymbol{\zeta}}+\tilde{\boldsymbol{\zeta}}^\top\tilde{\mathbf{S}}^d\bar{\boldsymbol{\zeta}}+\tilde{\boldsymbol{\zeta}}^\top\big(\mathbf{I}_{n_f}\otimes \mathbf{S}\big)\tilde{\boldsymbol{\zeta}}\\
&\ \ -{k_\zeta}\tilde{\boldsymbol{\zeta}}^\top(\boldsymbol{\mathcal{H}}^{\sigma(t)}\otimes
\mathbf{I}_{l})\tilde{\boldsymbol{\zeta}}-{\gamma_\zeta}\sqrt{
\tilde{\boldsymbol{\zeta}}^\top
(\boldsymbol{\mathcal{H}}^{\sigma(t)}\otimes
\mathbf{I}_{l})
\tilde{\boldsymbol{\zeta}}
}\\
&\leq \tilde{\boldsymbol{\zeta}}^\top\tilde{\mathbf{S}}^d\tilde{\boldsymbol{\zeta}}+\tilde{\boldsymbol{\zeta}}^\top\tilde{\mathbf{S}}^d\bar{\boldsymbol{\zeta}}+\tilde{\boldsymbol{\zeta}}^\top\big(\mathbf{I}_{n_f}\otimes \mathbf{S}\big)\tilde{\boldsymbol{\zeta}}\\
&\leq \tilde{\boldsymbol{\zeta}}^\top\tilde{\mathbf{S}}^d\tilde{\boldsymbol{\zeta}}+\frac{\|\tilde{\boldsymbol{\zeta}}^\top\tilde{\mathbf{S}}^d\|^2}{2}+\frac{\|\bar{\boldsymbol{\zeta}}\|^2}{2} +{\frac{1}{2}\lambda_{\max}(\mathbf{S}+\mathbf{S}^\top)}\|\tilde{\boldsymbol{\zeta}}\|^2.
\end{aligned}
\end{equation}
Here, we have used the Cauchy--Schwarz inequality. By using the inequalities $\tilde{\boldsymbol{\zeta}}^\top\tilde{\mathbf{S}}^d\tilde{\boldsymbol{\zeta}} \leq \|\tilde{\mathbf{S}}^d\|_F\|\tilde{\boldsymbol{\zeta}}\|^2$ and $\|\tilde{\boldsymbol{\zeta}}^\top\tilde{\mathbf{S}}^d\|^2\leq \|\tilde{\mathbf{S}}^d\|_F^2\|\tilde{\boldsymbol{\zeta}}\|^2$ together with the fact that $V_1 = \frac{1}{2}\|\tilde{\mathbf{S}}^d\|_F^2$, we obtain the following
\begin{equation}
\begin{aligned}
\dot{V}_2 &\leq \bigg( \sqrt{2V_1(t)}+V_1(t)+{\frac{1}{2}\lambda_{\max}(\mathbf{S}+\mathbf{S}^\top)}\bigg)\|\tilde{\boldsymbol{\zeta}}\|^2+\frac{\|\bar{\boldsymbol{\zeta}}\|^2}{2}.
\end{aligned}
\end{equation}
Noting that $\dot{V}_1 \leq 0$ and that $\|\boldsymbol{\zeta}(t)\| \leq e^{\|\mathbf{S}\|T_1}\|\boldsymbol{\zeta}(0)\|$ for all $t \in [0,T_1)$, it follows that
\begin{equation}
\dot{V}_2 \leq  \varpi_1 V_2 + \varpi_2,
\end{equation}
where $\varpi_1 = 2\big( \sqrt{2V_1(0)}+V_1(0)+{\frac{1}{2}\lambda_{\max}(\mathbf{S}+\mathbf{S}^\top)}\big)$ and $\varpi_2 = \frac{n_f}{2}e^{2\|\mathbf{S}\|T_1}\|\boldsymbol{\zeta}(0)\|^2$ are two constants. Hence, we conclude that $\tilde{\boldsymbol{\zeta}}_i(t)$ remains bounded over the interval $[0,T_1)$. Furthermore, from $\eqref{eq:observer_error_d}$, it follows that $\tilde{\mathbf{h}}_i(t)$ remains bounded on $[0,T_1)$.

\textbullet\ \textbf{Part 3:} Finite-time convergence of $\tilde{\boldsymbol{\zeta}}_i(t)$ and $\tilde{\mathbf{h}}_i(t)$.

For all $t \geq T_1$, we have $\tilde{\mathbf{S}}_i(t)=\mathbf{0}$ and $\tilde{\mathbf{F}}_i(t) = \mathbf{0}$. Thus, for all $t \geq T_1$, \eqref{eq:observer_error_c}--\eqref{eq:observer_error_d} become
\begin{equation}
\begin{aligned}
\dot{\tilde{\boldsymbol{\zeta}}} &= \big(\mathbf{I}_{n_f}\otimes \mathbf{S}-{k_\zeta}(\boldsymbol{\mathcal{H}}^{\sigma(t)}\otimes
\mathbf{I}_{l})\big)\tilde{\boldsymbol{\zeta}}\\
& \quad-{\gamma_\zeta}(\mathbf{H}_f^{\sigma(t)}\otimes \mathbf{I}_{l})^\top\mathrm{sgn}\big((\mathbf{H}_f^{\sigma(t)}\otimes \mathbf{I}_{l})\tilde{\boldsymbol{\zeta}}\big)\\
&\quad-{\gamma_\zeta}(\boldsymbol{\mathcal{D}}_f^{\sigma(t)}\otimes \mathbf{I}_{l})^\top\mathrm{sgn}\big((\boldsymbol{\mathcal{D}}_f^{\sigma(t)}\otimes \mathbf{I}_{l})\tilde{\boldsymbol{\zeta}}\big),\\ 
\tilde{\mathbf{h}}&=\big(\mathbf{I}_{n_f}\otimes \mathbf{F}\big)\tilde{\boldsymbol{\zeta}},
\end{aligned}
\end{equation}
where $\tilde{\mathbf{h}}=[\tilde{\mathbf{h}}_{n_l+1}^\top,\dots,\tilde{\mathbf{h}}_n^\top]^\top$. Thus, the following holds
\begin{equation}
\begin{aligned}
\dot{V}_2 &\leq \frac{1}{2}\tilde{\boldsymbol{\zeta}}^\top\big(\mathbf{I}_{n_f}\otimes (\mathbf{S}+\mathbf{S}^\top)-2{k_\zeta}(\boldsymbol{\mathcal{H}}^{\sigma(t)}\otimes
\mathbf{I}_{l})\big)\tilde{\boldsymbol{\zeta}}\\
&\quad-{\gamma_\zeta}\sqrt{
\tilde{\boldsymbol{\zeta}}^\top
(\boldsymbol{\mathcal{H}}^{\sigma(t)}\otimes
\mathbf{I}_{l})
\tilde{\boldsymbol{\zeta}}
}.
\end{aligned}
\end{equation}
Under condition $\eqref{eq:DOB_a}$, for $t \in [t_r^A+\Delta_r^A,t_{r+1}^A)$, we have
\begin{equation}
\begin{aligned}
\dot{V}_2 \leq -\omega V_2 -{\gamma_\zeta} \sqrt{2\lambda_{\min}^S(\boldsymbol{\mathcal{H}}^{\sigma(t)})}V_2^{1/2}.
\end{aligned}
\end{equation}
Meanwhile, for $t \in [t_r^A,t_r^A+\Delta_r^A)$, $\eqref{eq:DOB_b}$ implies
\begin{equation}\label{eq:DOB_broken}
\begin{aligned}
\dot{V}_2 \leq \kappa V_2.
\end{aligned}
\end{equation}
Under condition~\eqref{eq:DOB_c}, by following the proof of Theorem~\ref{thm:single_PI}, it holds that $V_2(t) \leq V_2(0)e^{\epsilon(\omega+\kappa)-t(\omega-\frac{\omega+\kappa}{\tau_d})}$ and $V_2(t)\rightarrow 0$ exponentially as $t \to \infty$. Thus, there exists a finite time $T_{21}\ge0$ such that $V_2(t)$ enters the norm ball $\mathcal{B}=\{V_2\le1\}$ and remains therein for all $t\ge T_{21}$. Upon entering $\mathcal{B}$, we have $V_2 \leq V_2^{1/2}$ and thus
\begin{equation}\hspace{-5mm}
V_2^{1/2}(t) \leq
\begin{cases}
V_2^{1/2}(t_r^A) + \dfrac{\kappa}{2} (t - t_r^A),\\
\qquad \qquad \qquad t \in [t_r^A, t_r^A+\Delta_r^A),\\[1mm]
V_2^{1/2}(t_r^A+\Delta_r^A) \\
\quad- \dfrac{{\gamma_\zeta} \sqrt{2\lambda_{\min}^S(\boldsymbol{\mathcal{H}}^{\sigma(t)})}}{2} (t - t_r^A - \Delta_r^A),\\
\ \ \qquad \qquad t \in [t_r^A+\Delta_r^A, t_{r+1}^A).
\end{cases}
\end{equation}
Following the proof of Theorem~\ref{thm:single_recovery} with $\rho = {\gamma_\zeta}\sqrt{2\lambda_{\min}^{S}(\boldsymbol{\mathcal{H}}^{\sigma(t)})}$, condition~\eqref{eq:DOB_d} guarantees that $V_2(t)$ converges to zero in finite time $T_{22}$. Combining this with the asymptotic stability of $V_2(t)$, we conclude that the observer error $\tilde{\boldsymbol{\zeta}}_i(t)$ is finite-time stable with settling time $T_{DOB} = T_1+T_{21}+T_{22}$. Moreover, \eqref{eq:observer_error_d} implies the finite-time stability of $\tilde{\mathbf{h}}_i(t)$. This completes the proof.
\hfill$\blacksquare$
\end{pf}
\begin{remark}
\textcolor{blue}{For any $\tau_d>1$, conditions \eqref{eq:DOB_a}--\eqref{eq:DOB_d} are always feasible. For example, choose $\kappa>\lambda_{\max}(\mathbf{S}+\mathbf{S}^\top)$ to satisfy \eqref{eq:DOB_b}, then $\omega>\kappa/(\tau_d-1)$ to satisfy \eqref{eq:DOB_c}, and finally $k_\zeta$ and $\gamma_\zeta$ sufficiently large to satisfy \eqref{eq:DOB_a} and \eqref{eq:DOB_d}, which is possible since $\lambda_{\min}^{S}(\boldsymbol{\mathcal{H}}^{\sigma})>0$.}
\end{remark}
% \begin{remark}
% Since $\lambda_{\min}^{S}(\boldsymbol{\mathcal{H}}^{\sigma(t)})>0$, it can be seen that conditions \eqref{eq:DOB_a}--\eqref{eq:DOB_b} are always feasible. 
% \end{remark}
\begin{lemma}\label{lm:target_est}
Consider the estimation law~\eqref{eq:target_estimator} under Assumptions~\ref{assume:desired_formation}, \ref{duration}, \ref{assume:exo}, and \ref{assume:communicate}, with $\gamma > 0$ and the observer gains selected as in Lemma~\ref{lm:DOB}. Then, each follower estimates its desired trajectory in finite time, i.e., there exists a finite time $T>0$ such that $\hat{\mathbf{p}}^*_i(t)=\mathbf{p}^*_i(t)$ for all $i$ and all $t\ge T$.
\end{lemma}
\begin{pf}
By denoting $\tilde{\mathbf{p}}^*_i(t)=\hat{\mathbf{p}}^*_i(t)-\mathbf{p}^*_i(t)$, $\tilde{\mathbf{p}}^{F*}(t)=[(\tilde{\mathbf{p}}^*_{n_l+1})^\top,\dots,(\tilde{\mathbf{p}}^*_{n})^\top]^\top$, and $\tilde{\mathbf{p}}^{*}=[\mathbf{0}^\top_{2n_l},(\tilde{\mathbf{p}}^{F*})^\top]^\top$, \eqref{eq:target_estimator} can be rewritten as follows
\begin{equation}
\begin{aligned}
\dot{\tilde{\mathbf{p}}}^*(t) &= 
\begin{bmatrix}
\mathbf{0}_{2n_l}\\
\tilde{\mathbf{h}}
\end{bmatrix}
-\gamma\overline{\mathbf{Z}}\big(\mathbf{R}_{\mathcal{A}}^{\sigma(t)}\big)^\top\mathrm{sgn}(\mathbf{R}_{\mathcal{A}}^{\sigma(t)}\tilde{\mathbf{p}}^{*}).
\end{aligned}
\end{equation}
Consider the following common Lyapunov function {candidate} $V(t)=\frac{1}{2}(\tilde{\mathbf{p}}^{F*})^\top\tilde{\mathbf{p}}^{F*}=\frac{1}{2}(\tilde{\mathbf{p}}^{*})^\top\tilde{\mathbf{p}}^{*}$. Then, $\dot V \in \dot{\tilde V}$ almost everywhere, 
where
\begin{equation}
\begin{aligned}
\dot{\tilde{V}} = (\tilde{\mathbf{p}}^{*})^\top\tilde{\mathbf{h}}-\gamma\|\mathbf{R}_{\mathcal{A}}^{\sigma(t)}\tilde{\mathbf{p}}^{*}\|_1.
\end{aligned}
\end{equation}
Hence, $\dot{\tilde{V}}$ is a singleton and thus $\dot{V}=\dot{\tilde{V}}$. Applying the Cauchy--Schwarz inequality, it holds that $\dot{V} \leq \frac{1}{2}\|\tilde{\mathbf{p}}^{*}\|^2+\frac{1}{2}\|\tilde{\mathbf{h}}\|^2=V(t)+\frac{1}{2}\|\tilde{\mathbf{h}}\|^2$. From the boundedness of $\|\tilde{\mathbf{h}}(t)\|$ over the interval $[0,T_{DOB})$, it follows that $\|\tilde{\mathbf{p}}^{*}(t)\|$ is also bounded over the same interval. For $t \geq T_{DOB}$, we have
$\dot{V}\leq -\gamma\|\mathbf{R}_{\mathcal{A}}^{\sigma(t)}\tilde{\mathbf{p}}^{*}\|_1
\leq -\sqrt{2}\gamma{\underline{\lambda}}^{1/2}V^{1/2}$ during localizability-maintained attacks, whereas $\dot{V}\leq 0$ during localizability-broken attacks. Thus, by following the proof of Part~1 of Lemma~\ref{lm:DOB}, one can conclude that there exists a finite time $T\geq T_{DOB}$ such that $\tilde{\mathbf{p}}^{*}(T)=\mathbf{0}$ and $\tilde{\mathbf{p}}^{*}(t)=\mathbf{0}$ for all $t\geq T$. This completes the proof.
\hfill$\blacksquare$
\end{pf}
\subsection{Resilient Simultaneous Localization and Formation Control Design}
After successfully estimating the desired formation velocity $\mathbf{h}(t)$ and the desired trajectories $\mathbf{p}^{*}(t)$ within a finite time $T$, we now design the simultaneous localization and formation tracking control law for each follower. We propose the following RSLAF algorithm
\begingroup
\begin{equation}\label{eq:RSLAF}
\begin{aligned}
\dot{\hat{\mathbf{p}}}_i(t) &= \mathbf{u}_i(t)-k_P\mathcal{F}_i^{\sigma(t)}(\alpha(t),\hat{\mathbf{p}}(t)),\\
\mathbf{u}_i(t) &= \mathbf{h}_i(t)-{k_\delta}(\hat{\mathbf{p}}_i(t)-\hat{\mathbf{p}}_i^*(t))+k_\psi {\min\{e_{i}(t),\bar e_i\}}\psi_i(t),
\end{aligned}
\end{equation}
\endgroup
where $k_P,{k_\delta}>0$ are control gains, {$k_\psi>0$ is a perturbation gain,} and $e_i = \sum_{{(j,i,k)}\in \mathcal{A}^{\sigma(t)}}{|\alpha_{{jik}}-\alpha_{{jik}}^*|_{2\pi}} \geq 0 $ is the sum of the absolute angle errors associated with agent $i$. {Here, $|\cdot|_{2\pi}$ denotes the distance on the unit circle, i.e., the absolute value after shifting the argument by a multiple of $2\pi$ into $(-\pi,\pi]$. For example, $\alpha=359^\circ$ and $\alpha^*=1^\circ$ yield an error of $2^\circ$.} {For each follower $i$, the constant $\bar e_i=\min_{(j,i,k)\in\mathcal{A}}\min\{|\alpha^*_{jik}|_{2\pi},\,|\alpha^*_{jik}-\pi|_{2\pi}\}>0$ is computed from its own desired angles and measures how far they are from the degenerate values $0$ and $\pi$.} {In addition, the perturbation signal $\psi_i(t)=[\cos(\nu_i t),\sin(\nu_i t)]^\top\in \mathbb{R}^2$ satisfies $\|\psi_i(t)\|=1$, where the frequencies $\nu_i>0$ are distinct odd multiples of a common base frequency}. {The perturbation term is introduced to prevent the formation from becoming trapped at an unlocalizable configuration, i.e., a configuration containing a degenerate triangle at which $\mathbf{L}_{ff}^{\sigma(t)}(\alpha)$ is singular. {Moreover, the saturation at $\bar e_i$ makes the amplitude of follower $i$ equal to $k_\psi\bar e_i$ whenever it belongs to a degenerate triangle.} }

Let $\boldsymbol{\delta}_1 = \mathbf{p}^F - \mathbf{p}^{F*} \in \mathbb{R}^{2n_f}$ denote the tracking error of the followers, and let $\textcolor{blue}{\boldsymbol{\delta}_{2}} = \hat{\mathbf{p}}^F - \mathbf{p}^F\in \mathbb{R}^{2n_f}$ denote their position estimation error, where $\hat{\mathbf{p}}^F = [\hat{\mathbf{p}}_{n_l+1}^{\top},\dots,\hat{\mathbf{p}}_{n}^{\top}]^{\top}$. In addition, define {$\varphi_i(t)= k_\psi$}{$\min\{e_{i}(t),\bar e_i\}$}{$\psi_i(t)$} and $\boldsymbol{\varphi} = [\varphi_{n_l+1}^\top,\dots,\varphi_{n}^\top]^\top$. {By Lemmas~\ref{lm:DOB} and~\ref{lm:target_est}, there exists a finite time $T$ such that $\mathbf{h}_i(t)=\mathbf{h}(t)$ and $\hat{\mathbf{p}}^*_i(t)=\mathbf{p}^*_i(t)$ for all $t\geq T$. For notational simplicity, we shift the time origin to $T$, which can be ensured by running \eqref{eq:distributed_observer}--\eqref{eq:target_estimator} for more than $T$ seconds during the initialization stage.} From \eqref{eq:RSLAF}, the following can be obtained
\begingroup
\begin{equation}\label{eq:RSLAF_closed}
\begin{aligned}
\begin{bmatrix}
\dot{\boldsymbol{\delta}}_1\\
\textcolor{blue}{\dot{\boldsymbol{\delta}}_{2}}
\end{bmatrix}
=-\begin{bmatrix}
    {k_\delta}\mathbf{I}_{2n_f} &{k_\delta}\mathbf{I}_{2n_f}\\
    \mathbf{0} & k_P\mathbf{L}_{ff}^{\sigma(t)}(\alpha)\\
\end{bmatrix}
\begin{bmatrix}
\boldsymbol{\delta}_1\\
\textcolor{blue}{\boldsymbol{\delta}_{2}}
\end{bmatrix}
+
\begin{bmatrix}
\boldsymbol{\varphi}\\
\mathbf{0}
\end{bmatrix}.
\end{aligned}
\end{equation}
\endgroup
{Note that the right-hand side of \eqref{eq:RSLAF_closed} is continuous between switching instants. The following lemma introduces the constant used in the gain condition of our main theorem.}

{\begin{lemma}\label{lm:Le}
{Let $\mathbf{e}=[e_{n_l+1},\dots,e_n]^\top$, where each $e_i$ is computed as in \eqref{eq:RSLAF} with the full angle set $\mathcal{A}$. Moreover, let $q_i$ denote the number of triangles in $\mathcal{A}$ containing follower $i$, and let $d^*$ denote the minimum of $\|\mathbf{p}_i^*-\mathbf{p}_j^*\|$ over all pairs of agents $i,j$ belonging to a common triangle. Define}
\begin{equation}\label{eq:Le_bound}
{\bar L_e=\frac{2\sqrt{2}\,\pi}{d^*}\sqrt{\sum_{i=n_l+1}^{n}q_i^2}.}
\end{equation}
\textcolor{blue}{Then, $\|\boldsymbol{\varphi}(t)\|\leq k_\psi\bar L_e\|\boldsymbol{\delta}_1(t)\|$ holds whenever $\mathbf{p}_i(t)\neq\mathbf{p}_j(t)$ for all $i\neq j$, i.e., no two agents coincide.}
\end{lemma}
\begin{pf} See Appendix~\ref{app:Le}.
\hfill$\blacksquare$
\end{pf}
The bound $\bar L_e$ in Lemma~\ref{lm:Le} depends only on the target formation and the triangle structure, and can thus be computed offline. Let $\boldsymbol{\psi}$ stack the signals $\psi_i$, which satisfy $\dot{\psi}_i=\nu_i\mathcal{R}(\pi/2)\psi_i$. \textcolor{blue}{The following lemma shows that, under the perturbation, no active triangle can remain degenerate while the localization error $\textcolor{blue}{\boldsymbol{\delta}_{2}}$ is frozen at a nonzero bias.}}
\begin{lemma}\label{lm:nondeg}
{Consider \eqref{eq:RSLAF_closed} with a fixed mode $\sigma$, and let $(\boldsymbol{\delta}_1(t),\textcolor{blue}{\boldsymbol{\delta}_{2}}(t),\boldsymbol{\psi}(t))$, $t\in I$, be a solution on an interval $I$ of positive length, along which no two agents coincide and $\textcolor{blue}{\boldsymbol{\delta}_{2}}$ \textcolor{blue}{is a constant vector}. Then, no active triangle of $\sigma$ remains degenerate, i.e., has zero area, for all $t\in I$.}
\end{lemma}
\begin{pf} {See Appendix~\ref{app:nondeg}.}
\hfill$\blacksquare$
\end{pf}

\textcolor{blue}{Since DoS attacks make the closed-loop signals piecewise continuous, we use the following extension of Barbalat's lemma in the proof of Theorem~\ref{thm:RSLAF}.}
\textcolor{blue}{\begin{lemma}\cite{LIN2012JFI}\label{lm:barbalat_extended}
Assume that the piecewise continuous function $\phi:[0,\infty)\to\mathbb{R}$ is right-continuous and right-differentiable at each $t_r$, where $0 \leq t_1 < t_2 < \cdots$, $\lim_{r\to\infty} t_r = \infty$, and the interval between any two consecutive switching instants is bounded below by a positive constant. Further, assume that there exists a constant $\Gamma > 0$ such that $|D_{+}\phi(t)| \leq \Gamma$ for all $t \geq 0$. If the integral $\int_{0}^{\infty}\phi(t)\,dt$ exists, then $\lim_{t\to\infty}\phi(t) = 0$.
\end{lemma}}

% \begin{theorem}\label{thm:RSLAF}
% Consider Problem \ref{prob2}. Suppose that Assumptions~\ref{assume:ANTIP}, \ref{duration}, \ref{assume:exo}--{\ref{assume:nonsing}} hold, and that {the inter-agent distances remain bounded below by the constant $D>0$ used in Lemma~\ref{lm:Le}}. Under the simultaneous localization and formation algorithm (\ref{eq:RSLAF}), if $\mathbb{A}(\mathcal{V}, \mathcal{A}, \mathbf{p}^*)$ is localizable {and the gains satisfy ${k_\delta}>k_\psi L_e$ with $L_e$ defined in \eqref{eq:Le},} then $\mathbf{p}^F(t) \to \mathbf{p}^{F*}(t)$ and $\hat{\mathbf{p}}^F \to \mathbf{p}^F$ as $t \to \infty$.
% \end{theorem}
\begin{theorem}\label{thm:RSLAF}
Consider Problem~\ref{prob2}, and suppose that Assumptions~\ref{assume:desired_formation}, \ref{duration}, and \ref{assume:exo}--\ref{assume:nonsing} hold. Let the RSLAF algorithm~\eqref{eq:RSLAF} be implemented with $\gamma>0$, the observer gains selected as in Lemma~\ref{lm:DOB}, and $k_\delta>k_\psi\bar L_e$, where $\bar L_e$ is given in \eqref{eq:Le_bound}. If all inter-agent distances remain at least some constant $D>0$ for all $t\geq0$, then $\mathbf{p}^F(t) \to \mathbf{p}^{F*}(t)$ and $\hat{\mathbf{p}}^F(t) \to \mathbf{p}^F(t)$ as $t \to \infty$.
\end{theorem}
\begin{pf}
{The proof consists of three parts.

\textbullet\ \textbf{Part 1:} Boundedness of $\boldsymbol{\delta}_1$ and $\textcolor{blue}{\boldsymbol{\delta}_{2}}$.

Consider $V_1=\frac{1}{2}\textcolor{blue}{\boldsymbol{\delta}_{2}}^\top\textcolor{blue}{\boldsymbol{\delta}_{2}}$. From \eqref{eq:RSLAF_closed}, we have
\begin{equation}\label{eq:V1_RSLAF}
\dot V_1=-k_P\textcolor{blue}{\boldsymbol{\delta}_{2}}^\top\mathbf{L}_{ff}^{\sigma(t)}(\alpha)\textcolor{blue}{\boldsymbol{\delta}_{2}}\leq 0.
\end{equation}
Hence, $\|\textcolor{blue}{\boldsymbol{\delta}_{2}}(t)\|\leq\|\textcolor{blue}{\boldsymbol{\delta}_{2}}(0)\|$ for all $t\geq0$, and $\lim_{t\to\infty}V_1(t)$ exists. Next, consider the Lyapunov function candidate $V_2=\frac{1}{2}\boldsymbol{\delta}_1^\top\boldsymbol{\delta}_1$. By Lemma~\ref{lm:Le}, it follows from \eqref{eq:RSLAF_closed} that
\begin{equation}\label{eq:V2_RSLAF}
\dot V_2 \leq -(k_\delta-k_\psi\bar L_e)\|\boldsymbol{\delta}_1\|^2+k_\delta\|\boldsymbol{\delta}_1\|\|\textcolor{blue}{\boldsymbol{\delta}_{2}}\|.
\end{equation}
Since $k_\delta>k_\psi\bar L_e$, $\dot V_2<0$ whenever $\|\boldsymbol{\delta}_1\|>\frac{k_\delta}{k_\delta-k_\psi\bar L_e}\|\textcolor{blue}{\boldsymbol{\delta}_{2}}(0)\|$, which yields
\begin{equation}\label{eq:delta1_bound}
\|\boldsymbol{\delta}_1(t)\|\leq\max\Big\{\|\boldsymbol{\delta}_1(0)\|,\ \frac{k_\delta}{k_\delta-k_\psi\bar L_e}\|\textcolor{blue}{\boldsymbol{\delta}_{2}}(0)\|\Big\},\quad \forall t\geq 0.
\end{equation}
Moreover, \eqref{eq:V2_RSLAF} implies that the $\boldsymbol{\delta}_1$-subsystem is input-to-state stable with respect to $\textcolor{blue}{\boldsymbol{\delta}_{2}}$, so that $\boldsymbol{\delta}_1(t)\to\mathbf{0}$ whenever $\textcolor{blue}{\boldsymbol{\delta}_{2}}(t)\to\mathbf{0}$ \cite{Khalil2002nonlinear}. It therefore remains to show that $\textcolor{blue}{\boldsymbol{\delta}_{2}}(t)\to\mathbf{0}$.

\textbullet\ \textbf{Part 2:} Convergence of $\textcolor{blue}{\boldsymbol{\delta}_{2}}^\top\mathbf{L}_{ff}^{\sigma(t)}(\alpha)\textcolor{blue}{\boldsymbol{\delta}_{2}}$.

Denote $\phi(t) = k_P\textcolor{blue}{\boldsymbol{\delta}_{2}}^\top\mathbf{L}_{ff}^{\sigma(t)}(\alpha)\textcolor{blue}{\boldsymbol{\delta}_{2}}$. From \eqref{eq:V1_RSLAF}, we obtain the following
\begin{equation}
\int_{0}^{\infty}\phi(t)dt\leq V_1(0) - V_1(\infty) < \infty.
\end{equation}
In order to apply Lemma~\ref{lm:barbalat_extended}, we next show that $D_+\phi(t)$ remains bounded everywhere. From Part~1, $\textcolor{blue}{\boldsymbol{\delta}_{2}}$ and $\boldsymbol{\delta}_1$ are bounded, and so is $\mathbf{u}_i$, since $e_i$ and $\psi_i$ are bounded. The matrix $\mathbf{L}_{ff}^{\sigma(t)}(\alpha)$ consists of trigonometric elements, ensuring a uniformly bounded induced norm. Hence, $\textcolor{blue}{\dot{\boldsymbol{\delta}}_{2}}$ is bounded, and so is $2k_P \textcolor{blue}{\boldsymbol{\delta}_{2}}^\top \mathbf{L}_{ff}^{\sigma(t)}(\alpha) \textcolor{blue}{\dot{\boldsymbol{\delta}}_{2}}$.} To assess the boundedness of the remaining term $k_P \textcolor{blue}{\boldsymbol{\delta}_{2}}^\top \dot{\mathbf{L}}_{ff}^{\sigma(t)}(\alpha) \textcolor{blue}{\boldsymbol{\delta}_{2}}$, where $\dot{\mathbf{L}}_{ff}^{\sigma(t)}(\alpha)$ denotes the right-sided time derivative of $\mathbf{L}_{ff}^{\sigma(t)}(\alpha)$ evaluated along the continuous trajectory between switching instants, we need only verify the boundedness of $\frac{d}{dt}\sin(\alpha)$ and $\frac{d}{dt}\cos(\alpha)$. Note that the following holds
\begin{equation}
\sin(\alpha_{kij}) = \mathbf{g}_{ij}^\top \mathcal{R}(\pi/2)\mathbf{g}_{ik}, \quad \cos(\alpha_{kij}) = \mathbf{g}_{ij}^\top \mathbf{g}_{ik}.
\end{equation}
The derivative of the bearing vector $\mathbf{g}_{ij}$ can be obtained as follows \cite{Zhao2019tac}
$$\dot{\mathbf{g}}_{ij} = \frac{\mathbf{I} - \mathbf{g}_{ij}\mathbf{g}_{ij}^\top}{\Vert{}\mathbf{p}_j - \mathbf{p}_i\Vert{}} (\dot{\mathbf{p}}_j - \dot{\mathbf{p}}_i)= \frac{\mathbf{I} - \mathbf{g}_{ij}\mathbf{g}_{ij}^\top}{\Vert{}\mathbf{p}_j - \mathbf{p}_i\Vert{}} (\mathbf{u}_j - \mathbf{u}_i).$$
{Since $\mathbf{u}_i$ is bounded and $\|\mathbf{p}_j-\mathbf{p}_i\|\geq D$}, it follows that $\dot{\mathbf{g}}_{ij}$, $\frac{d}{dt}\sin(\alpha)$, and $\frac{d}{dt}\cos(\alpha)$ are all bounded. Therefore, by invoking Lemma~\ref{lm:barbalat_extended}, we can conclude that {$\phi(t)\to 0$. Since $\|\mathbf{L}_{ff}^{\sigma(t)}(\alpha)\textcolor{blue}{\boldsymbol{\delta}_{2}}\|^2\leq\|\mathbf{L}_{ff}^{\sigma(t)}(\alpha)\|\,\textcolor{blue}{\boldsymbol{\delta}_{2}}^\top\mathbf{L}_{ff}^{\sigma(t)}(\alpha)\textcolor{blue}{\boldsymbol{\delta}_{2}}$, it also follows that $\mathbf{L}_{ff}^{\sigma(t)}(\alpha)\textcolor{blue}{\boldsymbol{\delta}_{2}}\to\mathbf{0}$.

\textbullet\ \textbf{Part 3:} Convergence of $\textcolor{blue}{\boldsymbol{\delta}_{2}}$.

We show that $\textcolor{blue}{\boldsymbol{\delta}_{2}}(t)\to\mathbf{0}$ by contradiction. Suppose that $\|\textcolor{blue}{\boldsymbol{\delta}_{2}}(t)\|\to\mu>0$, where the limit exists since $V_1$ is nonincreasing. Let $\mathbf{x}=(\boldsymbol{\delta}_1,\textcolor{blue}{\boldsymbol{\delta}_{2}},\boldsymbol{\psi})$. Since the angles depend only on relative positions, for a fixed mode $\sigma$, \eqref{eq:RSLAF_closed} is then a time-invariant system in $\mathbf{x}$. 

\emph{(a) Limit solution.} By Corollary~\ref{cor:longest} and Assumption~\ref{assume:dwell}, localizability-maintained modes keep recurring, and each of them remains active for at least $\tau_s$. Since there are finitely many modes, by Dirichlet's principle, there exist a localizability-maintained mode $\bar\sigma$ and a sequence of time instants $\{t_\ell\}_{\ell\in\mathbb{N}}$ with $t_\ell\to\infty$ as $\ell\to\infty$ such that $\sigma(t)=\bar\sigma$ on $[t_\ell,t_\ell+\tau_s]$ for all $\ell\in\mathbb{N}$. Since $\mathbf{x}(t)$ is bounded by Part~1, the Bolzano--Weierstrass theorem implies that, along a subsequence (still denoted by $\{t_\ell\}$ for simplicity), $\mathbf{x}(t_\ell)\to\bar{\mathbf{x}}$ as $\ell\to\infty$. Let $\bar{\mathbf{x}}(t)=(\bar{\boldsymbol{\delta}}_1(t),\textcolor{blue}{\bar{\boldsymbol{\delta}}_{2}}(t),\bar{\boldsymbol{\psi}}(t))$, $t\in[0,\tau_s]$, denote the solution of \eqref{eq:RSLAF_closed} with $\sigma\equiv\bar\sigma$ starting from $\bar{\mathbf{x}}$, and let $\bar\alpha(t)$ denote the angles of the corresponding configuration. Since the angles are smooth functions of the relative positions of non-coinciding agents{, and $|\cdot|_{2\pi}$ and $\min\{\cdot,\bar e_i\}$ are Lipschitz}, the right-hand side of \eqref{eq:RSLAF_closed} with $\sigma\equiv\bar\sigma$ is locally Lipschitz on the set of collision-free states. Hence, continuous dependence of solutions on initial conditions \cite[Thm.~3.4]{Khalil2002nonlinear} yields
\begin{equation*}
\sup_{t\in[0,\tau_s]}\|\mathbf{x}(t_\ell+t)-\bar{\mathbf{x}}(t)\|\to0\quad\text{as }\ell\to\infty.
\end{equation*}
That is, on the windows $[t_\ell,t_\ell+\tau_s]$, the actual solution becomes arbitrarily close to the limit solution $\bar{\mathbf{x}}(t)$ as $\ell\to\infty$. Now fix $t\in[0,\tau_s]$. Since $\sigma=\bar\sigma$ on each window, $\phi(t_\ell+t)=k_P\textcolor{blue}{\boldsymbol{\delta}_{2}}^\top(t_\ell+t)\mathbf{L}_{ff}^{\bar\sigma}(\alpha(t_\ell+t))\textcolor{blue}{\boldsymbol{\delta}_{2}}(t_\ell+t)$ is a continuous function of $\mathbf{x}(t_\ell+t)$, and thus converges to $k_P\textcolor{blue}{\bar{\boldsymbol{\delta}}_{2}}^\top(t)\mathbf{L}_{ff}^{\bar\sigma}(\bar\alpha(t))\textcolor{blue}{\bar{\boldsymbol{\delta}}_{2}}(t)$ as $\ell\to\infty$. On the other hand, $\phi(t_\ell+t)\to0$ by Part~2, since $t_\ell+t\to\infty$. Hence, by the uniqueness of limits, $\textcolor{blue}{\bar{\boldsymbol{\delta}}_{2}}^\top(t)\mathbf{L}_{ff}^{\bar\sigma}(\bar\alpha(t))\textcolor{blue}{\bar{\boldsymbol{\delta}}_{2}}(t)=0$, implying that $\textcolor{blue}{\dot{\bar{\boldsymbol{\delta}}}_{2}}(t)=-k_P\mathbf{L}_{ff}^{\bar\sigma}(\bar\alpha(t))\textcolor{blue}{\bar{\boldsymbol{\delta}}_{2}}(t)=\mathbf{0}$. Consequently, $\textcolor{blue}{\bar{\boldsymbol{\delta}}_{2}}$ is a constant vector with $\|\textcolor{blue}{\bar{\boldsymbol{\delta}}_{2}}\|=\lim_{\ell\to\infty}\|\textcolor{blue}{\boldsymbol{\delta}_{2}}(t_\ell)\|=\mu>0$.

\emph{(b) Analysis of the limit solution $\bar{\mathbf{x}}(t)$.} By (a), $\mathbf{L}_{ff}^{\bar\sigma}(\bar\alpha(t))$ is singular for all $t\in[0,\tau_s]$. Since $\bar\sigma$ is localizability-maintained, this is possible only if at least one active triangle of the limit configuration is degenerate. The set of instants at which a given triangle is degenerate is closed, and the finitely many such sets cover $[0,\tau_s]$. Hence, at least one {active triangle is degenerate on} an interval of positive length. {Since $\textcolor{blue}{\bar{\boldsymbol{\delta}}_{2}}$ is constant, Lemma~\ref{lm:nondeg} proves that it is impossible.}

Therefore, $\textcolor{blue}{\boldsymbol{\delta}_{2}}(t)\to\mathbf{0}$. It then follows from Part~1 that $\boldsymbol{\delta}_1(t)\to\mathbf{0}$.
 This completes the proof.}
\hfill$\blacksquare$
\end{pf}
% {The following corollary gives a condition for collision avoidance.}
\begin{lemma}
Suppose that Assumptions~\ref{assume:desired_formation}, \ref{duration}, and \ref{assume:exo}--\ref{assume:nonsing} hold, that $\gamma>0$ and the observer gains are selected as in Lemma~\ref{lm:DOB}, and that $k_\delta>k_\psi\bar L_e$ with $\bar L_e$ given in \eqref{eq:Le_bound}. If there exists $D>0$ satisfying
\begin{equation}\label{eq:single_collision_RSLAF}
\begin{aligned}
&{\max\Big\{\|\boldsymbol{\delta}_1(0)\|,\ \frac{k_\delta}{k_\delta-k_\psi \bar L_e}\|\textcolor{blue}{\boldsymbol{\delta}_{2}}(0)\|\Big\}}\\
&\qquad\leq\frac{{\min_{i\neq j}\|\mathbf{p}_i^*-\mathbf{p}_j^*\|}-D}{\sqrt{n_f}},
\end{aligned}
\end{equation}
{then $\|\mathbf{p}_i(t)-\mathbf{p}_j(t)\|\geq D$ for all $i\neq j$ and all $t\geq0$.}
\end{lemma}
\begin{pf}
{Suppose, to the contrary, that two agents coincide at some time, and let $t^*$ be the first time at which two agents coincide. On $[0,t^*)$, Part~1 of the proof of Theorem~\ref{thm:RSLAF} yields \eqref{eq:delta1_bound}. Moreover, as in \eqref{eq:pi_pj_PI}, $\|\mathbf{p}_i(t)-\mathbf{p}_j(t)\|\geq\|\mathbf{p}_i^*-\mathbf{p}_j^*\|-\sqrt{n_f}\|\boldsymbol{\delta}_1(t)\|\geq D$ on $[0,t^*)$. By continuity, $\|\mathbf{p}_i(t^*)-\mathbf{p}_j(t^*)\|\geq D>0$, which contradicts the definition of $t^*$. Hence, no collision occurs, and the above bound holds for all $t\geq0$.}
\hfill$\blacksquare$
\end{pf}

\begin{remark}[Discussion on scalability]
In large-scale networks, checking localizability for every attacked topology may be computationally impractical. However, the bounds \eqref{eq:V_single_const_attack}, \eqref{eq:V_single_smc_attack}, and \eqref{eq:DOB_broken} derived for localizability-broken attacks hold for every topology. Hence, it suffices to verify localizability for a subset $\mathcal{S}\subseteq S$ containing the attack-free formation, compute ${\underline{\lambda}}$ and $\lambda_{\min}^{S}(\boldsymbol{\mathcal{H}}^{\sigma(t)})$ over $\mathcal{S}$, and treat all other topologies as localizability-broken, i.e., subject to the corresponding duration constraints. The analysis remains valid for any such $\mathcal{S}$, but a smaller $\mathcal{S}$ imposes duration constraints on more topologies and thus yields more conservative conditions on the admissible attacks.
\end{remark}
%%%%%%%%%%%%%%%%%%%%%%%%%%%%%%%%%%%%
\section{Simulation Results}\label{sec:sim}
\begin{figure}
%{\resizebox{3cm}{!}
\centering
\begin{tikzpicture}[
roundnode/.style={circle, draw=black, thick, minimum size=2mm,inner sep= 0.25mm},
squarednode/.style={rectangle, draw=red!60, fill=red!5, very thick, minimum size=5mm},
]
 \node[roundnode, draw=blue, text=blue]   (v3)   at   (0,0) {$3$};%   {\footnotesize $v_1$};
    \node[roundnode, draw=blue, text=blue]   (v4)   at   (1,0) {$4$};%   {\footnotesize $v_1$};
    \node[roundnode,draw=red, text=red]   (v1)   at   (0,1) {$1$};%   {\footnotesize $v_1$};
    \node[roundnode, draw=red, text=red]   (v2)   at   (1,1) {$2$};%   {\footnotesize $v_1$};
    \node[roundnode, draw=blue, text=blue] (v5) at (-0.5,0.133975) {$5$};
    \node[roundnode, draw=blue, text=blue] (v6) at (1.5, 0.133975) {$6$};
    \node[roundnode, draw=blue, text=blue] (v7) at (0.5, -0.866025) {$7$};
    \draw[-,thick] (v1)--(v2)--(v4)--(v3)--(v1);
    %\draw[-,thick] (v1)--(v5)--(v3);
    \draw[-,thick] (v1)--(v4);
    \draw[-,thick] (v3)--(v5)--(v1);
    \draw[-,thick] (v2)--(v6)--(v4);
    \draw[-,thick] (v3)--(v7)--(v4);
\end{tikzpicture} 
%    \subfloat[Non-infinitesimally bearing rigid framework.]{\includegraphics[width=0.45\columnwidth]{norigid.png}\label{norigid example}}\\
    \caption{Desired formation shape of $\mathbb{A}(\mathcal{V},\mathcal{A},\mathbf{p}^*)$.}
    \label{fig:desired_formation}
\end{figure}
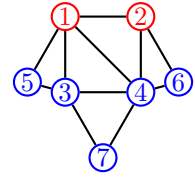
In this section, we present numerical examples to verify the effectiveness of the proposed methods in this paper. Consider a seven-agent formation in $\mathbb{R}^2$ with the leader–follower graph shown in Fig.~\ref{fig:desired_formation}. The formation consists of two leaders (agents 1 and 2) and five followers (agents 3–7). The desired formation shape of $\mathbb{A}(\mathcal{V},\mathcal{A},\mathbf{p}^*)$ satisfying Assumption~\ref{assume:desired_formation} is depicted in Fig.~\ref{fig:desired_formation}. The angle set is $\mathcal{A} = \{(2,4,1),(4,1,2),(1,2,4),(1,4,3),(4,3,1),(3,1,4),(1,3,5),\\(3,5,1),(5,1,3),(2,6,4),(6,4,2),(4,2,6),(3,4,7),(4,7,3),\\(7,3,4)\}$. The nonredundant angle set is $\bar{\mathcal{A}} = \{(2,4,1),\\(1,4,3),(1,3,5),(2,6,4),(3,4,7)\}$. The angle constraints are $\alpha^*_{241}=\alpha^*_{412}=\alpha^*_{143}=\alpha^*_{314}=\frac{\pi}{4},\alpha^*_{124}=\alpha^*_{431}=\frac{\pi}{2},\alpha^*_{264}=\alpha^*_{642}=\alpha^*_{135}=\alpha^*_{351}=\frac{5\pi}{12},\alpha^*_{426}=\alpha^*_{513}=\frac{\pi}{6},\alpha^*_{347}=\alpha^*_{473}=\alpha^*_{734}=\frac{\pi}{3}$. As a result, the following coefficient matrices can be obtained
\begin{equation*}
\scalebox{0.85}{$\displaystyle
\begin{aligned}
\mathbf{A}_4^{\blacktriangle 431}&=\begin{bmatrix}0&-0.7071\\0.7071&0\end{bmatrix},\
\mathbf{A}_3^{\blacktriangle 431}=\begin{bmatrix}0.7071&0.7071\\-0.7071&0.7071\end{bmatrix},\\
\mathbf{A}_1^{\blacktriangle 431}&=\begin{bmatrix}-0.7071&0\\0&-0.7071\end{bmatrix},\
\mathbf{A}_4^{\blacktriangle 412}=\begin{bmatrix}0.5000&-0.5000\\0.5000&0.5000\end{bmatrix},\\
\mathbf{A}_1^{\blacktriangle 412}&=\begin{bmatrix}0.5000&0.5000\\-0.5000&0.5000\end{bmatrix},\
\mathbf{A}_2^{\blacktriangle 412}=\begin{bmatrix}-1&0\\0&-1\end{bmatrix},\\
\mathbf{A}_1^{\blacktriangle 135}&=\begin{bmatrix}0.1294&-0.4830\\0.4830&0.1294\end{bmatrix},\
\mathbf{A}_3^{\blacktriangle 135}=\begin{bmatrix}0.8365&0.4830\\-0.4830&0.8365\end{bmatrix},\\
\mathbf{A}_5^{\blacktriangle 135}&=\begin{bmatrix}-0.9659&0\\0&-0.9659\end{bmatrix},\
\mathbf{A}_6^{\blacktriangle 642}=\begin{bmatrix}0.2500&-0.9330\\0.9330&0.2500\end{bmatrix},\\
\mathbf{A}_4^{\blacktriangle 642}&=\begin{bmatrix}0.2500&0.9330\\-0.9330&0.2500\end{bmatrix},\
\mathbf{A}_2^{\blacktriangle 642}=\begin{bmatrix}-0.5000&0\\0&-0.5000\end{bmatrix},\\
\mathbf{A}_4^{\blacktriangle 473}&=\begin{bmatrix}0.4330&-0.7500\\0.7500&0.4330\end{bmatrix},\
\mathbf{A}_7^{\blacktriangle 473}=\begin{bmatrix}0.4330&0.7500\\-0.7500&0.4330\end{bmatrix},\\
\mathbf{A}_3^{\blacktriangle 473}&=\begin{bmatrix}-0.8660&0\\0&-0.8660\end{bmatrix}.
\end{aligned}$}
\end{equation*}

The following matrices can thus be obtained
\begingroup
\small
\begin{equation*}
\begin{aligned}
&\mathbf{R}_{\mathcal{A}}=\\
&
\begin{bmatrix}
\mathbf{A}_1^{\blacktriangle 431} &\mathbf{0} &\mathbf{A}_3^{\blacktriangle 431} &\mathbf{A}_4^{\blacktriangle 431} &\mathbf{0} &\mathbf{0} &\mathbf{0}\\
\mathbf{A}_1^{\blacktriangle 412} &\mathbf{A}_2^{\blacktriangle 412} &\mathbf{0} &\mathbf{A}_4^{\blacktriangle 412} &\mathbf{0} &\mathbf{0} &\mathbf{0}\\
\mathbf{A}_1^{\blacktriangle 135} &\mathbf{0} &\mathbf{A}_3^{\blacktriangle 135} &\mathbf{0} &\mathbf{A}_5^{\blacktriangle 135} &\mathbf{0} &\mathbf{0}\\
\mathbf{0} &\mathbf{A}_2^{\blacktriangle 642}
&\mathbf{0} &\mathbf{A}_4^{\blacktriangle 642} &\mathbf{0} &\mathbf{A}_6^{\blacktriangle 642} &\mathbf{0}\\
\mathbf{0} &\mathbf{0} &\mathbf{A}_3^{\blacktriangle 473} &\mathbf{A}_4^{\blacktriangle 473} &\mathbf{0} &\mathbf{0} &\mathbf{A}_7^{\blacktriangle 473}
\end{bmatrix},
\end{aligned}
\end{equation*}
\endgroup
\begin{equation*}
\boldsymbol{\mathcal{L}}_f =
\scalebox{0.9}{$\displaystyle
\begin{bmatrix}
3 &-1 &-1 &0 &-1\\
-1 &3 &0 &-1 &-1\\
-1 &0 &1 &0 &0\\
0 &-1 &0 &1 &0\\
-1 &-1 &0 &0 &2
\end{bmatrix}$},\quad
\boldsymbol{\mathcal{D}}_f=\mathrm{diag}(1,1,1,1,0).
\end{equation*}
It can be seen that removing any of the above-listed triangles destroys the rigidity of the angularity, i.e., any attack on the formation constitutes a localizability-broken attack. Hence, we obtain the eigenvalues ${\underline{\lambda}}=0.0819$ and $\lambda_{\min}^{S}(\boldsymbol{\mathcal{H}}^{\sigma(t)})=0.6972$. Note that, since the leaders are assumed to track their desired trajectories ideally, attacks on the communication link between the leaders $e_1=(1,2)=(2,1)$ are not considered.
\subsection{Resilient Formation Tracking With Constant-Velocity Leaders}
\begin{figure}
% \vspace{-5mm}
\centering
\includegraphics[width=\linewidth]{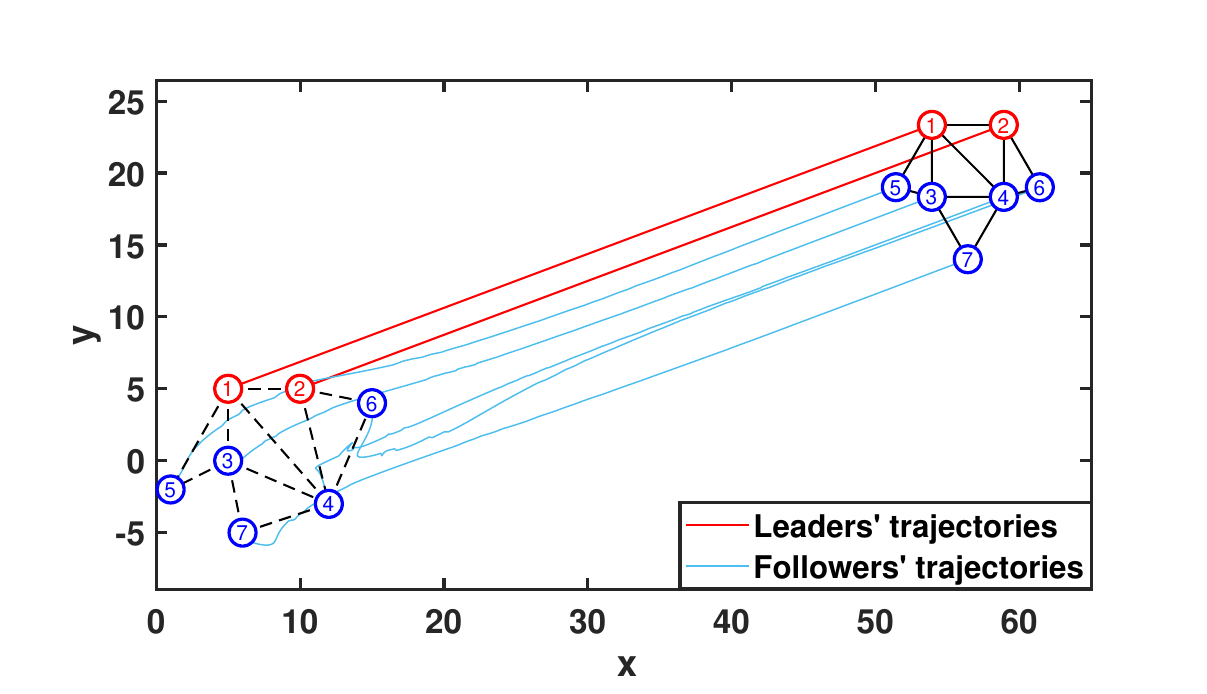}
(a)

\begin{minipage}[t]{0.49\linewidth}
\centering
\includegraphics[width=\linewidth]{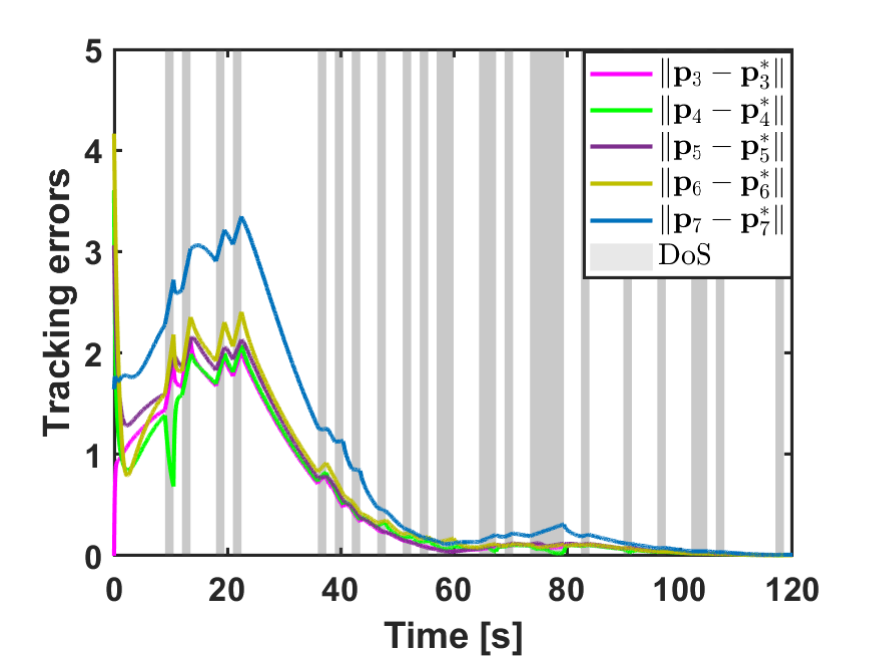}

(b)
\end{minipage}
\hfill
\begin{minipage}[t]{0.49\linewidth}
\centering
\includegraphics[width=\linewidth]{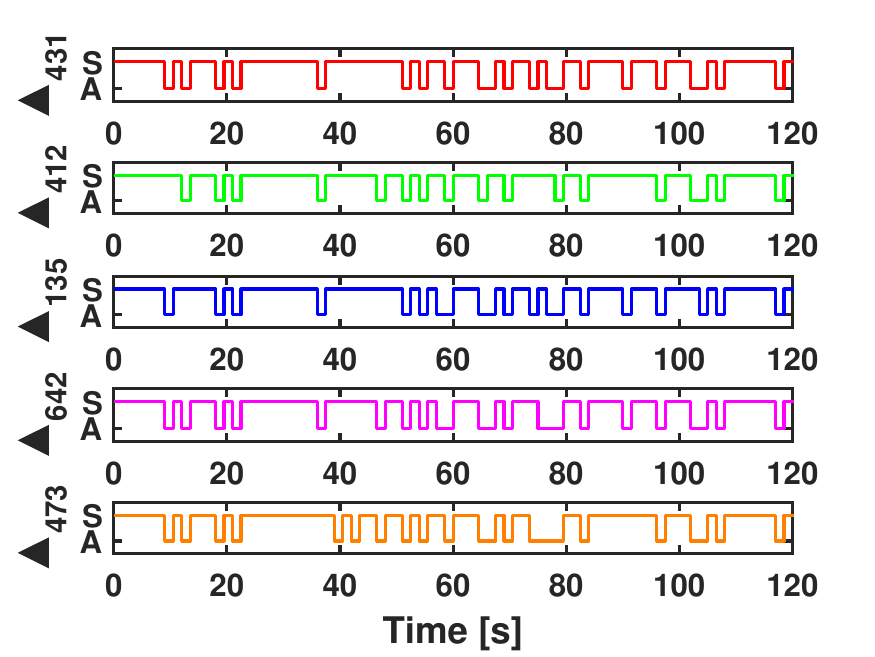}

(c)
\end{minipage}
\caption{Simulation 1: Resilient formation tracking with constant-velocity leaders. (a) The trajectories of the followers and leaders after 120 seconds are colored in blue and red, respectively. (b) The tracking errors of each follower. (c) Attack sequence for different triangles $\blacktriangle{ijk}$, where A and S denote the attacked and safe statuses, respectively.}\label{fig:PI}
\end{figure}

Simulations with moving leaders traveling at a constant velocity are conducted with $\mathbf{h} = [0.3,0.2]^\top$. The leaders are initially positioned at $\mathbf{p}_1(0) = [5,5]^\top$ and $\mathbf{p}_2(0) = [10,5]^\top$. We choose
$k_P = 1.5$ and $k_I = \omega = 0.08 < {\underline{\lambda}}$. It can be verified that these
parameter values satisfy \eqref{eq:main_a} and \eqref{eq:main_b},
respectively. Therefore, by substituting the corresponding values into \eqref{eq:main_c}, we obtain $\tau_d > 1+\frac{2(\sqrt{2}-1)}{2.5-\sqrt{0.25+4}}\approx 2.8895$. In other words, the proposed control law \eqref{eq:single_PI} {ensures resilient tracking of the desired formation against any DoS attack for which the fraction of time during which localizability is broken does not exceed} $\frac{1}{\tau_d}\approx34.61\%$. The simulation results are shown in Fig.~\ref{fig:PI}. The tracking errors of the followers over $120$~s are presented in Fig.~\ref{fig:PI}(b). Under ANTIP, the corresponding attack status of each triangle is depicted in Fig.~\ref{fig:PI}(c). It can be observed that the proposed control law~\eqref{eq:single_PI} successfully ensures asymptotic tracking of the desired follower positions.

\subsection{Resilient Formation Tracking With Maneuvering Leaders}
\begin{figure}
% \vspace{-5mm}
\centering
\includegraphics[width=\linewidth]{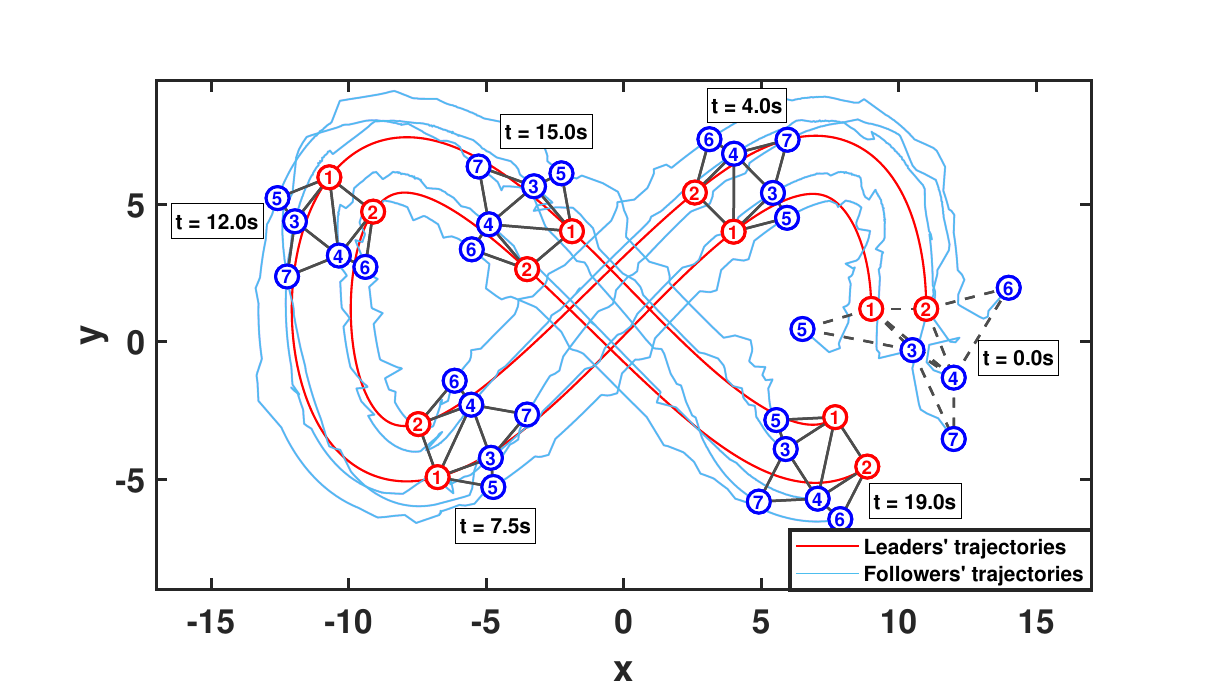}
(a)

\begin{minipage}[t]{0.49\linewidth}
\centering
\includegraphics[width=\linewidth]{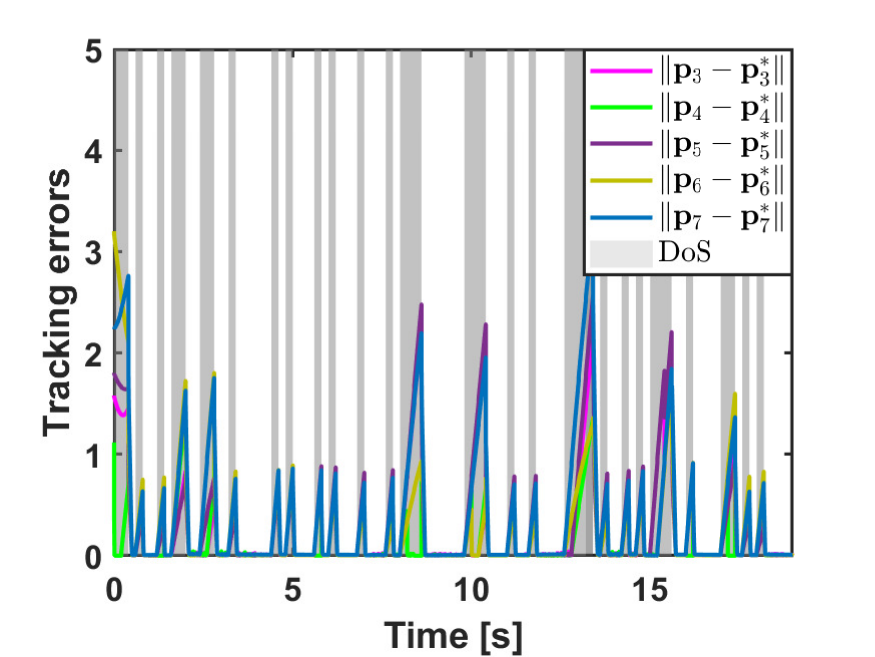}

(b)
\end{minipage}
\hfill
\begin{minipage}[t]{0.49\linewidth}
\centering
\includegraphics[width=\linewidth]{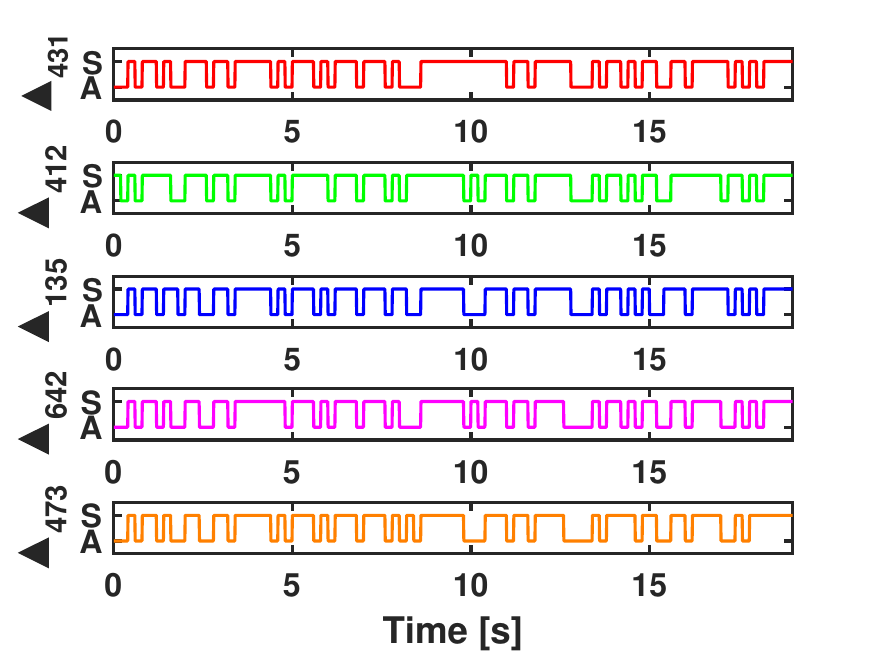}

(c)
\end{minipage}
\caption{Simulation 2: Resilient formation tracking with maneuvering leaders. (a) The trajectories of the followers and leaders after 19 seconds are colored in blue and red, respectively. (b) The tracking errors of each follower. (c) Attack sequence for different triangles $\blacktriangle{ijk}$, where A and S denote the attacked and safe statuses, respectively.}\label{fig:SMC}
\end{figure}

For the maneuvering-leader scenario, leaders are initialized at $\mathbf{p}_1(0) = [9,1.2]^\top$ and $\mathbf{p}_2(0) = [11,1.2]^\top$, and their velocity is given by
\begin{equation*}
\begin{aligned}
\mathbf{v}^L(t) &=  \mathbf{1}_2 \otimes \underbrace{\begin{bmatrix} -3 \cos(0.3t),\, 3 \cos(0.6t) \end{bmatrix}^\top}_{\mathbf{h}(t)}\\
    &+\underbrace{0.05\sin(1.5t)}_{s(t)}\mathbf{I}_{4}(\mathbf{p}^L-\mathbf{1}_{2}\otimes\bar{\mathbf{p}}^L)\\
    &+\underbrace{0.3 \frac{\cos(0.3t) \left( 1 + 2\sin^2(0.3t) \right)}{\cos^2(0.3t) + \cos^2(0.6t)}}_{\theta(t)}\bar{\mathbf{J}}_{2}(\mathbf{p}^L-\mathbf{1}_{2}\otimes\bar{\mathbf{p}}^L),
\end{aligned}
\end{equation*}
where $\bar{\mathbf{p}}^L = \frac{1}{2}(\mathbf{p}_1(t)+\mathbf{p}_2(t))$. An upper bound on the desired followers' velocities can be chosen as $\beta = 5$. With the attack characteristic $\tau_d = 3$, the control law \eqref{eq:single_SMC} is adopted with $\gamma = 100$, satisfying $\gamma > \frac{\tau_d}{\tau_d - 1} \beta \sqrt{\frac{2n}{{\underline{\lambda}}}}$. The simulation results over $19$ seconds are shown in Fig.~\ref{fig:SMC}. As shown in Fig.~\ref{fig:SMC}(b), the followers’ tracking errors deviate from zero during the localizability-broken attack intervals but rapidly converge back to the origin during the subsequent localizability-maintained intervals under the proposed control law \eqref{eq:single_SMC}.

\begin{figure*}[t]
\centering
\begin{minipage}[b]{0.6\linewidth}
\centering
\includegraphics[width=\linewidth]{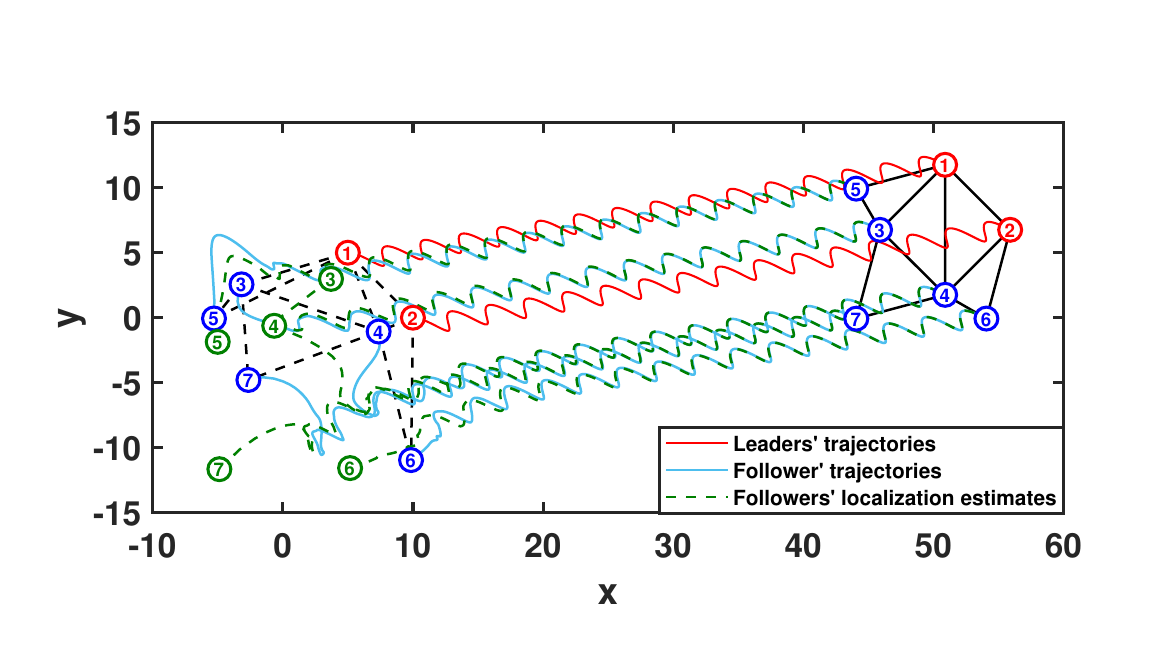}

(a)
\end{minipage}

\vspace{1em} % Adds vertical space between rows
\begin{minipage}[b]{0.25\linewidth}
\centering
\includegraphics[width=\linewidth]{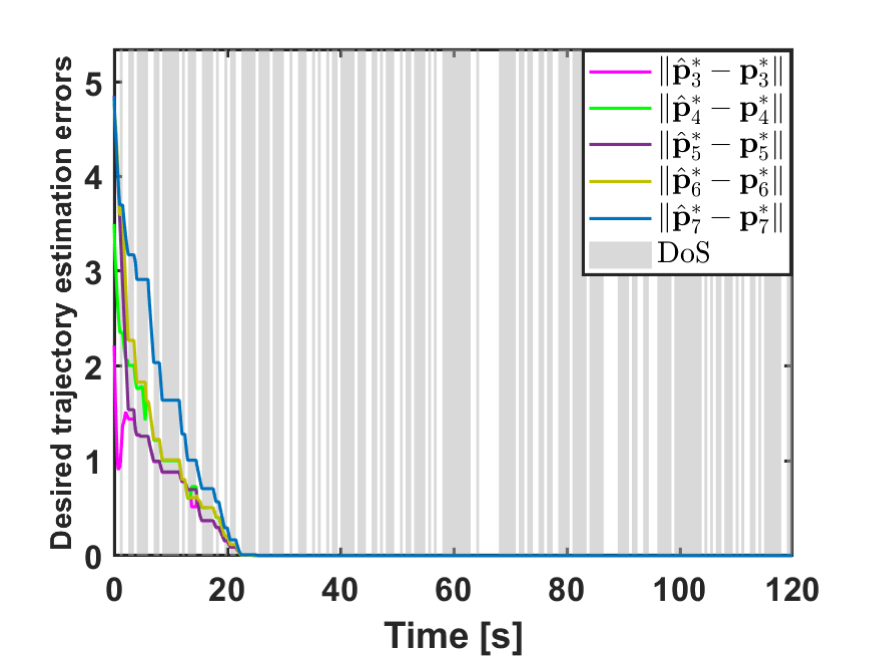}

(b)
\end{minipage}%
\hfill
\begin{minipage}[b]{0.25\linewidth}
\centering
\includegraphics[width=\linewidth]{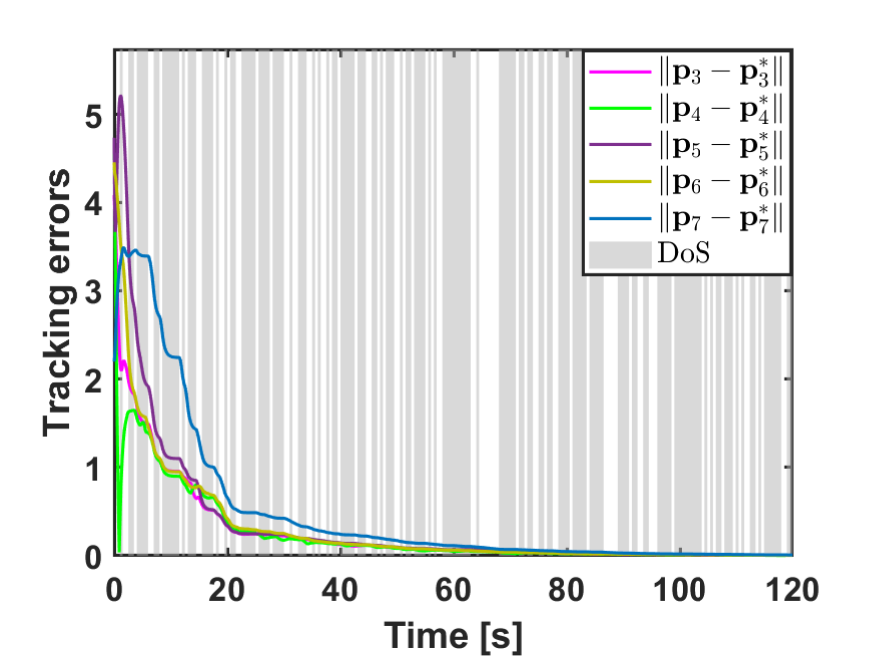}

(c)
\end{minipage}%
\hfill
\begin{minipage}[b]{0.25\linewidth}
\centering
\includegraphics[width=\linewidth]{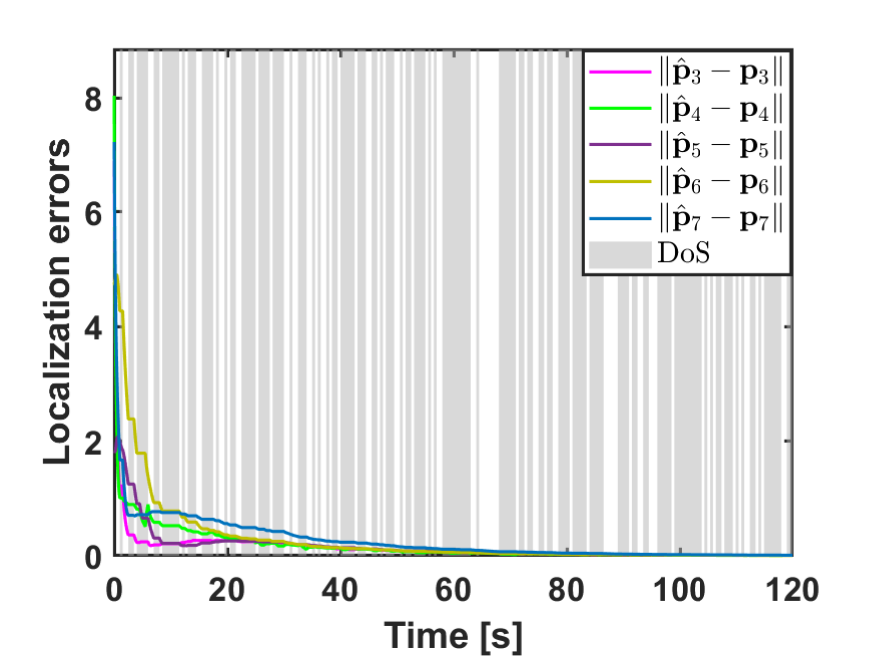}

(d)
\end{minipage}%
\hfill
\begin{minipage}[b]{0.25\linewidth}
\centering
\includegraphics[width=\linewidth]{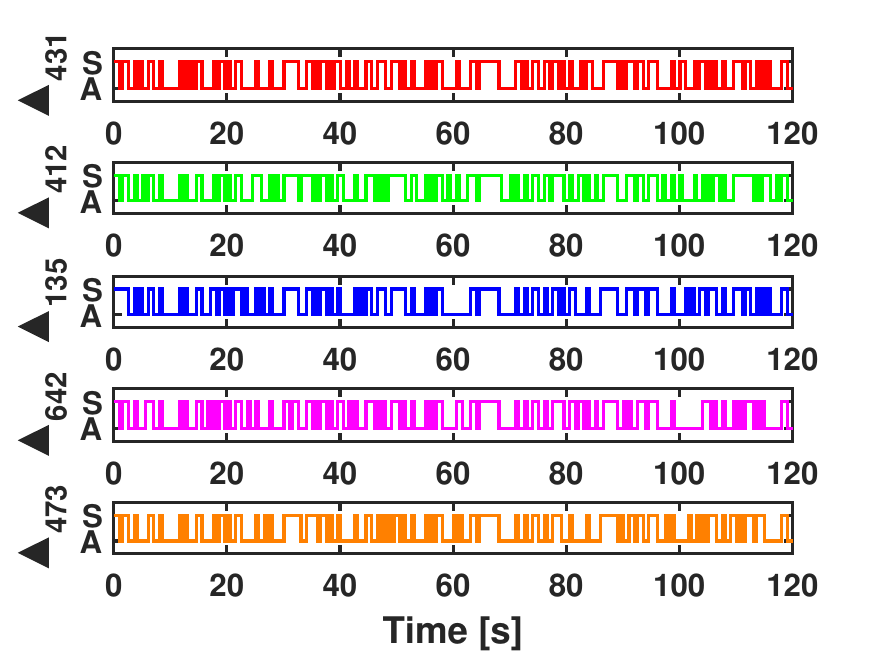}

(e)
\end{minipage}
\caption{Simulation 3, Case 1: Resilient simultaneous localization and formation control. (a) Trajectories of the followers and leaders, together with the followers' localization estimates, after 120 seconds, shown in blue, red, and green, respectively. (b) Target formation estimation errors. (c) Tracking errors of the followers. (d) Localization errors of the followers. (e) Attack sequence for the triangles $\blacktriangle{ijk}$, where A and S denote attacked and safe states, respectively.}
\label{fig:RSLAF}
\end{figure*}
\subsection{Resilient Simultaneous Localization and Formation Control}
In this simulation, the leaders are initialized at $\mathbf{p}_1(0)=[5,5]^\top$ and $\mathbf{p}_2(0)=[10,0]^\top$, while the system matrices of the exosystem are selected as
\begin{equation*}
\mathbf{S} =
\begin{bmatrix}
0 & 0 & 0.1 & 0\\
0 & 0 & 0 & 0\\
0 & 0 & 0 & -0.8\\
0 & 0 & 0.8 & 0
\end{bmatrix},\quad
\mathbf{F} =
\begin{bmatrix}
1 & 0 & -0.5 & 0\\
0 & 1 & 0.5 & 0
\end{bmatrix}.
\end{equation*}
% The control gains are chosen as {$\gamma_S=\gamma_F=k_\zeta=\gamma_\zeta=3$}, and $\gamma = {k_\delta} = k_P = 1$. The perturbation in \eqref{eq:RSLAF} is designed as $\psi_i(t) = [0.5\cos(200\pi t),0.5\sin(200\pi t)]^\top$ for all $i$. With the above control parameters, \eqref{eq:cond_distributed_observer} is satisfied with $\omega = 3$, $\kappa = 1$, and $\tau_d > \frac{4}{3}$. The simulation results are depicted in Fig.~\ref{fig:RSLAF}. As shown in Fig.~\ref{fig:RSLAF}(b), the desired trajectory of each follower is estimated within a finite time. Moreover, Fig.~\ref{fig:RSLAF}(c)--(d) shows that both the formation tracking errors and the localization errors asymptotically converge to the origin. Thus, Problem~\ref{prob2} is solved.
The control gains are chosen as {$\gamma_S=\gamma_F=k_\zeta=\gamma_\zeta=3$}, $\gamma=k_P=1$, and {$k_\delta=2$}. {The perturbation in \eqref{eq:RSLAF} is designed with $k_\psi=0.12$ and $\psi_i(t)=[\cos(\nu_i t),\sin(\nu_i t)]^\top$, where $(\nu_3,\nu_4,\nu_5,\nu_6,\nu_7)=(0.5,1.5,2.5,3.5,4.5)$~rad/s are distinct odd multiples of the base frequency $0.5$~rad/s, and $(\bar e_3,\bar e_4,\bar e_5,\bar e_6,\bar e_7)=(\pi/3,\pi/4,5\pi/12,5\pi/12,\pi/3)$ are computed from the desired angles. From \eqref{eq:Le_bound}, $\bar L_e\approx12.8$ for this formation, so that $k_\delta=2>k_\psi\bar L_e$.} With the above control parameters, \eqref{eq:cond_distributed_observer} is satisfied with $\omega = 3$, $\kappa = 1$, and $\tau_d > \frac{4}{3}$. The simulation results are depicted in Fig.~\ref{fig:RSLAF}. As shown in Fig.~\ref{fig:RSLAF}(b), the desired trajectory of each follower is estimated within a finite time. Moreover, Fig.~\ref{fig:RSLAF}(c)--(d) show that both the formation tracking errors and the localization errors asymptotically converge to the origin.

\begin{figure}
% \vspace{-5mm}
\centering
\includegraphics[width=\linewidth]{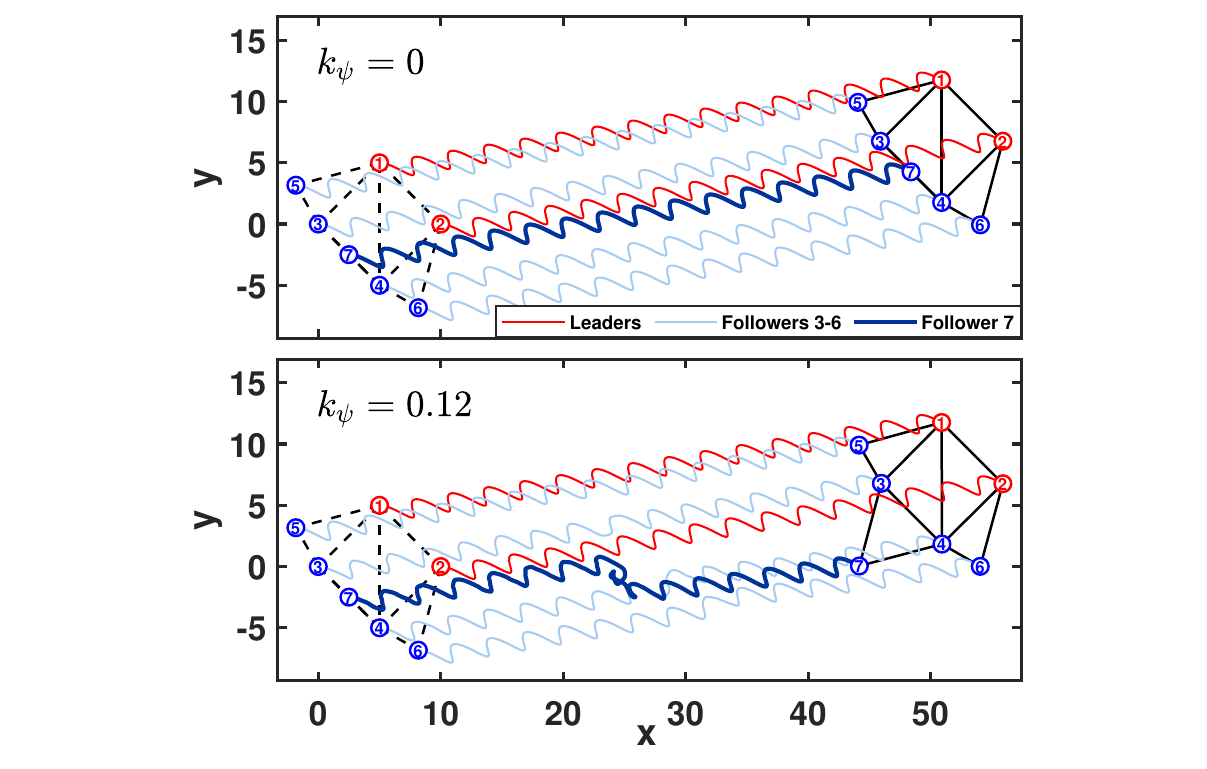}
(a)

\begin{minipage}[t]{0.49\linewidth}
\centering
\includegraphics[width=\linewidth]{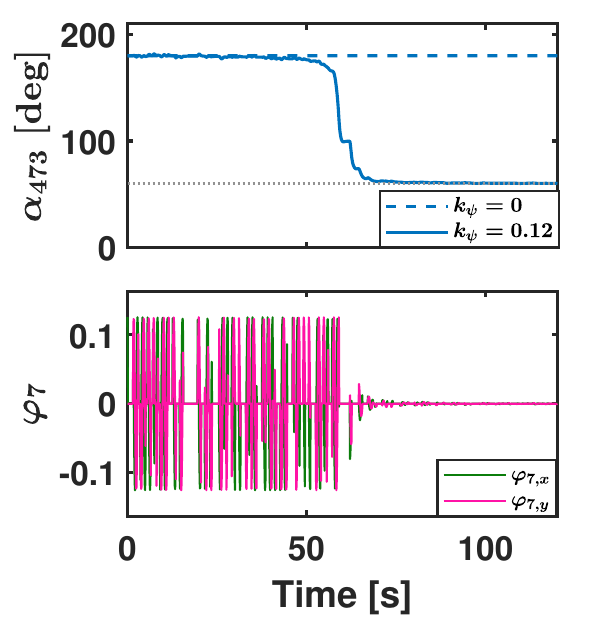}

(b)
\end{minipage}
\hfill
\begin{minipage}[t]{0.49\linewidth}
\centering
\includegraphics[width=\linewidth]{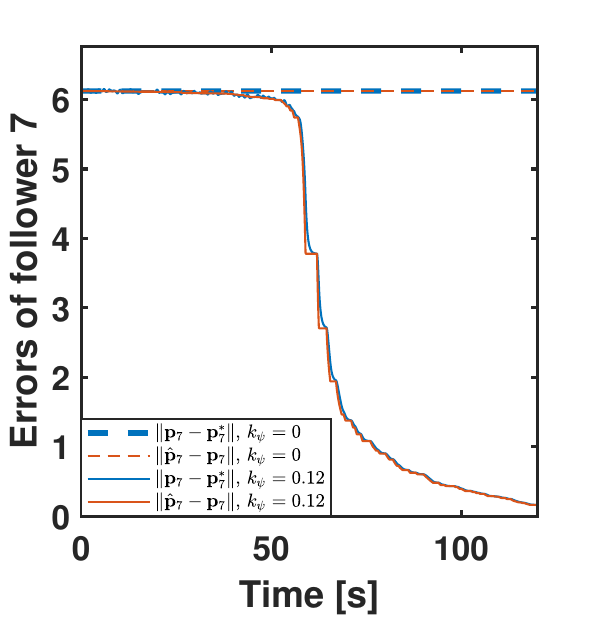}

(c)
\end{minipage}
\caption{\textcolor{blue}{Simulation 3, Case 2: Escape from a degenerate configuration. (a) Trajectories over 120 seconds with $k_\psi=0$ (top) and $k_\psi=0.12$ (bottom), where the leaders, followers 3--6, and follower 7 are colored in red, light blue, and dark blue, respectively. (b) Angle $\alpha_{473}$ of follower 7 (top) and perturbation $\boldsymbol{\varphi}_7$ for $k_\psi=0.12$ (bottom). (c) Tracking error $\|\mathbf{p}_7-\mathbf{p}_7^*\|$ and localization error $\|\hat{\mathbf{p}}_7-\mathbf{p}_7\|$ of follower 7, where dashed and solid lines correspond to $k_\psi=0$ and $k_\psi=0.12$, respectively.}}\label{fig:perturbation}
\end{figure}

\textcolor{blue}{An additional simulation is conducted to examine the role of the perturbation in~\eqref{eq:RSLAF} when an active triangle is degenerate. All settings are the same as in the previous simulation, except that followers 3--6 start at their desired positions and follower~7 starts at the midpoint of followers 3 and 4, so that $\blacktriangle{347}$ is collinear at $t=0$. Moreover, $\hat{\mathbf{p}}_i(0)=\hat{\mathbf{p}}_i^*(0)=\mathbf{p}_i^*(0)$ for all followers, and all estimates of the distributed observer are initialized at their true values. Two runs with $k_\psi=0$ and $k_\psi=0.12$ are conducted under the same DoS sequence, and the results are shown in Fig.~\ref{fig:perturbation}. With $k_\psi=0$, follower~7 stays on the segment between followers 3 and 4, $\alpha_{473}$ remains at $\pi$, and both errors of follower~7 stay at $6.12$. With $k_\psi=0.12$, the saturated perturbation $\boldsymbol{\varphi}_7$ pushes follower~7 out of the collinear position. Subsequently, $\alpha_{473}$ converges to $\pi/3$, $\boldsymbol{\varphi}_7$ vanishes, and both errors converge to zero despite DoS attacks. This verifies Lemma~\ref{lm:nondeg}.}
\section{Conclusion}\label{sec:conclude}
In this paper, resilient angle-constrained formation tracking and localization under DoS attacks were investigated. An ANTIP protocol was proposed to preserve the well-definedness of angularity localizability, followed by a classification of DoS attacks into localizability-maintained and localizability-broken attacks. Two resilient control laws were developed for constant and time-varying leader velocities. In addition, an RSLAF scheme with a finite-time distributed observer was further developed using inter-agent angle measurements and communications. Theoretical analysis and numerical simulations verified the convergence and resilience of the proposed methods. Future research includes the consideration of agent dynamic models and formation control with relative localization based on local distance, angle, or bearing measurements.
% \clearpage
\appendix
\section{Proof of Lemma~\ref{lm:Le}}\label{app:Le}
Consider any collision-free configuration $\mathbf{p}$ at $t=0$, and let $\mathbf{p}^*=\mathbf{p}^*(0)$ and $\boldsymbol{\delta}_1=\mathbf{p}^F-\mathbf{p}^{F*}$. Denote $\mathbf{z}_{ij}=\mathbf{p}_j-\mathbf{p}_i$, $\mathbf{z}^*_{ij}=\mathbf{p}^*_j-\mathbf{p}^*_i$, the bearings $\mathbf{g}_{ij}=\mathbf{z}_{ij}/\|\mathbf{z}_{ij}\|$, $\mathbf{g}^*_{ij}=\mathbf{z}^*_{ij}/\|\mathbf{z}^*_{ij}\|$, and let $\measuredangle(\cdot,\cdot)\in[0,\pi]$ denote the unsigned angle between two nonzero vectors. Unlike the interior angle $\alpha_{kij}\in[0,2\pi)$ in \eqref{eq:angle_def}, which is measured counterclockwise, $\measuredangle(\cdot,\cdot)$ has no orientation and is symmetric in its arguments. First, we introduce the relationship between angle errors and bearing errors. By \eqref{eq:angle_def}, $\mathbf{g}_{ij}=\mathcal{R}(\alpha_{kij})\mathbf{g}_{ik}$ and $\mathbf{g}^*_{ij}=\mathcal{R}(\alpha^*_{kij})\mathbf{g}^*_{ik}$. Since rotations preserve angles and $\measuredangle(\cdot,\cdot)$ satisfies the triangle inequality, we have
\begin{equation}\label{eq:angle_err}
\begin{aligned}
|\alpha_{kij}-\alpha^*_{kij}|_{2\pi}&=\measuredangle\big(\mathcal{R}(\alpha_{kij})\mathbf{g}_{ik},\mathcal{R}(\alpha^*_{kij})\mathbf{g}_{ik}\big)\\
&\leq\measuredangle(\mathbf{g}_{ij},\mathbf{g}^*_{ij})+\measuredangle\big(\mathcal{R}(\alpha^*_{kij})\mathbf{g}^*_{ik},\mathcal{R}(\alpha^*_{kij})\mathbf{g}_{ik}\big)\\
&=\measuredangle(\mathbf{g}_{ij},\mathbf{g}^*_{ij})+\measuredangle(\mathbf{g}_{ik},\mathbf{g}^*_{ik}).
\end{aligned}
\end{equation}
In other words, each angle error is bounded by the bearing errors of its two edges. Second, we relate the bearing errors to the position errors. Since $\sin x\geq 2x/\pi$ on $[0,\pi/2]$, $\|\mathbf{g}_{ij}-\mathbf{g}^*_{ij}\|=2\sin\frac{\measuredangle(\mathbf{g}_{ij},\mathbf{g}^*_{ij})}{2}\geq\frac{2}{\pi}\measuredangle(\mathbf{g}_{ij},\mathbf{g}^*_{ij})$. Moreover, $\|\mathbf{g}_{ij}-\mathbf{g}^*_{ij}\|\leq2\|\mathbf{z}_{ij}-\mathbf{z}^*_{ij}\|/\|\mathbf{z}^*_{ij}\|$ by the Dunkl--Williams inequality \cite{dunkl1964simple}. Since the leaders' errors vanish, $\|\mathbf{z}_{ij}-\mathbf{z}^*_{ij}\|\leq\|\mathbf{p}_j-\mathbf{p}^*_j\|+\|\mathbf{p}_i-\mathbf{p}^*_i\|\leq\sqrt{2}\|\boldsymbol{\delta}_1\|$, and $\|\mathbf{z}^*_{ij}\|\geq d^*$ for $i,j$ in a common triangle. Hence,
\begin{equation*}
\measuredangle(\mathbf{g}_{ij},\mathbf{g}^*_{ij})\leq\pi\frac{\|\mathbf{z}_{ij}-\mathbf{z}^*_{ij}\|}{\|\mathbf{z}^*_{ij}\|}\leq\frac{\sqrt{2}\,\pi}{d^*}\|\boldsymbol{\delta}_1\|.
\end{equation*}
Together with \eqref{eq:angle_err}, it follows that $|\alpha_{kij}-\alpha^*_{kij}|_{2\pi}\leq\frac{2\sqrt{2}\pi}{d^*}\|\boldsymbol{\delta}_1\|$.
Follower $i$ is the vertex of exactly one angle in each of the $q_i$ triangles containing it. Hence, $e_i\leq\frac{2\sqrt{2}\pi}{d^*}q_i\|\boldsymbol{\delta}_1\|$ and $\|\mathbf{e}\|\leq\bar L_e\|\boldsymbol{\delta}_1\|$ for all such $\mathbf{p}$. Since {$\|\varphi_i\|=k_\psi\min\{e_i,\bar e_i\}\leq k_\psi e_i$}, we have $\|\boldsymbol{\varphi}(t)\|\leq k_\psi\|\mathbf{e}(\mathbf{p}(t))\|$, since $e_i$ in \eqref{eq:RSLAF} sums nonnegative terms over $\mathcal{A}^{\sigma(t)}\subseteq\mathcal{A}$. It remains to relate $\|\mathbf{e}(\mathbf{p}(t))\|$ to $\bar L_e$, since the above bound was derived with the target formation frozen at $\mathbf{p}^*(0)$. Under Assumption~\ref{assume:exo}, the target formation only translates with the common velocity $\mathbf{h}(t)$, i.e., $\mathbf{p}^*(t)=\mathbf{p}^*(0)+\mathbf{1}_n\otimes\int_0^t\mathbf{h}(\tau)d\tau$. Shifting all agents by $-\int_0^t\mathbf{h}(\tau)d\tau$ places the leaders at $\mathbf{p}^{L*}(0)$ and preserves all inter-agent distances and angles, and $\boldsymbol{\delta}_1(t)$ remains unchanged. Hence, whenever no two agents coincide, the shifted configuration satisfies the conditions above, which yields $\|\mathbf{e}(\mathbf{p}(t))\|\leq\bar L_e\|\boldsymbol{\delta}_1(t)\|$. Therefore, $\|\boldsymbol{\varphi}(t)\|\leq k_\psi\bar L_e\|\boldsymbol{\delta}_1(t)\|$, which completes the proof.
\hfill$\blacksquare$
\section{Proof of Lemma~\ref{lm:nondeg}}\label{app:nondeg}
{Suppose, to the contrary, that an active triangle $\blacktriangle{ijk}$ remains degenerate for all $t\in I$. By ANTIP, the three angles of $\blacktriangle{ijk}$ are measured, and each equals $0$ or $\pi$. Let $i$ be any follower vertex of $\blacktriangle{ijk}$. Since the desired angles at vertex $i$ differ from $0$ and $\pi$ by at least $\bar e_{i}$, we have $e_{i}\geq\bar e_{i}$, and the perturbation of follower $i$ equals $k_\psi\bar e_{i}\psi_{i}$ on $I$. Let $t_0$ denote the left endpoint of $I$. Since $\textcolor{blue}{\boldsymbol{\delta}_{2}}$ is constant, $\hat{\mathbf{p}}_{i}-\mathbf{p}_{i}$ is constant. Hence, \eqref{eq:RSLAF} and $\dot{\mathbf{p}}^*_{i}=\mathbf{h}$ give $\dot{\mathbf{y}}_{i}=-{k_\delta}\mathbf{y}_{i}+k_\psi\bar e_{i}\psi_{i}$, where $\mathbf{y}_{i}=\mathbf{p}_{i}-\mathbf{p}^*_{i}+\hat{\mathbf{p}}_{i}-\mathbf{p}_{i}$. Using $\dot\psi_{i}=\nu_{i}\mathcal{R}(\pi/2)\psi_{i}$, the solution of this linear system is
\begin{equation*}
\begin{aligned}
\mathbf{y}_{i}(t)&=e^{-{k_\delta}(t-t_0)}\mathbf{y}_{i}(t_0)+k_\psi\bar e_{i}\int_{t_0}^{t}e^{-{k_\delta}(t-s)}\psi_{i}(s)\,ds\\
&=\mathbf{w}^{\mathrm{sine}}_{i}(t)+e^{-{k_\delta}(t-t_0)}\big(\mathbf{y}_{i}(t_0)-\mathbf{w}^{\mathrm{sine}}_{i}(t_0)\big),
\end{aligned}
\end{equation*}
where $\mathbf{w}^{\mathrm{sine}}_{i}(t)=k_\psi\bar e_{i}\big({k_\delta}\mathbf{I}_2+\nu_{i}\mathcal{R}(\pi/2)\big)^{-1}\psi_{i}(t)=\mathcal{R}\big(\nu_{i}(t-t_0)\big)\mathbf{w}^{\mathrm{sine}}_{i}(t_0)$ is a circular motion at frequency $\nu_{i}$ with radius $\|\mathbf{w}^{\mathrm{sine}}_{i}\|=k_\psi\bar e_{i}/\sqrt{{k_\delta^2}+\nu_{i}^2}>0$. Moreover, $\mathbf{p}^*_{i}$ and all leaders undergo the same translation $\int_{t_0}^t\mathbf{h}(\tau)d\tau$ over $[t_0,t]$. Subtracting this translation from all agents, which affects neither angles nor areas, makes $\mathbf{p}^*_{i}$ and all leader positions constant. Hence,
\begin{equation*}
\mathbf{p}_{i}(t)=\mathbf{w}^{\mathrm{const}}_{i}+\mathbf{w}^{\mathrm{sine}}_{i}(t)+e^{-{k_\delta}(t-t_0)}\mathbf{w}^{\mathrm{decay}}_{i},
\end{equation*}
where $\mathbf{w}^{\mathrm{const}}_{i}=\mathbf{p}^*_{i}-(\hat{\mathbf{p}}_{i}-\mathbf{p}_{i})$ and $\mathbf{w}^{\mathrm{decay}}_{i}=\mathbf{y}_{i}(t_0)-\mathbf{w}^{\mathrm{sine}}_{i}(t_0)$ are constant vectors.
Since the leaders have already reached their desired positions, $\blacktriangle{ijk}$ contains at least one follower, and three cases arise. Consider the first case in which $i$ and $j$ are followers with $\nu_{i}<\nu_{j}$, and $k$ is a leader. The area of $\blacktriangle{ijk}$ can be obtained as
\begin{equation*}
\begin{aligned}
&2\operatorname{area}(\blacktriangle{ijk})=\mathbf{z}_{ki}^\top\mathcal{R}(\pi/2)\mathbf{z}_{kj}\\
&=(\mathbf{w}^{\mathrm{const}}_{i}-\mathbf{p}_k)^\top\mathcal{R}(\pi/2)(\mathbf{w}^{\mathrm{const}}_{j}-\mathbf{p}_k)\\
&\quad+(\mathbf{w}^{\mathrm{const}}_{i}-\mathbf{p}_k)^\top\mathcal{R}(\pi/2)\mathbf{w}^{\mathrm{sine}}_{j}(t)\\
&\quad+\big(\mathbf{w}^{\mathrm{sine}}_{i}(t)\big)^\top\mathcal{R}(\pi/2)(\mathbf{w}^{\mathrm{const}}_{j}-\mathbf{p}_k)\\
&\quad+\big(\mathbf{w}^{\mathrm{sine}}_{i}(t)\big)^\top\mathcal{R}(\pi/2)\mathbf{w}^{\mathrm{sine}}_{j}(t)+\text{decaying terms},
\end{aligned}
\end{equation*}
where the decaying terms are the products involving $e^{-{k_\delta}(t-t_0)}\mathbf{w}^{\mathrm{decay}}_{i}$ or $e^{-{k_\delta}(t-t_0)}\mathbf{w}^{\mathrm{decay}}_{j}$, i.e., constants or sinusoids of frequency $\nu_{i}$ or $\nu_{j}$ multiplied by $e^{-{k_\delta}(t-t_0)}$ or $e^{-2{k_\delta}(t-t_0)}$. The first term is constant, and the second and third terms are sinusoids of frequencies $\nu_{j}$ and $\nu_{i}$, respectively. In particular, the fourth term equals
\begin{align*}
&\big(\mathbf{w}^{\mathrm{sine}}_{i}(t_0)\big)^\top\mathcal{R}\big(-\nu_{i}(t-t_0)\big)\mathcal{R}(\pi/2)\mathcal{R}\big(\nu_{j}(t-t_0)\big)\mathbf{w}^{\mathrm{sine}}_{j}(t_0)\\
&=\big(\mathbf{w}^{\mathrm{sine}}_{i}(t_0)\big)^\top\mathcal{R}\big(\pi/2+(\nu_{j}-\nu_{i})(t-t_0)\big)\mathbf{w}^{\mathrm{sine}}_{j}(t_0),
\end{align*}
which is a sinusoid of frequency $\nu_{j}-\nu_{i}$ with amplitude $\|\mathbf{w}^{\mathrm{sine}}_{i}\|\|\mathbf{w}^{\mathrm{sine}}_{j}\|$. Since $\nu_{i}$ and $\nu_{j}$ are odd multiples of the base frequency, $\nu_{j}-\nu_{i}$ is a positive even multiple of it. Hence, $\nu_{j}-\nu_{i}\neq0$, and $\nu_{j}-\nu_{i}$ differs from both $\nu_{i}$ and $\nu_{j}$. Thus, the fourth term is the only term of frequency $\nu_{j}-\nu_{i}$ in $\operatorname{area}(\blacktriangle{ijk})$. Because complex exponential functions with distinct exponents are linearly independent on any interval of positive length, $\operatorname{area}(\blacktriangle{ijk})=0$ on $I$ would require $\|\mathbf{w}^{\mathrm{sine}}_{i}\|\|\mathbf{w}^{\mathrm{sine}}_{j}\|=0$. This is impossible, since both radii are positive. The other two cases follow in the same way.
\begin{itemize}
\item If $i$, $j$, and $k$ are followers with $\nu_{i}<\nu_{j}<\nu_{k}$, the term $-\big(\mathbf{w}^{\mathrm{sine}}_{i}(t)\big)^\top\mathcal{R}(\pi/2)\mathbf{w}^{\mathrm{sine}}_{k}(t)$ is the only term of frequency $\nu_{k}-\nu_{i}$, since this frequency exceeds $\nu_{j}-\nu_{i}$ and $\nu_{k}-\nu_{j}$ and, being an even multiple, differs from $\nu_{i}$, $\nu_{j}$, and $\nu_{k}$. Its amplitude $\|\mathbf{w}^{\mathrm{sine}}_{i}\|\|\mathbf{w}^{\mathrm{sine}}_{k}\|$ is positive.
\item If $i$ is the only follower, the only term of frequency $\nu_{i}$ is $\big(\mathbf{w}^{\mathrm{sine}}_{i}(t)\big)^\top\mathcal{R}(\pi/2)(\mathbf{p}_j-\mathbf{p}_k)$, whose amplitude $\|\mathbf{w}^{\mathrm{sine}}_{i}\|\|\mathbf{p}_j-\mathbf{p}_k\|$ is positive.
\end{itemize}
This contradiction completes the proof.}
\hfill$\blacksquare$
\bibliographystyle{abbrv}
%\newpage
\bibliography{ref}

\end{document}